%% file: paper-060725.tex
\documentclass[prd,aps,showpacs,floats,floatfix,superscriptaddress]{revtex4}

\usepackage[dvips]{graphicx}

\usepackage{longtable}
\usepackage{dcolumn,epsfig}

\def\laq{\raise 0.4ex\hbox{$<$}\kern -0.8em\lower 0.62ex\hbox{$\sim$}}
\def\gaq{\raise 0.4ex\hbox{$>$}\kern -0.7em\lower 0.62ex\hbox{$\sim$}}

\newcommand{\beq}{\begin{equation}}
\newcommand{\eeq}{\end{equation}}
\newcommand{\bea}{\begin{eqnarray}} 
\newcommand{\eea}{\end{eqnarray}}
\newcommand{\ba}{\begin{array}}
\newcommand{\ea}{\end{array}}

\newcommand{\pa}{\partial} 
\newcommand{\vphi}{\varphi} 

\newcommand{\eqref}[1]{(\ref{#1})}

\newcommand{\ket}[1]{{#1}}
\newcommand{\braket}[2]{{\langle #1,#2 \rangle}}

\def\np{({\bf n}\cdot{\bf p})}
\def\pp{{\bf p}^2}
\def\ppp{({\bf p}^2)}

\newcommand{\mytextrm}[1]{{}}

\newlength{\sizeonefig}
\newlength{\sizetwofig}
\newlength{\sizeonefigb}
\newlength{\sizetwofigb}
\begin{document}

\title{Detection template families for gravitational waves from the
final stages of binary--black-hole inspirals: Nonspinning case}

\author{Alessandra Buonanno} 

\affiliation{Institut d'Astrophysique de Paris (GReCO, FRE 2435 du CNRS), 
98$^{\rm bis}$ Boulevard Arago, 
75014 Paris, France}

\affiliation{Theoretical Astrophysics, California 
Institute of Technology, Pasadena, CA 91125}

\author{Yanbei Chen} 

\affiliation{Theoretical Astrophysics, California 
Institute of Technology, Pasadena, CA 91125}

\author{Michele Vallisneri}

\affiliation{Theoretical Astrophysics, California 
Institute of Technology, Pasadena, CA 91125}

\begin{abstract}
We investigate the problem of detecting gravitational waves from
binaries of nonspinning black holes with masses $m=5\mbox{--}20
M_\odot$, moving on quasicircular orbits, which are arguably the most
promising sources for first-generation ground-based detectors.  We
analyze and compare all the currently available post--Newtonian
approximations for the relativistic two-body dynamics; for these
binaries, different approximations predict different waveforms.
We then construct examples of detection template families that embed
all the approximate models, and that could be used to detect the true
gravitational-wave signal (but not to characterize accurately its
physical parameters).  We estimate that the fitting factor for our
detection families is $\gtrsim 0.95$ (corresponding to an event-rate loss $\lesssim 15$\%) and we estimate that the discretization of the
template family, for $\sim 10^4$ templates, increases the loss to
$\lesssim 20$\%.
\end{abstract}

\maketitle

\section{Introduction}
\label{sec1}

A network of broadband ground-based laser interferometers, aimed at
detecting gravitational waves (GWs) in the frequency band
10--10$^3$ Hz, is currently beginning operation and, hopefully, will
start the first science runs within this year (2002). This network
consists of the British--German GEO, the American Laser Interferometer
Gravitational-wave Observatory (LIGO), the Japanese TAMA and the
Italian--French VIRGO (which will begin operating in
2004)~\cite{Inter}.

The first detection of gravitational waves with LIGO and VIRGO
interferometers is likely to come from binary black-hole systems where
each black hole has a mass~\footnote{These are
binaries formed either from massive main-sequence progenitor binary
stellar systems (field binaries), or from capture processes in
globular clusters or galactic centers (capture binaries).} of a few $M_\odot$, and the
total mass is roughly in the range $10\mbox{--}40
M_{\odot}$~\cite{FK}, and where the orbit is quasicircular (it is
generally assumed that gravitational radiation reaction will
circularize the orbit by the time the binary is close to the final
coalescence~\cite{LW}).  It is easy to see why. Assuming for
simplicity that the GW signal comes from a quadrupole-governed,
Newtonian inspiral that ends at a frequency outside the range of good
interferometer sensitivity, the signal-to-noise ratio S/N is
$\propto {\cal M}^{5/6}/d$ (See, e.g., Ref.~\cite{KT300}),
where ${\cal M}=M\eta^{3/5}$ is the \emph{chirp mass} (with $M = m_1 +
m_2$ the total mass and $\eta = m_1 m_2/M^2$), and $d$ is the distance
between the binary and the Earth.  Therefore, for a given
signal-to-noise detection threshold (see Sec.\ \ref{sec2}) and for
equal-mass binaries ($\eta=1/4$), the larger is the total mass, the
larger is the distance $d$ that we are able to probe. [In
Sec.~\ref{sec5} we shall see how this result is modified when we relax
the assumption that the signal ends outside the range of good
interferometer sensitivity.]

For example, a black-hole--black hole binary (BBH) of total mass $M=20
M_\odot$ at 100 Mpc gives (roughly) the same S/N as a
neutron-star--neutron-star binary (BNS) of total mass $M=2.8M_\odot$
at 20 Mpc. The expected measured-event rate scales as the third power
of the probed distance, although of course it depends also on the
system's coalescence rate per unit volume in the universe. To give
some figures, computed using LIGO-I's sensitivity specifications, if
we assume that BBHs originate from main-sequence
binaries~\cite{postnov}, the estimated detection rate per year is
$\lesssim \, 4 \times 10^{-3}\mbox{--}0.6$ at $100\, {\rm
Mpc}$~\cite{KNST,KT}, while if globular clusters are considered as
incubators of BBHs~\cite{PZ99} the estimated detection rate per year
is $\sim 0.04\mbox{--}0.6$ at $100\, {\rm Mpc}$~\cite{KNST,KT}; by
contrast, the BNS detection rate per year is in the range $3 \times
10^{-4}\mbox{--}0.3$ at $20 \,{\rm Mpc}$~\cite{KNST,KT}. The very
large cited ranges for the measured-event rates reflect the
uncertainty implicit in using population-synthesis techniques and
extrapolations from the few known galactic BNSs to evaluate the
coalescence rates of binary systems. [In a recent article
\cite{Miller}, Miller and Hamilton suggest that four-body effects in
globular clusters might enhance considerably the BBH coalescence rate,
brightening the prospects for detection with first-generation
interferometers; the BBHs involved might have relatively high
BH masses ($\sim 100 M_\odot$) and eccentric orbits, and they
will not be considered in this paper.]

The GW signals from standard comparable-mass BBHs with $M =
10\mbox{--}40 M_\odot$ contain only few ($50 \mbox{--} 800$) cycles in
the LIGO--VIRGO frequency band, so we might expect that the task of
modeling the signals for the purpose of data analysis could be
accomplished easily. However, the frequencies of best interferometer
sensitivity correspond to GWs emitted during the final stages of the
inspiral, where the post--Newtonian (PN) expansion~\cite{PN}, which
for compact bodies is essentially an expansion in the characteristic
orbital velocity $v/c$, begins to fail. It follows that these sources
require a very careful analysis. As the two bodies draw closer, and
enter the nonlinear, strong-curvature phase, the motion becomes
relativistic, and it becomes harder and harder to extract reliable
information from the PN series. For example, using the Keplerian
formula $v=(\pi M f_{\rm GW})^{1/3}$ [where $f_\mathrm{GW}$ is the GW
frequency] and taking $f_{\rm GW} = 153$ Hz [the LIGO-I
peak-sensitivity frequency] we get $v(M) = 0.14(M/M_\odot)^{1/3}$;
hence, for BNSs $v(2.8 M_\odot) = 0.2$, but for BBHs $v(20 M_\odot) =
0.38$ and $v(40 M_\odot) = 0.48$.

The final phase of the inspiral (at least when BH spins are
negligible) includes the transition from the adiabatic inspiral to the
plunge, beyond which the motion of the bodies is driven
(almost) only by the conservative part of the dynamics.  Beyond the
plunge, the two BHs merge, forming a single rotating BH in a very
excited state; this BH then eases into its final stationary Kerr
state, as the oscillations of its quasinormal modes die out. In this
phase the gravitational signal will be a superposition of
exponentially damped sinusoids (ringdown waveform).  For nonspinning
BBHs, the plunge starts roughly at the innermost stable circular orbit
(ISCO) of the BBH. At the ISCO, the GW frequency [evaluated in the
Schwarzschild test-mass limit as $f_{\rm GW}^{\rm ISCO}(M) \simeq
0.022/M$] is $f_{\rm GW}^{\rm ISCO}(20 M_\odot) \simeq 220\, {\rm Hz}$
and $f_{\rm GW}^{\rm ISCO}(30 M_\odot) \simeq 167\, {\rm Hz}$. These
frequencies are well inside the LIGO and VIRGO bands.

The data analysis of inspiral, merger (or plunge), and ringdown of
compact binaries was first investigated by Flanagan and
Hughes~\cite{FH}, and more recently by Damour, Iyer and
Sathyaprakash~\cite{DIS3}.  Flanagan and Hughes \cite{FH} model the
inspiral using the standard quadrupole prediction (see, e.g.,
Ref.~\cite{KT300}), and assume an ending frequency of $0.02/M$ (the
point where, they argue, PN and numerical-relativity predictions start
to deviate by $\sim 5\%$~\cite{C}). They then use a crude argument to
estimate upper limits for the total energy radiated in the merger
phase ($\sim 0.1 M$) and in the ringdown phase ($\sim 0.03 M$) of
maximally-spinning--BBH coalescences.  Damour, Iyer and
Sathyaprakash~\cite{DIS3} study the nonadiabatic PN-resummed model 
for non spinning BBHs of Refs.~\cite{BD1,BD2,EOB3PN}, where the plunge can be seen as a natural continuation of the inspiral~\cite{BD2} rather than a separate phase; the total radiated energy is $0.007M$ in the merger and $0.007M$ in the ringdown~\cite{BD3}. (All these values for the energy should be also
compared with the value, $0.025\mbox{--}0.03 M$, estimated recently in
Ref.~\cite{BBCLT01} for the plunge and ringdown for non spinning BBHs.)  
When we deal with nonadiabatic models, we too shall choose not to 
separate the various phases. Moreover, because the ringdown phase does 
not give a significant contribution to the signal-to-noise ratio for $M \leq 200
M_\odot$~\cite{FH,DIS3}, we shall not include it in our
investigations.

BHs could have large spins: various
studies~\cite{kidder,apostolatos} have shown that when this is the
case, the time evolution of the GW phase and amplitude during the
inspiral will be significantly affected by spin-induced modulations
and irregularities. These effects can become dramatic, if the
two BH spins are large and are not aligned or
antialigned with the orbital angular momentum. There is a considerable
chance that the analysis of interferometer data, carried out without
taking into account spin effects, could miss the signals from
spinning BBHs altogether. We shall tackle the crucial issue of spin in
a separate paper~\cite{BCV}.

The purpose of the present paper is to discuss the problem of
the failure of the PN expansion during the last stages of inspiral for
nonspinning BHs, and the possible ways to deal with this failure. This
problem is known in the literature as the intermediate binary black
hole (IBBH) problem~\cite{BCT}. Despite the considerable progress made by
the numerical-relativity community in recent
years~\cite{C,TB,PTC,GGB}, a reliable estimate of the waveforms
emitted by BBHs is still some time ahead (some results for the plunge
and ringdown waveforms were obtained very recently \cite{BBCLT01}, but
they are not very useful for our purposes, because they do not include
the last stages of the inspiral before the plunge, and their initial
data are endowed with large amounts of spurious GWs).  To tackle the 
delicate issue of the late orbital evolution
of BBHs, various nonperturbative analytical approaches to that
evolution (also known as PN resummation methods) have been
proposed~\cite{DIS1,BD1,BD2,EOB3PN}.

The main features of PN resummation methods can be summarized as
follows: (i) they provide an analytic (gauge-invariant) resummation of
the orbital energy function and gravitational flux function 
(which, as we shall see in Sec.~\ref{sec3}, are the two
crucial ingredients to compute the gravitational waveforms in the
adiabatic limit); (ii) they can describe the motion of the bodies (and
provide the gravitational waveform) beyond the adiabatic
approximation; and (iii) in principle they can be extended to higher
PN orders. More importantly, they can provide initial
dynamical data for the two BHs at the beginning of the plunge
(such as their positions and momenta), which can be used (in
principle) in numerical relativity to help build the initial
gravitational data (the metric and its time derivative) and
then to evolve the full Einstein equations through the merger
phase. However, these resummation methods are based on some
assumptions that, although plausible, have not been proved: for
example, when the orbital energy and the gravitational flux functions
are derived in the comparable-mass case, it is assumed that they are
smooth deformations of the analogous quantities in the test-mass
limit. Moreover, in the absence of both exact solutions and
experimental data, we can test the robustness and reliability of the
resummation methods only by internal convergence tests.

In this paper we follow a more conservative point of view. We shall
maintain skepticism about waveforms emitted by BBH with $M =
10\mbox{--}40 M_\odot$ and evaluated from PN calculations,  
as well as all other waveforms ever computed for the late
BBH inspiral and plunge, and we shall develop families of search
templates that incorporate this skepticism. More specifically, we shall
be concerned only with detecting BBH GWs, and not with extracting
physical parameters, such as masses and spins, from the measured
GWs. The rationale for this choice is twofold. First, detection is the
more urgent problem at a time when GW interferometers are about to
start their science runs; second, a viable detection strategy must be
constrained by the computing power available to process a very long
stream of data, while the study of detected signals to evaluate
physical parameters can concentrate many resources on a small stretch
of detector output. In addition, as we shall see in
Sec.~\ref{sec6}, and briefly discuss in Sec.~\ref{subsec6.5}, the
different PN methods will give different parameter estimations for
the same waveform, making a full parameter extraction fundamentally
difficult.

This is the strategy that we propose: we guess (and hope) that the
conjunction of the waveforms from all the post--Newtonian models
computed to date spans a region in signal space that includes (or
almost includes) the true signal. We then choose a \emph{detection} 
(or \emph{effective}) template family that approximates very well 
all the PN expanded and resummed models (henceforth denoted as 
\emph{target models}). If our
guess is correct, the \emph{effectualness}~\cite{DIS1} of the
effective model in approximating the targets (i.e., its capability of
reproducing their signal shapes) should be indicative of its
effectualness in approximating the true signals.  Because our goal is
the \emph{detection} of BBH GWs, we shall not require the detection 
template family to be \emph{faithful} \cite{DIS1} (i.e., to have a
small bias in the estimation of the masses). 

As a backup strategy, we require the detection template family 
to embed the targets in a signal space of higher dimension (i.e., with more
parameters), trying to guess the functional directions in which the
true signals might lie with respect to the targets (of course, this
guess is rather delicate!). So, the detection template families constructed
in this paper cannot be guaranteed to capture the true signal, but
they should be considered as indications. 

This paper is organized as follows. In Sec.~\ref{sec2}, we
briefly review the theory of matched-filtering GW detections, which
underlies the searches for GWs from inspiraling binaries. Then in
Secs.~\ref{sec3}, \ref{sec4}, and \ref{sec5} we present the target
models and give a detailed analysis of the differences between them,
both from the point of view of the orbital dynamics and of the
gravitational waveforms. More specifically, in Sec.~\ref{sec3} we
introduce the two-body adiabatic models, both PN expanded and
resummed; in Sec.~\ref{sec4} we introduce nonadiabatic approximations
to the two-body dynamics; and in Sec.~\ref{sec5} we discuss the signal-to-noise 
ratios obtained for the various two-body  models. Our proposals for the
detection template families are discussed in the Fourier domain in
Sec.~\ref{sec6}, and in the time domain in Sec.~\ref{sec7}, where we
also build the mismatch metric~\cite{Sub,O} for the template banks and use it
to evaluate the number of templates needed for
detection. Section~\ref{sec8} summarizes our conclusions.

Throughout this paper we adopt the LIGO noise curve given in
Fig.~\ref{noise} and Eq.\ \eqref{ligoI}, and used also in
Ref.~\cite{DIS3}. Because the noise curve anticipated for VIRGO [see
Fig.~\ref{noise}] is quite different (both at low frequencies, and in
the location of its peak-sensitivity frequency) our results cannot be
applied naively to VIRGO. We plan to repeat our study for VIRGO in the
near future.

\section{The theory of matched-filtering signal detection}
\label{sec2}

The technique of matched-filtering detection for GW signals is based
on the systematic comparison of the measured detector output $\ket{s}$
with a bank of theoretical \emph{signal templates} $\{\ket{u_i}\}$
that represent a good approximation to the class of physical signals
that we seek to measure.  This theory was developed by many authors
over the years, who have published excellent expositions
\cite{Finn,Finnchernoff,Wainstein,DIS1,O,FH,Davis,Oppenheim,Hancock,
Cutlerflanagan,
SatDhu1,SatDhu2,Sathya,DIS2}. In
the following, we summarize the main results and equations that are
relevant to our purposes, and we establish our notation.

\subsection{The statistical theory of signal detection}
\label{subsec2.1}

The detector output $\ket{s}$ consists of noise $\ket{n}$ and possibly
of a true gravitational signal $\ket{h_i}$ (part of a family
$\{\ket{h_i}\}$ of signals generated by different sources for
different source parameters, detector orientations, and so on).
Although we may be able to characterize the properties of the noise in
several ways, each separate \emph{realization} of the noise is
unpredictable, and it might in principle fool us by hiding a physical
signal (hence the risk of a \emph{false dismissal}) or by simulating
one (\emph{false alarm}).  Thus, the problem of signal detection is
essentially probabilistic. In principle, we could try to evaluate the
conditional probability $P(h|s)$ that the measured signal $s$ actually
contains one of the $h_i$. In practice, this is inconvenient, because
the evaluation of $P(h|s)$ requires the knowledge of the \emph{a
priori} probability that a signal belonging to the family $\{h_i\}$ is
present in $s$.

What we can do, instead, is to work with a \emph{statistic} (a
functional of $s$ and of the $h_i$) that (for different realizations
of the noise) will be distributed around low values if the physical
signal $\ket{h_i}$ is absent, and around high value if the signal is
present. Thus, we shall establish a \emph{decision rule} as follows
\cite{Wainstein}: we will claim a detection if the value of a
statistic (for a given instance of $\ket{s}$ and for a specific
$\ket{h_i}$) is higher than a predefined threshold. We can then study
the probability distribution of the statistic to estimate the
probability of false alarm and of false dismissal. The steps involved
in this statistical study are easily laid down for a generic model of
noise, but it is only in the much simplified case of \emph{normal
noise} that it is possible to obtain manageable formulas; and while
noise will definitely \emph{not} be normal in a real detector, the
Gaussian formulas can still provide useful guidelines for the
detection problems. Eventually, the statistical analysis of detector
search runs will be carried out with numerical Montecarlo techniques
that make use of the measured characteristics of the noise. So
throughout this paper we shall always assume Gaussian noise.

The statistic that is generally used is based on the symmetric inner
product $\braket{g}{h}$ between two real signals $\ket{g}$ and
$\ket{h}$, which represents essentially the cross-correlation between
$g$ and $h$, weighted to emphasize the correlation at the frequencies
where the detector sensitivity is better. We follow Cutler and
Flanagan's conventions \cite{Cutlerflanagan} and define
\begin{equation}
\braket{g}{h} = 2 \int_{-\infty}^{+\infty} \frac{\tilde{g}^*(f)
\tilde{h}(f)}{S_n(|f|)} df =
4 \, \mathrm{Re} \int_{0}^{+\infty}
\frac{\tilde{g}^*(f) \tilde{h}(f)}{S_n(f)} df,
\label{eq:innerproduct}
\end{equation}
where $S_n(f)$, the one-sided \emph{noise power spectral density}, is
given by
\begin{equation}
\overline{\tilde{n}^*(f_1) \tilde{n}(f_2)} = \frac{1}{2} \delta(f_1-f_2) S_n(f_1) \quad
\mbox{for $f_1>0$},
\label{eq:defsn}
\end{equation}
and $S_n(f_1) = 0$ for $f_1 < 0$. We then define the
\emph{signal-to-noise ratio} $\rho$ (for the measured signal $\ket{s}$
after filtering by $\ket{h_i}$), as
\begin{equation}
\rho(h_i) = \frac{\braket{s}{h_i}}{\mathrm{rms} \, \braket{n}{h_i}} =
\frac{\braket{s}{h_i}}{\sqrt{\braket{h_i}{h_i}}},
\end{equation}
where the equality follows because $\overline{\braket{h_i}{n}
\braket{n}{h_i}} = \braket{h_i}{h_i}$ (see, e.g.,
\cite{Wainstein}). In the case of Gaussian noise, it can be proved
that this filtering technique is \emph{optimal}, in the sense that it
maximizes the probability of correct detection for a given probability
of false detection.

In the case when $\ket{s} = \ket{n}$, and when noise is Gaussian, it
is easy to prove that $\rho$ is a normal variable with a mean of zero
and a variance of one.  If instead $\ket{s} = \ket{h_i} + \ket{n}$,
then $\rho$ is a normal variable with mean $\sqrt{\braket{h_i}{h_i}}$
and unit variance.  The \emph{threshold} $\rho_*$ for detection is set
as a tradeoff between the resulting false-alarm probability,
\begin{equation}
\mathcal{F} = \sqrt{\frac{1}{2 \pi}} \int_{\rho_*}^{+\infty} e^{-\rho^2/2} d \rho =
\frac{1}{2} \mathrm{erfc} \, (\rho_* / \sqrt{2})
\label{eq:falsealarm}
\end{equation}
(where $\mathrm{erfc}$ is the \emph{complementary error
function}~\cite{Abramowitz}), and the probability of correct detection
\begin{equation}
\mathcal{D} = \frac{1}{2} \mathrm{erfc} \, [(\rho_* - \sqrt{\braket{h_i}{h_i}})/ \sqrt{2}]
\end{equation}
(the probability of false dismissal is just $1-\mathcal{D}$).

\subsection{Template families and extrinsic parameters}
\label{subsec2.2}

We can now go back to the initial strategy of comparing the measured
signal against a bank of $\mathcal{N}_i$ templates $\{\ket{u_i}\}$
that represent a plurality of sources of different types and physical
parameters.  For each stretch $\ket{s}$ of detector output, we shall
compute the signal-to-noise ratio
$\braket{s}{u_i}/\sqrt{\braket{u_i}{u_i}}$ for all the $\ket{u_i}$,
and then apply our rule to decide whether the physical signal
corresponding to any one of the $\ket{u_i}$ is actually present within
$\ket{s}$ \cite{KT300}.  Of course, the threshold $\rho_*$ needs to be
adjusted so that the probability $\mathcal{F}_\mathrm{tot}$ of false
alarm \emph{over all the templates} is still acceptable. Under the
assumption that all the inner products $\braket{n}{u_i}$ of the
templates with noise alone are statistically independent variables
[this hypothesis entails $\braket{u_i}{u_j} \simeq 0$],
$\mathcal{F}_\mathrm{tot}$ is just $1-(1-\mathcal{F})^{\mathcal{N}_i}
\sim \mathcal{N}_i \mathcal{F}$.  If the templates are not
statistically independent, this number is an upper limit on the false
alarm rate.  However, we first need to note that, for any template
$\ket{u_i}$, there are a few obvious ways (parametrized by the
so-called \emph{extrinsic parameters}) of changing the signal shape
that do not warrant the inclusion of the modified signals as separate
templates \footnote{Parameters that are not extrinsic are known as
\emph{intrinsic}. This nomenclature was introduced by Owen \cite{O}, but the
underlying concept had been present in the data-analysis literature
for a long time (see, e.g., \cite{Wainstein}). Sathyaprakash
\cite{Sathya} draws the same distinction between \emph{kinematical}
and \emph{dynamical} parameters.}

The extrinsic parameters are the signal \emph{amplitude}, \emph{phase}
and \emph{time of arrival}. Any true signal $\ket{h}$ can be written
in all generality as
\begin{equation}
h(t) = \mathcal{A}_h a_h[t-t_h] \cos [\Phi_h(t-t_h) + \phi_h],
\label{eq:generic}
\end{equation}
where $a_h(t) = 0$ for $t < 0$, where $\Phi_h(0) = 0$, and where
$a_h(t)$ is normalized so that $\braket{h}{h} =
\mathcal{A}_h^2$. While the template bank $\{\ket{u_i}\}$ must contain
signal shapes that represent all the physically possible functional
forms $a(t)$ and $\Phi(t)$, it is possible to modify our search
strategy so that the variability in $\mathcal{A}_h$, $\phi_h$ and $t_h$ is
automatically taken into account without creating additional
templates.

The signal amplitude is the simplest extrinsic parameter. It is
expedient to \emph{normalize} the templates $\ket{u_i}$ so that
$\braket{u_i}{u_i} = 1$, and $\rho(\ket{u_i}) = \braket{s}{u_i}$.
Indeed, throughout the rest of this paper we shall always assume
normalized templates.  If $\ket{s}$ contains a scaled version $h_i
= \mathcal{A} u_i$ of a template $\ket{u_i}$ (here $\mathcal{A}$ is
known as the signal \emph{strength}), then $\overline{\rho(u_i)} =
\mathcal{A}$. However, the statistical distribution of $\rho$ is the
same \emph{in the absence of the signal}. Then the problem of detection
signals of known shape and unknown amplitude is easily solved by using
a single normalized template and the same threshold $\rho_*$ as used
for the detection of completely known signals \cite{Wainstein}.
Quite simply, the stronger an actual signal, the easier it will be to
reach the threshold.

We now look at phase, and we try to match $\ket{h}$ with a continuous
one-parameter subfamily of templates $u(\phi_t;t) = a_h(t) \cos
[\Phi_h(t) + \phi_t]$. It turns out that for each time signal shape
$\{a(t),\Phi(t)\}$, we need to keep in our template bank only two
copies of the corresponding $\ket{u_i}$, for $\phi_t = 0$ and $\phi_t
= \pi/2$, and that the signal to noise of the detector output
$\ket{s}$ against $\ket{u_i}$, for the best possible value of
$\phi_t$, is automatically found as \cite{Wainstein}
\begin{equation}
\rho_\phi = \max_{\phi_t} \braket{s}{u_i(\phi_t)} =
\sqrt{\left|\braket{s}{u_i(0)}\right|^2 +
\left|\braket{s}{u_i(\pi/2)}\right|^2}\,,
\end{equation}
where $u_i(0)$ and $u_i(\pi/2)$ have been orthonormalized.  The
statistical distribution of the phase-maximized statistic $\rho_\phi$,
for the case of (normal) noise alone, is the \emph{Raleigh
distribution}~\cite{Wainstein}
\begin{equation}
p_0(\rho_\phi) = \rho_\phi e^{-\rho_\phi^2/2},
\end{equation}
and the false-alarm probability for a threshold $\rho_{\phi*}$ is just
\begin{equation}
\mathcal{F} = e^{-\rho_{\phi*}^2 / 2}.
\label{eq:newfalsealarm}
\end{equation}

Throughout this paper, we will find it useful to consider inner
products that are maximized (or minimized) with respect to the phases
of \emph{both} templates and reference signals. In particular, we shall
follow Damour, Iyer and Sathyaprakash in making a distinction between
the \emph{best match} or \emph{maxmax match}
\begin{equation}
\label{eq:maxmaxd}
\mathrm{maxmax} \braket{h}{u_i} =
\max_{\phi_h} \max_{\phi_t}
\braket{h(\phi_h)}{u_i(\phi_t)}, 
\end{equation} 
which represents the most favorable combination of phases between the
signals $\ket{h}$ and $\ket{u_i}$, and the \emph{minmax match}
\begin{equation}
\mathrm{minmax} \braket{h}{u_i} = \min_{\phi_h} \max_{\phi_t}
\braket{h(\phi_h)}{u_i(\phi_t)}, 
\end{equation} 
which represents the safest estimate in the realistic situation, where
we cannot choose the phase of the physical measured signal, but only
of the template used to match the signal. Damour, Iyer and
Sathyaprakash [see Appendix B of Ref.~\cite{DIS1}] show that both
quantities are easily computed as
\begin{equation}
\left(
\begin{array}{c} 
\mathrm{maxmax} \\
\mathrm{minmax}
\end{array} \right) =
\left\{
\frac{A+B}{2} \pm
\left[
\left(\frac{A-B}{2}\right)^2 + C^2
\right]^{1/2}
\right\}^{1/2},
\end{equation}
where
\begin{eqnarray}
A &=& \braket{h(0)}{u_i(0)}^2 + \braket{h(0)}{u_i(\pi/2)}^2, \\
B &=& \braket{h(\pi/2)}{u_i(0)}^2 + \braket{h(\pi/2)}{u_i(\pi/2)}^2,
\\
C &=& \braket{h(0)}{u_i(0)}\braket{h(\pi/2)}{u_i(0)} + \\
& & \braket{h(0)}{u_i(\pi/2)}\braket{h(\pi/2)}{u_i(\pi/2)}. \nonumber
\end{eqnarray}
In these formulas we have assumed that the two bases
$\{\ket{h(0)},\ket{h(\pi/2)}\}$ and
$\{\ket{u_i(0)},\ket{u_i(\pi/2)}\}$ have been orthonormalized.

The \emph{time of arrival} $t_h$ is an extrinsic parameter because the
signal to noise for the normalized, time-shifted template
$\ket{u(t-t_0)}$ against the signal $\ket{s}$ is just
\begin{equation}
\braket{s}{u(t_0)} = 4 \, \mathrm{Re} \int_0^{+\infty}
\frac{\tilde{s}^*(f) \tilde{u}(f)}{S_n(f)} e^{i 2 \pi f t_0} df,
\label{eq:finteg}
\end{equation} 
where we have used a well-known property of the Fourier transform of
time-shifted signals. These integrals can be computed at the same time
for all the time of arrivals $\{t_0\}$, using a \emph{fast Fourier
transform} technique that requires $\sim N_s \log N_s$ operations
(where $N_s$ is the number of the samples that describe the signals)
as opposed to $\sim N_s^2$ required to compute all the integrals
separately \cite{Schutz}. Then we can look for the optimal $t_0$ that
yields the maximum signal to noise.

We now go back to adjusting the threshold $\rho_*$ for a search over a
vast template bank, using the estimate \eqref{eq:newfalsealarm} for
the false-alarm probability.  Assuming that the statistics $\rho_\phi$
for each signal shape \emph{and} starting time are independent, we
require that
\begin{equation}
e^{-\rho_{\phi*}^2/2} \simeq \frac{\mathcal{F}_\mathrm{tot}}{N_\mathrm{times}
N_\mathrm{shapes}},
\end{equation}
or
\begin{equation}
\rho_* \simeq \sqrt{2 (\log N_\mathrm{times} + \log N_\mathrm{shapes}
- \log \mathcal{F}_\mathrm{tot})}.
\label{eq:falses}
\end{equation}
It is generally assumed that $N_\mathrm{times} \sim 3 \times 10^{10}$ (equivalent to
templates displaced by 0.01 s over one year~\cite{Cutler,FH}) 
and that the false-alarm probability $\mathcal{F}_\mathrm{tot} \sim 10^{-3}$. 
Using these values, we find that an increase of $\rho_*$ by
about $\sim 3\%$ is needed each time we increase $N_\mathrm{shapes}$ by one
order of magnitude. So there is a tradeoff between the improvement in
signal-to-noise ratio obtained by using more signal shapes and the
corresponding increase in the detection threshold for a fixed
false-alarm probability. 

\subsection{Imperfect detection and discrete families of templates}
\label{subsec2.3}

There are two distinct reasons why the detection of a physical signal
$\ket{h}$ by matched filtering with a template bank $\{\ket{u_i}\}$
might result in signal-to-noise ratios lower than the optimal
signal-to-noise ratio,
\begin{equation}
\rho_\mathrm{opt} = \sqrt{\braket{h}{h}}.
\label{eq:optsn}
\end{equation}
First, the templates, understood as a \emph{continuous} family
$\{u(\lambda^A)\}$ of functional shapes indexed by one or more
\emph{intrinsic parameters} $\lambda^A$ (such as the masses, spins,
etc.), might give an unfaithful representation of $\ket{h}$,
introducing errors in the representation of the phasing or the
amplitude. The loss of signal to noise due to unfaithful templates is
quantified by the \emph{fitting factor} $\mathrm{FF}$, introduced by
Apostolatos \cite{ApostolatosFF}, and defined by
\begin{equation}
\label{eq:ffd}
\mathrm{FF}(\ket{h},\ket{u(\lambda^A)}) = \frac{\max_{\lambda^A} 
\braket{h}{u(\lambda^A)}}{\sqrt{\braket{h}{h}}}.
\end{equation}
In general, we will be interested in the FF of the continuous template
bank in representing a \emph{family} of physical signals $\{
\ket{h(\theta^A)} \}$, dependent upon one or more physical parameters
$\theta^A$: so we shall write $\mathrm{FF}(\theta^A) =
\mathrm{FF}(\ket{h(\theta^A)},\ket{u(\lambda^A)})$.  Although it is
convenient to index the template family by the same physical
parameters $\theta^A$ that characterize $\ket{h(\theta^A)}$, this is
by no means necessary; the template parameters $\lambda^A$ might be a
different number than the physical parameters (indeed, this is
desirable when the $\theta^A$ get to be very many), and they might not
carry any direct physical meaning. Notice also that the value of
the $\mathrm{FF}$ will depend on the parameter range chosen to
maximize the $\lambda^A$.

The second reason why the signal-to-noise will be degraded with
respect to its optimal value is that, even if our templates are
perfect representations of the physical signals, in practice we will
not adopt a continuous family of templates, but we will be limited to
using a discrete bank $\{\ket{u_i} \equiv
\ket{u(\lambda^A_i)}\}$. This loss of signal to noise depends on how
finely templates are laid down over parameter space \cite{SatDhu1,SatDhu2,Sathya}; a
notion of metric in template space (the \emph{mismatch metric}
\cite{O,Sub,SO}) can be used to guide the disposition of templates so that
the loss (in the perfect-template abstraction) is limited to a fixed,
predetermined value, the \emph{minimum match} $\mathrm{MM}$,
introduced in Refs.~\cite{SatDhu1,O}, and defined by
\beq
\mathrm{MM}=
\min_{\hat{\lambda}^A} \max_{\lambda^A_i} \braket{u(\hat{\lambda}^A)}{u(\lambda^A_i)} 
= \min_{\hat{\lambda}^A} \max_{\Delta \lambda^A_i}
\braket{u(\hat{\lambda}^A)}{u(\hat{\lambda}^A + \Delta \lambda^A_i)},
\eeq
where $\Delta \lambda^A_i \equiv \lambda^A_i - \hat{\lambda}^A$.  The
\emph{mismatch metric} $g_{BC}(\hat{\lambda}^A)$ for the template space
$\{u(\lambda^A)\}$ is obtained by expanding the inner product (or
\emph{match}) $\braket{u(\hat{\lambda}^A)}{u(\hat{\lambda}^A+\Delta
\lambda^A)}$ about its maximum of 1 at $\Delta
\lambda^A = 0$:
\beq
\braket{u(\hat{\lambda}^A)}{u(\hat{\lambda}^A+\Delta
\lambda^A)}
= M(\hat{\lambda}^A,\hat{\lambda}^A + \Delta \lambda^A) 
= 1 + \frac{1}{2}
\left. \frac{\partial^2 M}{\partial \Delta \lambda^B \partial \Delta
\lambda^C} \right|_{\hat{\lambda}^A} \Delta \lambda^B \Delta \lambda^C
+ \, \cdots, 
\eeq
so the \emph{mismatch} $1 - \mathrm{M}$ between
$\ket{u(\hat{\lambda}^A)}$ and the nearby template
$\ket{u(\hat{\lambda}^A+\Delta \lambda^A)}$ can be seen as the square
of the proper distance in a differential manifold indexed by the
coordinates $\lambda^A$ \cite{O},
\begin{equation}
1 - M(\hat{\lambda}^A,\hat{\lambda}^A+\Delta \lambda^A) = g_{BC} \Delta \lambda^B
\Delta \lambda^C, 
\end{equation}
where
\begin{equation}
g_{BC} = - \frac{1}{2} \left. \frac{\partial^2 M}{\partial \Delta
\lambda^B \partial \Delta \lambda^C} \right|_{\hat{\lambda}^A}.
\end{equation}
If, for simplicity, we lay down the $n$-dimensional discrete template
bank $\{\ket{u(\lambda^A_i)}\}$ along a hypercubical grid of cellsize
$dl$ in the metric $g_{AB}$ (a grid in which all the templates on
nearby corners have a mismatch of $dl$ with each other), the minimum
match occurs when $\hat{\lambda}^A$ lies exactly at the center of one
of the hypercubes: then $1 - \mathrm{MM} = n (dl/2)^2$. Conversely,
given MM, the volume of the corresponding hypercubes is given by
$V_{\mathrm{MM}} = (2 \sqrt{(1-\mathrm{MM})/n})^n$. The number of
templates required to achieve a certain MM is obtained by integrating
the proper volume of parameter space within the region of physical
interest, and then dividing by $V_\mathrm{MM}$:
\begin{equation}
\mathcal{N}[g,\mathrm{MM}] = \frac{\int \sqrt{|g|} d\lambda^A}{\left(2
\sqrt{[1-\mathrm{MM}]/n}\right)^n}.
\label{eq:metric}
\end{equation}
In practice, if the metric is not constant over parameter
space it will not be possible to lay down the templates on an exact
hypercubical grid of cellsize $dl$, so $\mathcal{N}$ will be somewhat
higher than predicted by Eq.\ \eqref{eq:metric}. However, we estimate
that this number should be correct within a factor of two, which is
adequate for our purposes.

In the worst possible case, the combined effect of unfaithful modeling
($\mathrm{FF} < 1$) and discrete template family ($\mathrm{MM} < 1$)
will degrade the optimal signal to noise by a factor of about
$\mathrm{FF} + \mathrm{MM} - 1$. This estimate for the total
signal-to-noise loss is exact when, in the space of signals, the two
segments that join $\ket{h(\hat{\theta}^A)}$ to its projection
$\ket{u(\hat{\lambda}^A)}$ and $\ket{{u}(\hat{\lambda}^A)}$ to the
nearest discrete template $\ket{{u}(\hat{\lambda}^A_i)}$ can be
considered orthogonal:
\begin{equation}
\braket{h(\theta^A) - u(\hat{\lambda}^A)}{{u}(\hat{\lambda}^A) - u(\hat{\lambda}^A_i)} \simeq 0.
\end{equation}
This assumption is generally very accurate if $\mathrm{FF}$ and
$\mathrm{MM}$ are small enough, as in this paper; so we will adopt
this estimate. However, it is possible to be more precise, by defining
an \emph{external metric} $g^\mathrm{E}_{AB}$~\cite{Sub,CA} that
characterizes directly the mismatch between $\ket{h(\hat{\theta}^A)}$ and a
template $\ket{u(\hat{\lambda}^A + \Delta \lambda^A)}$ that is
displaced with respect to the template $u(\hat{\lambda}^A)$ that is
yields the maximum match with $\ket{h(\hat{\theta}^A)}$.

Since the strength of gravity-wave signals scales as the inverse of
the distance \footnote{The amplitude of the measured gravity-wave
signals depends not only on the actual distance to the source, but
also on the reciprocal orientation between the detector and the
direction of propagation of the waves. A combination of several
detectors will be needed, in general, to evaluate the distance to a
gravity-wave source starting from the signal-to-noise ratio alone.},
the matched-filtering scheme, with a chosen signal-to-noise threshold
$\rho_*$, will allow the reliable detection of a signal $\ket{h}$,
characterized by the signal strength $\mathcal{A}_{d_0} =
\sqrt{\braket{h}{h}}$ at the distance $d_0$, out to a maximum distance
\begin{equation}
\frac{d_\mathrm{max}}{d_0} = \frac{\mathcal{A}_{d_0}}{\rho_*}.
\end{equation}

If we assume that the measured GW events happen with a homogeneous
event rate throughout the accessible portion of the universe, then the
detection rate will scale as $d^3_\mathrm{max}$. It follows that the
use of unfaithful, discrete templates $\{\ket{u_i}\}$ to detect the
signal $\ket{h}$ will effectively reduce the signal strength, and
therefore $d_\mathrm{max}$, by a factor $\mathrm{FF} + \mathrm{MM} -
1$. This loss in the signal-to-noise ratio can also be seen as an
increase in the detection threshold $\rho_*$ necessary to achieve the
required false-alarm rate, because the imperfect templates introduce
an element of uncertainty. In either case, the detection rate will be
reduced by a factor $(\mathrm{FF} + \mathrm{MM} - 1)^3$.

\subsection{Approximations for detector noise spectrum and gravitational-wave signal}
\label{subsec2.4}

For LIGO-I we use the analytic fit to the noise power spectral density
given in Ref.~\cite{DIS3}, and plotted in Fig.\ \ref{nsd}:
\begin{figure}[t]
\begin{center}
\epsfig{file = 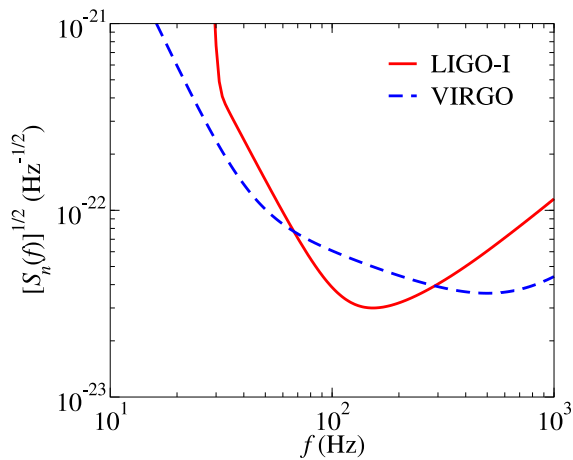,width=\sizeonefig}
\caption{\label{noise}Square root of the noise spectral density
$\sqrt{S_n(f)}$ versus frequency $f$, for LIGO-I [Eq.~(\ref{ligoI})],
and VIRGO (from Tab.\ IV of Ref.~\cite{DIS3}).\label{nsd}}
\end{center}
\end{figure}
\beq
\frac{S_n(f)}{{\rm Hz}^{-1}} = 9.00 \times 10^{-46}\,
\left [ \left ( 4.49\frac{f}{f_0} \right )^{-56}
+ 0.16 \left ( \frac{f}{f_0} \right )^{-4.52} + 0.52 + 0.32 \left ( \frac{f}{f_0} \right )^{2}
\right ]\,,
\label{ligoI}
\eeq
where $f_0 = 150$ Hz. The first term in the square brackets represents
seismic noise, the second and third, thermal noise, and the fourth,
photon shot noise.

Throughout this paper, we shall compute BBH waveforms in the
quadrupole approximation (we shall compute the phase evolution of the
GWs with the highest possible accuracy, but we shall omit all
harmonics higher than the quadrupole, and we shall omit
post--Newtonian corrections to the amplitude; this is a standard
approach in the field, see, e.g., \cite{PN}). The signal received at
the interferometer can then be written as \cite{KT300,Finnchernoff}
\begin{equation}
h(t) = \frac{\Theta}{d_\mathrm{L}} M \eta (\pi M f_\mathrm{GW})^{2/3}
\cos \varphi_\mathrm{GW},
\label{eq:quadapp}
\end{equation}
where $f$ and $\varphi_\mathrm{GW}$ are the instantaneous GW frequency
and phase at the time $t$, $d_\mathrm{L}$ is the \emph{luminosity
distance}, $M$ and $\eta$ are respectively the BBH total mass $m_1 +
m_2$ and the dimensionless mass ratio $m_1 m_2 / M^2$, and where we
have taken $G = c = 1$. The coefficient $\Theta$ depends on the
inclination of the BBH orbit with respect to the plane of the sky, and
on the polarization and direction of propagation of the GWs with
respect to the orientation of the interferometer. Finn and Chernoff
\cite{Finnchernoff} examine the distribution of $\Theta$, and show
that $\Theta_\mathrm{max} = 4$, while $\mathrm{rms}\, \Theta =
8/5$. We shall use this last value when we compute optimal
signal-to-noise ratios.  The waveform given by Eq.\
\eqref{eq:quadapp}, after dropping the factor $\Theta M
\eta/d_\mathrm{L}$, is known as \emph{restricted waveform}.
\begin{table*}
\begin{tabular}{l|l|l|l}
\hline \hline
model & shorthand & evolution equation & section \\
\hline
\begin{minipage}[t]{0.3\textwidth}\flushleft adiabatic model with Taylor-expanded energy $\mathcal{E}(v)$ and flux $\mathcal{F}(v)$ \end{minipage} & T$(n\mathrm{PN},m\mathrm{PN};\hat{\theta})$ & energy-balance equation & Sec.\ \ref{subsec3.1} \\ \hline
\begin{minipage}[t]{0.3\textwidth}\flushleft adiabatic model with Pad\'e-expanded energy $\mathcal{E}(v)$ and flux $\mathcal{F}(v)$  \end{minipage} & P$(n\mathrm{PN},m\mathrm{PN};\hat{\theta})$ & energy-balance equation & Sec.\ \ref{subsec3.2} \\ \hline
\begin{minipage}[t]{0.3\textwidth}\flushleft adiabatic model with 
Taylor-expanded energy $\mathcal{E}(v)$ and flux $\mathcal{F}(v)$ in the stationary-phase approximation \end{minipage} & SPA$(n\mathrm{PN} \equiv m\mathrm{PN})$ & \begin{minipage}[t]{0.3\textwidth}\flushleft energy-balance equation in the freq.\ domain \end{minipage} & Sec.\ \ref{subsec6.7} \\ \hline
\begin{minipage}[t]{0.3\textwidth}\flushleft nonadiabatic Hamiltonian model with Taylor-expanded GW flux  \end{minipage} & HT$(n\mathrm{PN},m\mathrm{PN};\hat{\theta})$  & Hamilton equations & Sec.\ \ref{subsec4.1} \\ \hline
\begin{minipage}[t]{0.3\textwidth}\flushleft nonadiabatic Hamiltonian model with Pad\'e-expanded GW flux  \end{minipage} & HP$(n\mathrm{PN},m\mathrm{PN};\hat{\theta})$ & Hamilton equations & Sec.\ \ref{subsec4.1} \\ \hline
\begin{minipage}[t]{0.3\textwidth}\flushleft nonadiabatic Lagrangian model  \end{minipage} & L$(n\mathrm{PN},m\mathrm{PN})$ & $\mathbf{F} = m \mathbf{a}$ & Sec.\ \ref{subsec4.2} \\ \hline
\begin{minipage}[t]{0.3\textwidth}\flushleft nonadiabatic effective-one-body model with Taylor-expanded GW flux  \end{minipage} & ET$(n\mathrm{PN},m\mathrm{PN};\hat{\theta};\tilde{z}_1,\tilde{z}_2)$ & eff.\ Hamilton equations & Sec.\ \ref{subsec4.3} \\ \hline
\begin{minipage}[t]{0.3\textwidth}\flushleft nonadiabatic effective-one-body model with Pad\'e-expanded GW flux \end{minipage}& EP$(n\mathrm{PN},m\mathrm{PN};\hat{\theta};\tilde{z}_1,\tilde{z}_2)$ & eff.\ Hamilton equations & Sec.\ \ref{subsec4.3} \\
\hline \hline \end{tabular}
\caption{\label{tablemodels}Post--Newtonian models of two-body dynamics defined in this paper. The notation X$(n\mathrm{PN},m\mathrm{PN};\hat{\theta})$ denotes the model X, with terms up to order $n$PN for the conservative dynamics, and with terms up to order $m$PN for radiation-reaction effects; for $m \geq 3$ we also need to specify the arbitrary flux parameter $\hat{\theta}$ (see Sec.~\ref{subsec3.1}); for $n \geq 3$, the effective-one-body models need also two additional parameters $\tilde{z}_1$ and $\tilde{z}_2$ (see Sec.~\ref{subsubsec4.3.1}).}
\end{table*}

\section{Adiabatic models}
\label{sec3}

We turn, now, to a discussion of the currently available mathematical
models for the inspiral of BBHs. Table \ref{tablemodels} shows a list
of the models that we shall consider in this paper, together with the
shorthands that we shall use to denote them. We begin in this section
with adiabatic models.  BBH adiabatic models treat the orbital
inspiral as a quasistationary sequence of circular orbits, indexed by
the invariantly defined velocity
\begin{equation}
v=(M\dot{\varphi})^{1/3} = (\pi M f_{\rm
GW})^{1/3}.
\end{equation}
The evolution of the inspiral (and in particular of the
orbital phase $\varphi$) is completely determined by the
\emph{energy-balance equation}
\beq
\frac{d {\cal E}(v)}{d t} = - {\cal F}(v),
\label{3.1}
\eeq
This equation relates the time derivative of the energy function
${\cal E}(v)$ (which is given in terms of the total relativistic
energy ${\mathcal E}_\mathrm{tot}$ by $\mathcal{E} = {\mathcal
E}_\mathrm{tot} - m_1 - m_2$, and which is conserved in absence of
radiation reaction) to the gravitational flux (or luminosity) function
${\cal F}(v)$. Both functions are known for quasicircular orbits as a
PN expansion in $v$. It is easily shown that Eq.~(\ref{3.1}) is
equivalent to the system (see, e.g., Ref.~\cite{DIS1})
\beq
\frac{d \varphi_{\rm GW}}{d t} = \frac{2v^3}{M}\,, \quad \quad  
\frac{d v}{dt} = - \frac{{\cal F}(v)}{M\,d{\cal E}(v)/dv}.
\label{3.2}
\eeq
In accord with the discussion around Eq.~\eqref{eq:quadapp}, we shall
only consider the \emph{restricted waveform} $h(t)=v^2 \cos
\varphi_{\rm GW}(t)$, where the GW phase $\varphi_{\rm GW}$ is twice
the orbital phase $\varphi$.

\subsection{Adiabatic PN expanded models}
\label{subsec3.1}
The equations of motion for two compact bodies at 2.5PN order were
first derived in Refs.~\cite{DD}. The 3PN equations of motion have
been obtained by two separate groups of researchers: Damour,
Jaranowski and Sch\"afer~\cite{DJS} used the Arnowitt--Deser--Misner
(ADM) canonical approach, while Blanchet, Faye and de
Andrade~\cite{DBF} worked with the PN iteration of the Einstein
equations in the harmonic gauge. Recently Damour and
colleagues~\cite{DJSd}, working in the ADM formalism and applying
dimensional regularization, determined uniquely the \emph{static
parameter} that enters the 3PN equations of motion~\cite{DJS,DBF} and
that was until then unknown. In this paper we shall adopt their value
for the static parameter. Thus at present the energy function ${\cal
E}$ is known up to 3PN order.

The gravitational flux emitted by compact binaries was first computed
at 1PN order in Ref.~\cite{1PN}. It was subsequently determined at 2PN
order with a formalism based on multipolar and post--Minkowskian
approximations, and, independently, with the direct integration of the
relaxed Einstein equations~\cite{2PN}. Nonlinear effects of tails at
2.5PN and 3.5PN orders were computed in
Refs.~\cite{2.5PNand3.5PN}. More recently, Blanchet and colleagues
derived the gravitational-flux function for quasicircular orbits up to
3.5PN order~\cite{BIJ,BFIJ}. However, at 3PN order~\cite{BIJ,BFIJ} the
gravitational-flux function depends on an arbitrary parameter
$\hat{\theta}$ that could not be fixed in the regularization scheme
used by these authors.
\begin{figure}[t]
\begin{center}
\epsfig{file=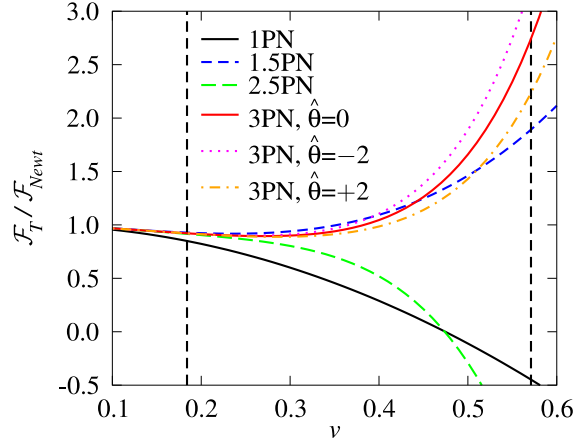,width=\sizeonefig,angle=0}
\end{center}
\caption{\label{Fig1}
Normalized flux function ${\cal F}_{T_N}/{\cal F}_{\rm Newt}$ versus $v$, at different PN orders for equal-mass binaries, $\eta=0.25$. Note that the 1.5PN and 2PN flux, and the 3PN and
3.5PN flux, are so close that they cannot be distinguished in these
plots. The two long-dashed vertical lines correspond to $v \simeq
0.18$ and $v \simeq 0.53$; they show the velocity range that
corresponds to the LIGO frequency band $40 \leq f_{\rm GW} \leq 240$
Hz for BBHs with total mass in the range $10\mbox{--}40 M_{\odot}$.}
\end{figure}

\subsubsection*{PN energy and flux}
Denoting by ${\cal E}_{T_{N}}$ and ${\cal F}_{T_{N}}$ the $N^{\rm
th}$-order Taylor approximants (T-approximants) to the energy and the
flux functions, we have
\bea
\label{BE}
{\cal E}_{T_{2N}}(v) \equiv {\cal E}_{\rm Newt}(v)\,\sum_{k=0}^N\,{\cal E}_k(\eta)\,
v^{2k}\,,\\
{\cal F}_{T_{N}}(v) \equiv {\cal F}_{\rm Newt}(v)\,\sum_{k=0}^N\,{\cal F}_k(\eta)\,
v^k\,,
\eea
where ``Newt'' stands for Newtonian order, and the subscripts $2N$ and $N$
stand for post$^{2N}$--Newtonian and post$^N$--Newtonian order. The
quantities in these equations are 
\bea
{\cal E}_{\rm Newt}(v) = -\frac{1}{2}\eta\,v^2\,, 
\quad \quad 
{\cal F}_{\rm Newt}(v) = \frac{32}{5}\eta^2\,v^{10}\,, 
\eea
\beq
{\cal E}_0(\eta) = 1\,, \quad \quad 
{\cal E}_1(\eta) = -\frac{3}{4}-\frac{\eta}{12}\,, \quad \quad
{\cal E}_2(\eta) = -\frac{27}{8}+\frac{19}{8}\eta-\frac{\eta^2}{24}\,, 
\eeq
\beq
{\cal E}_3(\eta) = -\frac{675}{64} + \left (\frac{34445}{576}-\frac{205}{96}\pi^2 \right )\eta
-\frac{155}{96}\eta^2-\frac{35}{5184}\eta^3\,,
\eeq
\beq
\label{f1}
{\cal F}_0(\eta) = 1\,, \quad \quad 
{\cal F}_1(\eta) =0 \,, \quad \quad 
{\cal F}_2(\eta) = -\frac{1247}{336} -\frac{35}{12}\eta \,, \quad \quad 
{\cal F}_3(\eta) = 4\pi\,, \\
\eeq
\beq
\label{f2}
{\cal F}_4(\eta) = -\frac{44711}{9072}+\frac{9271}{504}\eta+\frac{65}{18}\eta^2\,, \quad \quad 
{\cal F}_5(\eta) = -\left ( \frac{8191}{672} + \frac{583}{24}\eta \right )\pi\,, 
\eeq
\bea
\label{f3}
{\cal F}_6(\eta) &=& \frac{6643739519}{69854400}
+\frac{16}{3}\,\pi^2 -\frac{1712}{105}\,\gamma_E-\frac{856}{105}\,\log (16 v^2) 
 + \nonumber \\ && 
\left (-\frac{2913613}{272160}+\frac{41}{48}\,\pi^2-\frac{88}{3}\,\hat{\theta} \right )\eta 
-\frac{94403}{3024}\,\eta^2 - \frac{775}{324}\,\eta^3 \,,
\eea
\beq
\label{f4}
{\cal F}_7(\eta) = \left (-\frac{16285}{504}+\frac{214745}{1728}\eta+\frac{193385}{3024}\eta^2 \right )
\,\pi\,.
\eeq
Here $\eta = m_1 m_2 / (m_1+m_2)^2$, $\gamma_E$ is Euler's gamma,
and $\hat{\theta}$ is the arbitrary 3PN flux
parameter~\cite{BIJ,BFIJ}. From Tab.\ I of Ref.~\cite{BFIJ} we read
that the extra number of GW cycles accumulated by the PN terms of a
given order decreases (roughly) by an order of magnitude when we
increase the PN order by one. Hence, we find it reasonable to expect
that at 3PN order the parameter $\hat{\theta}$ should be of order
unity, and we choose as typical values $\hat{\theta} =0,\pm2$.
(Note for v3 of this paper on gr-qc: Eqs.\ (39) and (41) are now revised as per Ref.\ \cite{errata};
the parameter $\hat{\theta}$ has been determined to be 1039/4620 \cite{thetapar}.)
\begin{figure}[t]
\begin{center}
\epsfig{file = 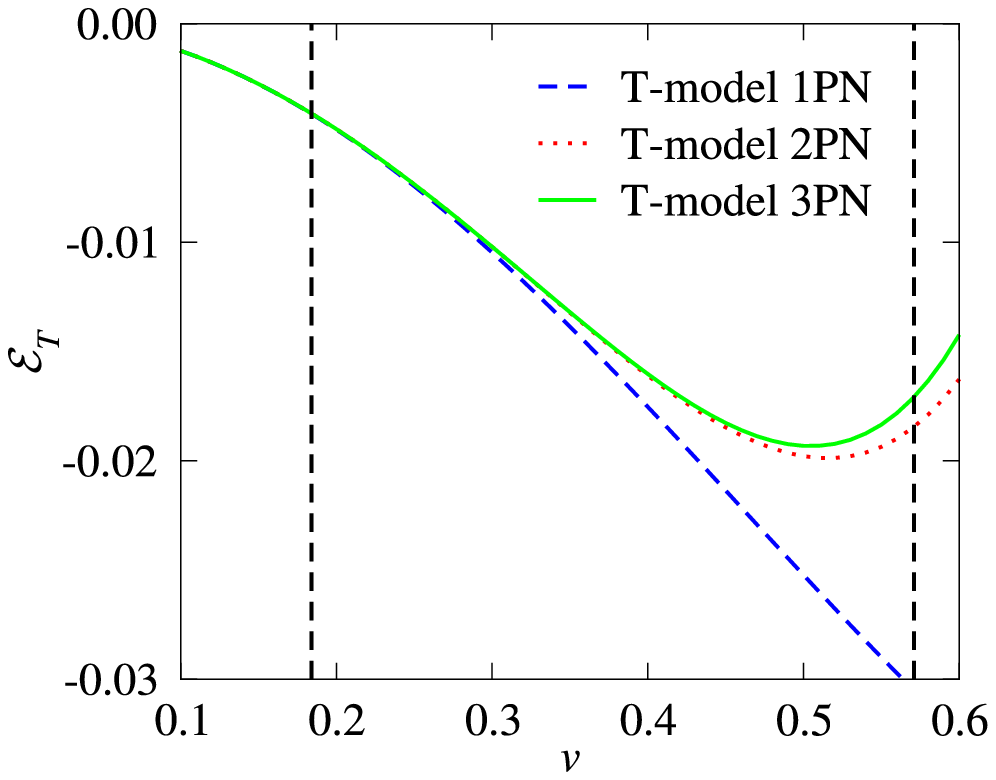,width=\sizetwofig}
\hspace{0.5cm}
\epsfig{file = 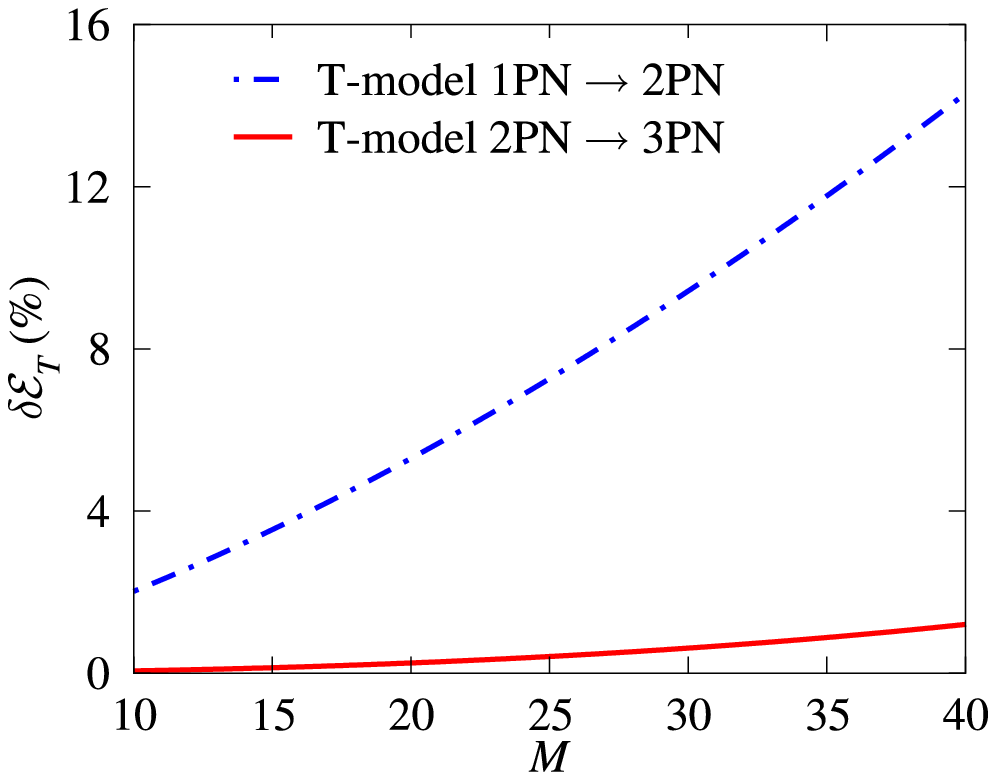,width=\sizetwofig}
\caption{In the left panel, we plot the energy function ${\cal
E}_{T_N}$ versus $v$, at different PN orders, for $\eta=0.25$. The two
long-dashed vertical lines in the left figure correspond to $v \simeq
0.18$ and $v \simeq 0.53$; they show the velocity range that
corresponds to the LIGO frequency band $40 \leq f_{\rm GW} \leq 240$
Hz, for BBHs with total mass in the range $10\mbox{--}40 M_{\odot}$. In
the right panel, we plot the percentage difference $\delta {\cal
E}_{T_N} = 100\, |({\cal E}_{T_{N+1}}- {\cal E}_{T_N})/{\cal
E}_{T_N}|$ versus the total mass $M$, for $N=1,2$, at the LIGO-I
peak-sensitivity GW frequency, $f_\mathrm{peak} = 153$ Hz [note:
$v_\mathrm{peak} = (\pi M f_\mathrm{peak})^{1/3}$].\label{Fig2}}
\end{center}
\end{figure}

In Fig.~\ref{Fig1} we plot the normalized flux ${\mathcal
F}_{T_N}/{\mathcal F}_{\rm Newt}$ as a function of $v$ at various PN
orders for the equal mass case $\eta=0.25$.  To convert $v$ to a GW
frequency we can use
\beq
f_{\rm GW} \simeq 3.2 \times 10^4 \left ( \frac{20 M_\odot}{M} \right ) v^3.
\label{eq:ffrequency}
\eeq
The two long-dashed vertical lines in Fig.~\ref{Fig1} correspond to $v
\simeq 0.18$ and $v \simeq 0.53$; they show the velocity range that
corresponds to the LIGO frequency band $40 \leq f_{\rm GW} \leq 240$
Hz for BBHs with total mass in the range $10\mbox{--}40 M_{\odot}$.
At the LIGO-I peak-sensitivity frequency, which is $153$ Hz according
to our noise curve, and for a (10+10)$M_\odot$ BBH,
we have $v\simeq 0.362$; and the percentage difference between subsequent PN
orders is ${\rm Newt} \rightarrow 1{\rm PN}: -58\%$; 
$1{\rm PN} \rightarrow 1.5{\rm PN}: +142\%$;
$1.5{\rm PN} \rightarrow 2{\rm PN}: -0.2\%$; $2{\rm PN}
\rightarrow 2.5{\rm PN}: -34\%$; $2.5{\rm PN} \rightarrow 3{\rm PN}
(\hat{\theta}=0): +43\%$; $3{\rm PN} \rightarrow 3.5{\rm PN}
(\hat{\theta}=0): +0.04\%$. The percentage difference between the
3PN fluxes with $\hat{\theta} = \pm 2$ is $\sim 7\%$. It is
interesting to notice that while there is a big difference between the
1PN and 1.5PN orders, and between the 2PN and 2.5PN orders, the 3PN
and 3.5PN fluxes are rather close. Of course this observation is
insufficient to conclude that the PN sequence is converging at 3.5PN
order. 

In the left panel of Fig.~\ref{Fig2}, we plot the T-approximants for
the energy function versus $v$, at different PN orders, while in the
right panel we plot (as a function of the total mass $M$, and at the
LIGO-I peak-sensitivity GW frequency $f_\mathrm{peak} = 153$ Hz) the
percentage difference of the energy function between T-approximants to
the energy function of successive PN orders. We note that the 1PN and
2PN energies are distant, but the 2PN and 3PN energies are quite
close.
\begin{table*} 
\begin{tabular}{r||c|c|c||c|c||c||c|c|c}
\hline \hline
\multicolumn{1}{c||}{$M$} &
\multicolumn{5}{c||}{$f_{\rm GW}$(Hz) at MECO} &
\multicolumn{4}{c}{$f_{\rm GW}$(Hz) at ISCO} \\
\multicolumn{1}{c||}{}
  & \multicolumn{1}{c|}{T (1PN)}
  & \multicolumn{1}{c|}{T (2PN)}
  & \multicolumn{1}{c||}{T (3PN)}
  & \multicolumn{1}{c|}{P (2PN)}
  & \multicolumn{1}{c||}{P (3PN)} 
  & \multicolumn{1}{c||}{H (1PN)}
  & \multicolumn{1}{c|}{E (1PN)}
  & \multicolumn{1}{c|}{E (2PN)}
  & \multicolumn{1}{c}{E (3PN)} \\
\hline
$(5+5)M_\odot$   & 3376   &  886  & 832 & 572 & 866 & 183 & 446  &  473  & 570 \\
$(10+10)M_\odot$ & 1688   &  442  & 416 & 286 & 433 & 92  & 223  &  236  & 285 \\
$(15+15)M_\odot$ & 1125   &  295  & 277 & 191 & 289 & 61  & 149  &  158  & 190 \\
$(20+20)M_\odot$ & 844    &  221  & 208 & 143 & 216 & 46  & 112  &  118  & 143 \\
\hline \hline
\end{tabular}
\caption{\label{MECOT}
Location of the MECO/ISCO.
The first six columns show the GW frequency at the Maximum binding Energy for Circular Orbits (MECO), computed using the T- and P-approximants to the energy function; the remaining columns show the GW frequency at the Innermost Stable Circular Orbit (ISCO), computed using the H-approximant to the energy, and using the EOB improved Hamiltonian (\protect\ref{himpr}) with $\tilde{z}_1=\tilde{z}_2=0$. For the H-approximant the ISCO exists only at 1PN order.
}
\end{table*}

\subsubsection*{Definition of the models}
The evolution equations (\ref{3.2}) for the adiabatic inspiral lose
validity (the inspiral ceases to be adiabatic) a little before $v$
reaches $v_{\rm MECO}^{T_N}$, where MECO stands for
Maximum--binding-Energy Circular Orbit~\cite{LB,DGG}.  This $v_{\rm
MECO}^{T_N}$ is computed as the value of $v$ at which $d {\cal
E}_{T_N}(v)/dv =0$. In building our adiabatic models we evolve Eqs.\
\eqref{3.2} right up to $v_\mathrm{MECO}$ and stop there. We shall
refer to the frequency computed by setting $v = v_\mathrm{MECO}$ in
Eq.\ \eqref{eq:ffrequency} as the \emph{ending frequency} for these
waveforms, and in Tab.\ \ref{MECOT} we show this frequency for some BH
masses.  However, for certain binaries, the 1PN and 2.5PN flux
functions can go to zero before $v=v_{\rm MECO}^{T_N}$ [see Fig.~\ref{Fig1}]. In those cases we choose as the ending
frequency the value of $f = v^3/(\pi M)$ where ${\cal F}(v)$ becomes
10\% of $\mathcal{F}_\mathrm{Newt}(v)$. [When using the 2.5PN flux, our 
choice of the ending frequency differs from the one used 
in Ref.~\cite{DIS3}, where the authors stopped the evolution at the 
GW frequency corresponding to the Schwarzschild innermost stable circular 
orbit. For this reason there are some differences between our overlaps 
and theirs.]

We shall refer to the models discussed in this section as ${\rm
T}(n{\rm PN},m{\rm PN})$, where $n$PN ($m$PN) denotes the maximum PN
order of the terms included for the energy (the flux). We shall
consider $(n{\rm PN}, m{\rm PN}) = (1, 1.5), (2, 2), (2, 2.5)$ and
$(3, 3.5, \hat{\theta})$ [at 3PN order we need to indicate also a choice
of the arbitrary flux parameter $\hat{\theta}$].
\begin {table*} 
\begin {tabular}{l|rr|rrrr|rr}
\hline \hline
\multicolumn{1}{c|}{$N$} & \multicolumn{8}{c}{$\left <{\rm T}_N,{\rm T}_{N+1} \right >$} \\
     & \multicolumn{2}{c}{$(5+20)M_\odot$}
     & \multicolumn{4}{c}{$(10+10)M_\odot$}
     & \multicolumn{2}{c}{$(15+15)M_\odot$}\\
\hline
 0 & 0.432 &   & 0.553 & (0.861, & 19.1, & 0.241) & 0.617 \\
 1 & 0.528 & [0.638]  & 0.550 & (0.884, & 22.0, & 0.237) & 0.645 & [0.712] \\
 2 ($\hat{\theta} =+2$) & 0.482 & [0.952]  & 0.547 & (0.841, & 18.5, & 0.25) & 0.563 & [0.917] \\
 2 ($\hat{\theta} =-2$) & 0.457 & [0.975] & 0.509 & (0.821, & 18.7, & 0.241) & 0.524 & [0.986] \\
\hline \hline
\end {tabular}
\caption{\label{CauchyT}Test for the Cauchy convergence of the
T-approximants. The values quoted are maxmax matches obtained by
maximizing with respect to the extrinsic parameters, but not to the
intrinsic parameters (i.e., the matches are computed for T waveforms
with the same masses, but different PN orders). Here we define ${\rm
T}_0={\rm T}(0, 0)$, ${\rm T}_1={\rm T}(1, 1.5)$, ${\rm T}_2 = {\rm
T}(2, 2.5)$, ${\rm T}_3 = {\rm T}(3, 3.5, \hat{\theta})$.  In the
Newtonian case, ${\rm T}_0=(0, 0)$, the MECO does not exist and we
stop the integration of the balance equation at $v=1$.  The values
in brackets, ``[...],'' are obtained by setting ${\rm T}_2 = {\rm T}(2, 2)$
instead of ${\rm T}(2, 2.5)$; the values in parentheses, ``(...),'' are obtained
by maximizing with respect to the extrinsic \emph{and} intrinsic
parameters, and they are shown together with the ${\rm T}_{N+1}$
parameters $M$ and $\eta$ where the maxima are achieved. In all cases
the integration of the equations is started at a GW frequency of $20$
Hz.}
\end {table*}

\subsubsection*{Waveforms and matches}
In Tab.\ \ref{CauchyT}, for three typical choices of BBH masses, we
perform a convergence test using Cauchy's criterion~\cite{DIS1},
namely, the sequence ${\rm T}_N$ converges if and only if for each
$k$, $\braket{{\rm T}_N}{{\rm T}_{N+k}} \rightarrow 1$ as $N
\rightarrow \infty$. One requirement of this criterion is that
$\braket{{\rm T}_N}{{\rm T}_{N+1}} \rightarrow 1$ as $N \rightarrow
\infty$, and this is what we test in Tab.\ \ref{CauchyT}, setting
$\mathrm{T}_N \equiv \mathrm{T}(N, N+0.5)$.  The values quoted assume
maximization on the extrinsic parameters but not on the intrinsic
parameters.  [For the case $(10+10)M_\odot$, we show in parentheses
the maxmax matches obtained by maximizing with respect to the
intrinsic and extrinsic parameters, together with the intrinsic
parameters $M$ and $\eta$ of $\mathrm{T}_{N+1}$ where the maxima are
attained.] These results suggest that the PN expansion is far from
converging. However, the very low matches between $N=1$ and $N=2$, and
between $N=2$ and $N=3$, are due to the fact that the 2.5PN flux
goes to zero before the MECO can be reached.
If we redefine $\mathrm{T}_2$ as $\mathrm{T}(2, 2)$ instead of
$\mathrm{T}(2, 2.5)$, we obtain the higher values shown in brackets is
Tab.\ \ref{CauchyT}.
\begin{figure}[t]
\begin{center}
\epsfig{file =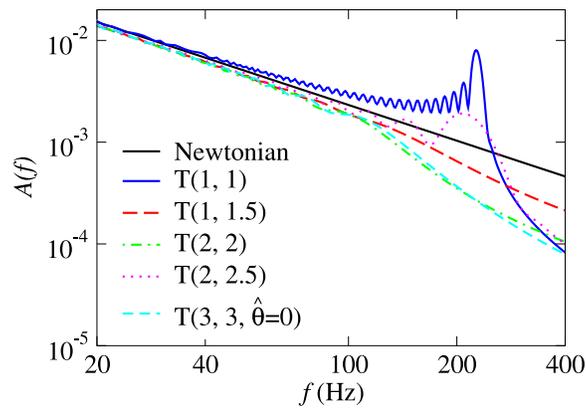,width=\sizeonefig,angle=0}
\end{center}
\caption{\label{AmplT}Frequency-domain amplitude versus frequency for
the T-approximated (restricted) waveforms, at different PN orders, for
a $(15 + 15 )M_\odot$ BBH. The T$(3,3.5,\hat{\theta}=0)$ curve, not plotted, is almost identical to the T$(3,3,\hat{\theta}=0)$ curve.}
\end{figure}

In Fig.~\ref{AmplT}, we plot the frequency-domain amplitude of the
T-approximated waveforms, at different PN orders, for a $(15 +
15)M_\odot$ BBH. The Newtonian amplitude, ${\cal A}_{\rm Newt}(f) =
f^{-7/6}$, is also shown for comparison. In the $\mathrm{T}(1, 1)$ and
$\mathrm{T}(2, 2.5)$ cases, the flux function goes to zero before
$v=v_{\rm MECO}^{T_N}$; this means that the radiation-reaction effects
become negligible during the last phase of evolution, so the binary is
able to spend many cycles at those final frequencies, skewing the
amplitude with respect to the Newtonian result. For $\mathrm{T}(2,
2)$, $\mathrm{T}(3, 3)$ and $\mathrm{T}(3, 3.5)$, the evolution is
stopped at $v=v_{\rm MECO}^{T_N}$, and, although $f_{\rm MECO}^{\rm
GW} \simeq 270\mbox{--}300$ Hz (see Tab.\ \ref{MECOT}) the amplitude
starts to deviate from $f^{-7/6}$ around $100$ Hz. This is a
consequence of the abrupt termination of the signal in the time
domain.

The effect of the arbitrary parameter $\hat{\theta}$ on the T
waveforms can be seen in Tab.\ \ref{VIII} in the intersection between
the rows and columns labeled T$(3,3.5,+2)$ and T$(3,3.5,-2)$. For
three choices of BBH masses, this table shows the maxmax matches
between the \emph{search} models at the top of the columns and the
\emph{target} models at the left end of the rows, \emph{maximized over
the mass parameters of the search models in the columns}. These
matches are rather high, suggesting that for the range of BBH masses 
we are concerned, the effect of changing
$\hat{\theta}$ is just a remapping of the BBH mass
parameters. Therefore, in the following we shall consider only the
case of $\hat{\theta} = 0$.

A quantitative measure of the difference between the $\mathrm{T}(2,
2)$, $\mathrm{T}(2, 2.5)$ and $\mathrm{T}(3, 3.5)$ waveforms can be
seen in Tab.\ \ref{VIa} in the intersection between the rows and
columns labeled T$(\ldots)$.  For four choices of BBH masses, this
table shows the maxmax matches between the search models in the
columns and the target models in the rows, maximized over the
search-model parameters $M$ and $\eta$; in the search, $\eta$ is
restricted to its physical range $0 < \eta \leq 1/4$, where $0$
corresponds to the test-mass limit, while $1/4$ is obtained in the
equal-mass case. These matches can be interpreted as the fitting
factors [see Eq.\ \eqref{eq:ffd}] for the projection of the target
models onto the search models.  For the case $\mathrm{T}(2, 2.5)$ the
values are quite low: if the $\mathrm{T}(3, 3.5)$ waveforms turned out
to give the true physical signals and if we used the $\mathrm{T}(2,
2.5)$ waveforms to detect them, we would lose $\sim 32\mbox{--}49 \%$
of the events. The model $\mathrm{T}(2, 2)$ would do match better,
although it would still not be very faithful.  Once more, the difference
between $\mathrm{T}(2, 2)$ and $\mathrm{T}(2, 2.5)$ is due to the fact
that the 2.5PN flux goes to zero before the BHs reach the MECO.

\subsection{Adiabatic PN resummed methods: Pad\'e approximants}
\label{subsec3.2}

The PN approximation outlined above can be used quite generally to
compute the shape of the GWs emitted by BNSs or BBHs, but it
\emph{cannot be trusted} in the case of binaries with comparable
masses in the range $M \simeq 10\mbox{--}40M_{\odot}$, because for 
these sources LIGO and VIRGO will detect the GWs emitted when the
motion is strongly relativistic, and the convergence of the PN series
is very slow. To cope with this problem, Damour, Iyer and
Sathyaprakash~\cite{DIS1} proposed a new class of models based on the
systematic application of Pad\'e resummation to the PN expansions of
${\cal E}(v)$ and ${\cal F}(v)$. This is a standard mathematical
technique used to accelerate the convergence of poorly converging or
even divergent power series.

If we know the function $g(v)$ only through its Taylor approximant
$G_N(v) = g_0 + g_1\,v + \cdots + g_N\,v^N \equiv T_N[g(v)]$, the 
central idea of Pad\'e resummation~\cite{BO} is the replacement of the
power series $G_N(v)$ by the sequence of rational functions
\beq
P_K^M[g(v)] = \frac{A_M(v)}{B_K(v)} \equiv
\frac{\sum_{j=0}^M a_j\,v^j}{\sum_{j=0}^K b_j\,v^j}\,,
\eeq
with $M+K=N$ and $T_{M+K}[P_K^M(v)]=G_N(v)$ (without loss of
generality, we can set $b_0=1$). We expect that for $M, K
\rightarrow + \infty$, $P_K^M[g(v)]$ will converge to $g(v)$ more
rapidly than $T_N[g(v)]$ converges to $g(v)$ for $N \rightarrow +
\infty$.

\subsubsection*{PN energy and flux}
\label{subsubsec3.2.1}

Damour, Iyer and Sathyaprakash~\cite{DIS1}, and then Damour, 
Sch\"afer and Jaranowski~\cite{EOB3PN}, proposed the following 
Pad\'e-approximated (P-approximated) 
${\cal E}_{P_N}(v)$ and ${\cal F}_{P_N}(v)$ (for $N=2,3$):
\bea
{\cal E}_{P_N} &=& \sqrt{1+2\eta\,\sqrt{1+e_{P_N}(v)}-1}-1\,, \\
{\cal F}_{P_N} &=& \frac{32}{5}\eta^2\,v^{10}\,
\frac{1}{1-v/v^{P_N}_{\rm pole}}\,f_{P_N}(v,\eta)\,,
\eea 
where
\beq e_{P_2}(v) = -v^2\,\frac{1+\frac{1}{3}\eta-\left (
4-\frac{9}{4}\eta+\frac{1}{9}\eta^2 \right )\,v^2}
{1+\frac{1}{3}\eta-\left ( 3-\frac{35}{12}\eta \right )\,v^2}\,, \eeq
\beq
e_{P_3}(v) = -v^2\,\frac{1-\left (1+ \frac{1}{3}\eta +w_3(\eta) \right)\,v^2-
\left (3-\frac{35}{12}\eta-\left ( 1+\frac{1}{3}\eta\right )\,w_3(\eta)\right )\,v^4}{
1-w_3(\eta)\,v^2}\,,
\eeq
\beq
w_3=\frac{40}{36-35\eta}\,\left [\frac{27}{10}+\frac{1}{16}
\left (\frac{41}{4}\pi^2-\frac{4309}{15}\right )\eta+\frac{103}{120}\eta^2
-\frac{1}{270}\eta^3\right ]\,,
\eeq
\beq
f_{P_2}(v) = \left(
1 + \frac{c_1\,v}{1 + \frac{c_2\,v}{1 + 
\ldots }}
\right)^{-1}
\quad \text{(up to $c_5$)},
\eeq
\beq
f_{P_3}(v) = \left (1 -\frac{1712}{105}\,v^6\,\log \frac{v}{v_{\rm MECO}^{P_2}}\right )\, 
\left(
1 + \frac{c_1\,v}{1 + \frac{c_2\,v}{1 + \ldots}}
\right)^{-1}
\text{(up to $c_7$)}.
\eeq
Here the dimensionless coefficients $c_i$ depend only on $\eta$. The 
$c_k$'s are explicit functions of the coefficients $f_k$ ($k =1, ... 5$),
\bea && c_1 = - f_1\,, \quad \quad c_2 = f_1 - \frac{f_2}{f_1}\,,
\quad \quad c_3 = \frac{f_1\,f_3 - f_2^2}{f_1\,(f_1^2 - f_2)}\,,\\ &&
c_4 = -\frac{f_1\,(f_2^3 + f_3^2 + f_1^2\,f_4 - f_2\,(2\,f_1\,f_3 +
f_4))} {(f_1^2 -f_2)\,(f_1\,f_3 - f_2^2)}\,,\\ && c_5 = -\frac{(f_1^2
- f_2)\,(-f_3^3 + 2f_2\,f_3\,f_4 - f_1\,f_4^2 - f_2^2\,f_5 +
f_1\,f_3\,f_5)}{(f_1\,f_3 - f_2^2)\, (f_2^3 + f_3^2 + f_1^2\,f_4 -
f_2\,(2\,f_1\,f_3 + f_4))}\,, \eea
where
\beq
f_k = {\cal F}_k - \frac{{\cal F}_{k-1}}{v^{P_2}_{\rm pole}} \,.
\eeq
Here ${\cal F}_k$ is given by Eqs.~(\ref{f1})--(\ref{f4}) [for $k=6$
and $k=7$, the term $-856/105 \log 16 v^2$ should be replaced by
$-856/105 \log 16 (v^{P_2}_{\rm MECO})^2$]. The coefficients $c_7$ and
$c_8$ are straightforward to compute, but we do not show them because
they involve rather long expressions. The quantity $v^{P_2}_{\rm
MECO}$ is the MECO of the energy function $e_{P_2}$ [defined by $d
e_{P_2}(v) / dv = 0$].
\begin{figure}[t]
\begin{center}
\epsfig{file = 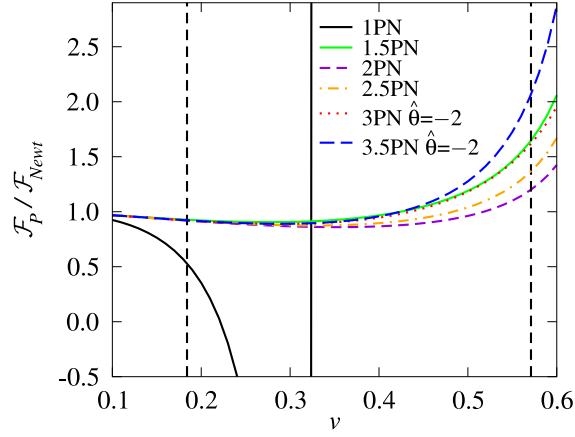,width=\sizetwofig}
\caption{Normalized flux function ${\cal F}_{P_N}/{\cal F}_{\rm Newt}$ versus
$v$, at different PN orders.  The two long-dashed vertical lines give
$v \simeq 0.18$ and $v \simeq 0.53$; they show the velocity range that
corresponds to the LIGO frequency band $40 \leq f_{\rm GW} \leq 240$
Hz for BBHs with total mass in the range $10\mbox{--}40 M_{\odot}$. Compare with Fig.\ \ref{Fig1}.\label{Fig3}}
\end{center}
\end{figure}
The quantity $v^{P_2}_{\rm pole}$, given by
\beq v^{P_2}_{\rm pole}=
\frac{1}{\sqrt{3}}\,\sqrt{\frac{1+\frac{1}{3}\eta}{1-\frac{35}{36}\eta}}\,,
\eeq is the pole of $e_{P_2}$, which plays an important role in the
scheme proposed by Damour, Iyer and Sathyaprakash~\cite{DIS1}. It is
used to augment the Pad\'e resummation of the PN expanded energy and
flux with information taken from the test-mass case, where the flux
(known analytically up to 5.5PN order) has a pole at the light
ring. Under the hypothesis of \emph{structural stability} \cite{DIS1},
the flux should have a pole at the light ring also in the
comparable-mass case. In the test-mass limit, the light ring
corresponds to the pole of the energy, so the analytic structure of
the flux is modified in the comparable-mass case to include
$v^{P_2}_{\rm pole}(\eta)$. At 3PN order, where the energy has no
pole, we choose (somewhat arbitrarily) to keep using the value
$v^{P_2}_{\rm pole}(\eta)$; the resulting 3PN approximation to the
test-mass flux is still very good.

In Fig.~\ref{Fig3}, we plot the P-approximants for
the flux function $\mathcal{F}_{P_N}$(v), at different PN orders. Note
that at 1PN order the P-approximant has a pole. At the LIGO-I
peak-sensitivity frequency, $153$ Hz, for a (10+10)$M_\odot$ BBH, the
value of $v$ is $\simeq 0.362$, and the percentage difference in
$\mathcal{F}_{P_N}(0.362)$, between successive PN orders is $1.5{\rm
PN} \rightarrow 2{\rm PN}: -8\%$; $2{\rm PN} \rightarrow 2.5{\rm PN}:
+2.2\%$; $2.5{\rm PN} \rightarrow 3{\rm PN}\, (\hat{\theta}=-2):
+3.6\%$; $3{\rm PN} \rightarrow 3.5{\rm PN}\, (\hat{\theta}=-2):
+0.58\%$. So the percentage difference decreases as we increase the PN
order. While in the test-mass limit it is known that the P-approximants
converge quite well to the known exact flux function (see Fig.~3 of
Ref.~\cite{DIS1}), in the equal-mass case we cannot be sure
that the same is happening, because the exact flux function is
unknown. (If we assume that the equal-mass flux function is a smooth
deformation of the test-mass flux function, with $\eta$ the
deformation parameter, then we could expect that the P-approximants
are converging.)
\begin{figure}[t]
\begin{center}
\epsfig{file = 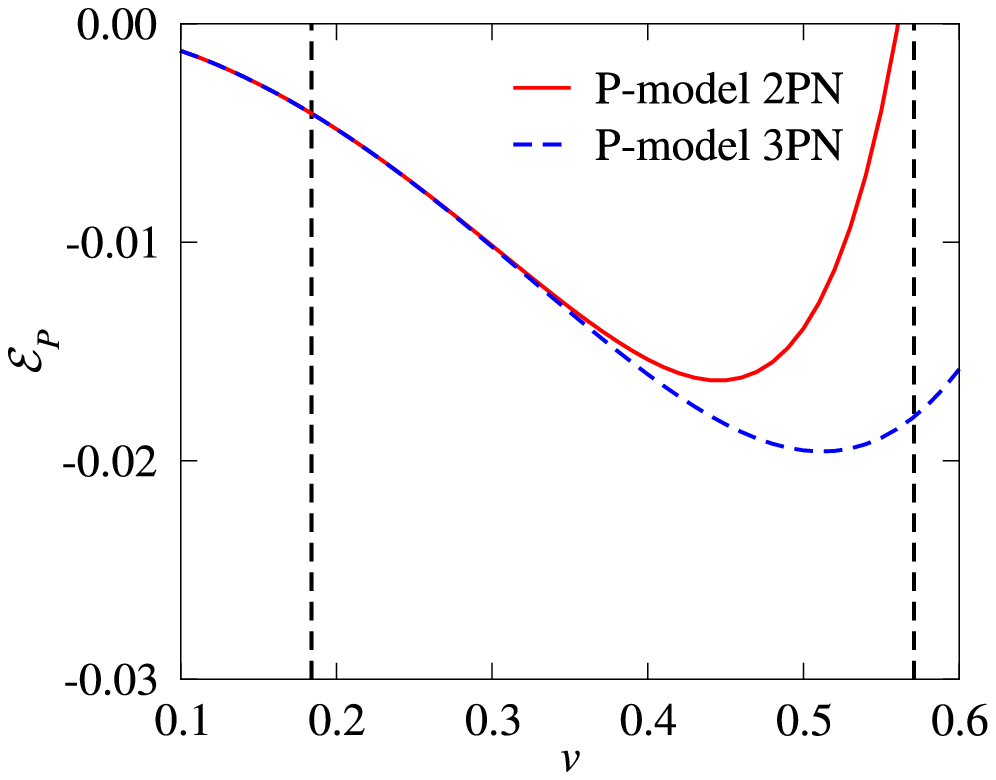,width=\sizetwofig}
\hspace{0.5cm}
\epsfig{file = 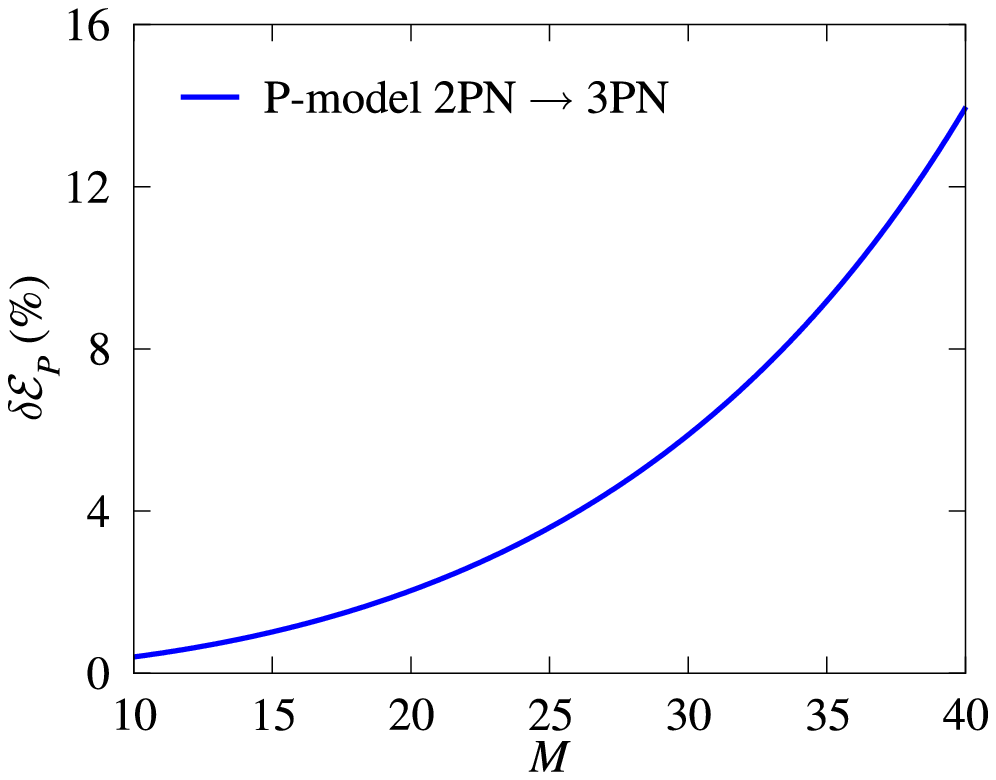,width=\sizetwofig}
\caption{In the left panel, we plot the energy function
${\cal E}_{P_N}$ versus $v$, at different PN orders.
In the right panel, we plot the percentage difference between 2PN and 3PN
P-approximants, $\delta {\cal E}_P(v_\mathrm{peak}) = 100\, |[{\cal
E}_{P_3}(v_\mathrm{peak}) - {\cal E}_{P_2}(v_\mathrm{peak})]/{\cal
E}_{P_2}(v_\mathrm{peak})|$ versus the total mass $M$, again evaluated at
the LIGO-I peak-sensitivity GW frequency $f_\mathrm{peak}=153$ Hz 
[note: $v_\mathrm{peak} = (\pi M f_\mathrm{peak})^{1/3}$].\label{Fig4}}
\end{center}
\end{figure}
In the left panel of Fig.~\ref{Fig4}, we plot the P-approximants to
the energy function as a function of $v$, at 2PN and 3PN orders; in the
right panel, we plot the percentage difference between 2PN and 3PN
P-approximants to the energy function, as a function of the total mass $M$,
evaluated at the LIGO-I peak-sensitivity GW frequency
$f_\mathrm{peak}=153$ Hz.
\begin{table*} 
\begin {tabular}{l|c|rrrr|c}
\hline \hline
\multicolumn{1}{c|}{$N$} & \multicolumn{6}{c}{$\left <P_N,P_{N+1} \right >$} \\
& \multicolumn{1}{c}{$(20+5)M_\odot$}
     & \multicolumn{4}{c}{$(10+10)M_\odot$}
     & \multicolumn{1}{c}{$(15+15)M_\odot$} \\
\hline
 2  ($\hat{\theta} =+2$) & 0.902  & 0.915 &(0.973, &20.5, &0.242) & 0.868 \\
 2  ($\hat{\theta} =-2$) & 0.931  & 0.955 &(0.982, &20.7, &0.236) & 0.923 \\
\hline \hline
\end {tabular}
\caption{\label{CauchyP}Test for the Cauchy convergence of the
P-approximants. The values quoted are maxmax matches obtained by
maximizing with respect to the extrinsic parameters, but not to the
intrinsic parameters (i.e., the matches are computed for P waveforms
with the same masses, but different PN orders). Here we define
$\mathrm{P}_2 = \mathrm{P}(2, 2.5)$, $\mathrm{P}_3 = \mathrm{P}(3,
3.5)$.  The values in parentheses are the maxmax matches obtained by
maximizing with respect to the extrinsic \emph{and} intrinsic
parameters, shown together with the $\mathrm{P}_{N+1}$ parameters $M$
and $\eta$ where the maxima are attained.  In all cases the
integration of the equations is started at a GW frequency of $20$ Hz.}
\end{table*}
\begin{figure}[t]
\begin{center}
\epsfig{file = 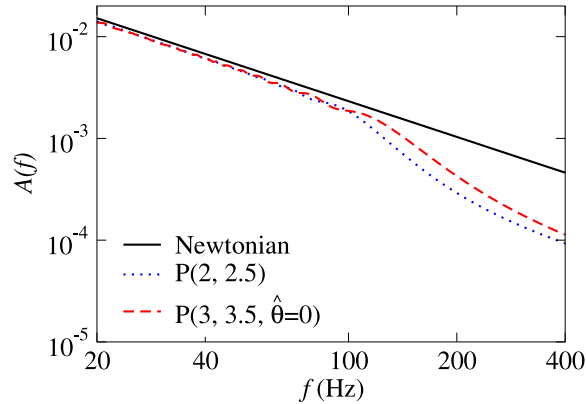,width=\sizeonefig,angle=0}
\end{center}
\caption{\label{AmplP}Frequency-domain amplitude versus frequency for
the P-approximated (restricted) waveform, at different PN orders, for
a $(15 + 15)M_\odot$ BBH.}
\end{figure}
\subsubsection*{Definition of the models}
\label{subsubsec3.2.2}

When computing the waveforms for P-approximant adiabatic models, the
integration of the Eqs.\ \eqref{3.2} is stopped at $v = v_{\rm
MECO}^{P_N}$, which is the solution of the equation $d {\cal
E}_{P_N}(v)/dv = 0$.  The corresponding GW frequency will be the
\emph{ending frequency} for these waveforms, and in Tab.\ \ref{MECOT}
we show this frequency for typical BBH masses. Henceforth, we shall
refer to the P-approximant models as ${\rm P}(n{\rm PN}, m{\rm PN})$,
and we shall consider $(n{\rm PN}, m{\rm PN}) = (2, 2.5), (3, 3.5,
\hat{\theta})$. [Recall that $n$PN and $m$PN are the maximum
post--Newtonian order of the terms included, respectively, in the
energy and flux functions $\mathcal{E}(v)$ and $\mathcal{F}(v)$; at
3PN order we need to indicate also a choice of the arbitrary flux
parameter $\hat{\theta}$.]

\subsubsection*{Waveforms and matches}
\label{subsubsec3.2.3}

In Tab.\ \ref{CauchyP}, for three typical choices of BBH masses, we
perform a convergence test using Cauchy's criterion~\cite{DIS1}. The
values are quite high, especially if compared to the same test for the
T-approximants when the 2.5PN flux is used, see
Tab.\ \ref{CauchyT}. However, as we already remarked, we do not have a
way of testing whether they are converging to the true limit.  In
Fig.~\ref{AmplP}, we plot the frequency-domain amplitude of the
P-approximated (restricted) waveform, at different PN orders, for a
$(15 + 15) M_\odot$ BBH. The Newtonian amplitude, ${\cal A}_{\rm
Newt}(f) = f^{-7/6}$, is also shown for comparison. At 2.5PN and 3.5PN
orders, the evolution is stopped at $v=v_{\rm MECO}^{P_N}$; although
$f_{\rm MECO}^{\rm GW} \simeq 190-290$ Hz (see Tab.\ \ref{MECOT}), the
amplitude starts to deviate from $f^{-7/6}$ around $100$ Hz, well
inside the LIGO frequency band. Again, this is a consequence of the
abrupt termination of the signal in the time domain.
\begin{figure}[t]
\begin{center}
\epsfig{file =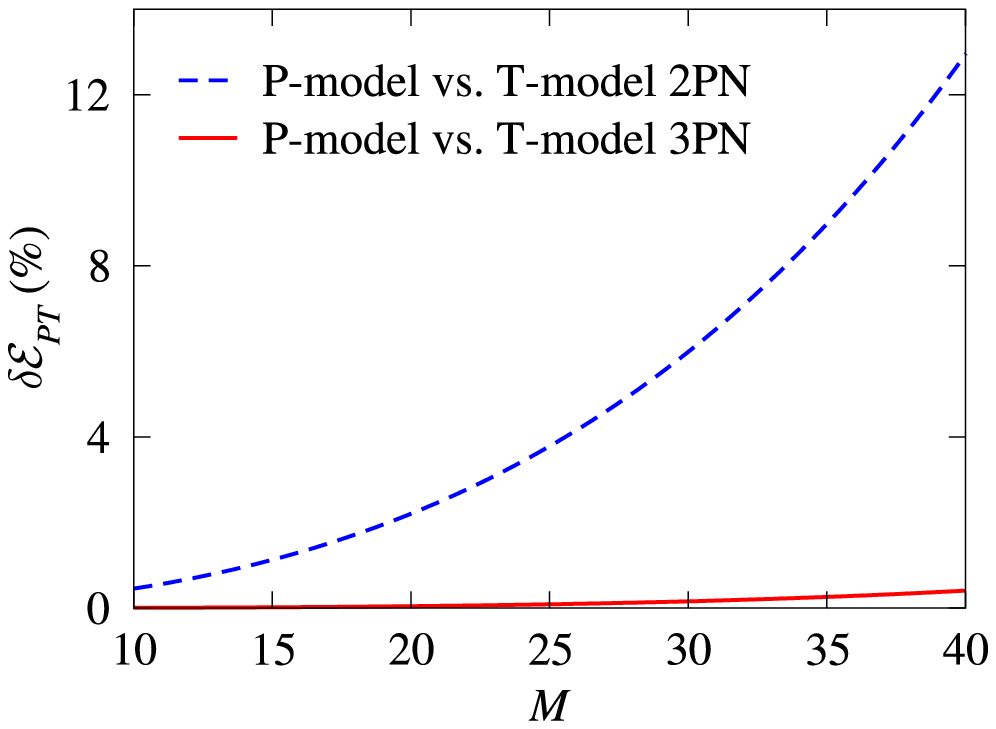,width=\sizeonefig}
\hspace{0.5cm}
\epsfig{file = 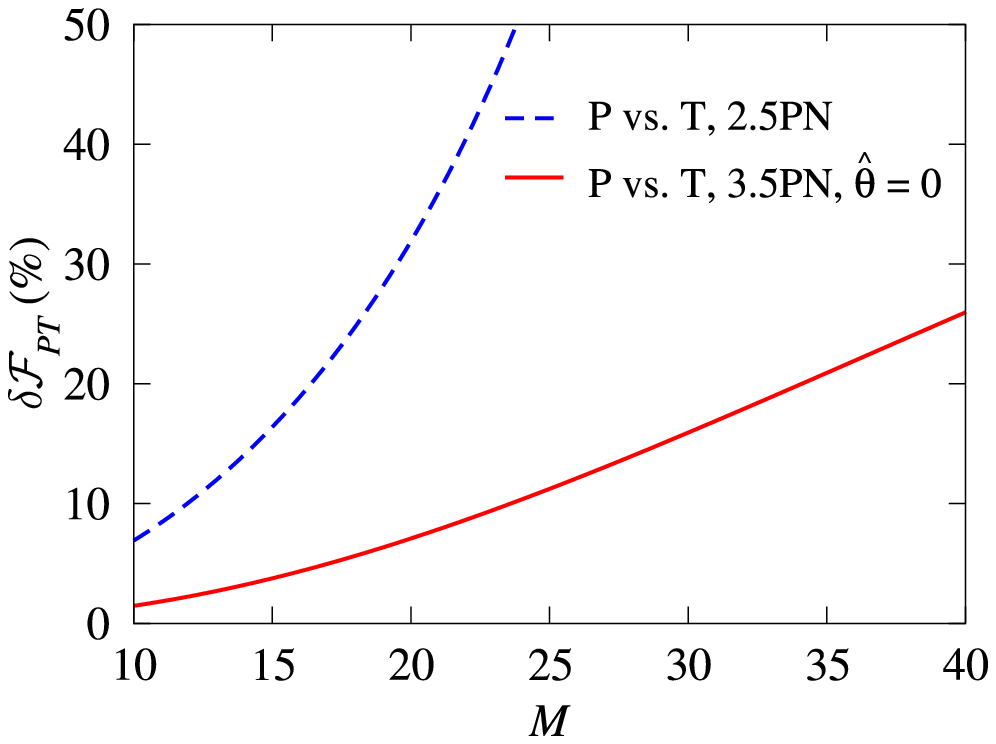,width=\sizetwofig}
\caption{In the left panel, we plot the percentage difference $\delta
{\cal E}_{PT}(v_\mathrm{peak}) = 100\, |[{\cal
E}_{P_N}(v_\mathrm{peak}) - {\cal E}_{T_N}(v_\mathrm{peak})]/{\cal
E}_{P_N}(v_\mathrm{peak})|$ versus the total mass $M$, for $N=2,3$, at
the LIGO-I peak-sensitivity GW frequency $f_\mathrm{peak}=153$ Hz
[note: $v_\mathrm{peak} = (\pi M f_\mathrm{peak})^{1/3}$]. In the
right panel, we plot the percentage difference between 2PN and 3PN
P-approximants, $\delta {\cal F}_P(v_\mathrm{peak}) = 100\, |[{\cal
F}_{P_3}(v_\mathrm{peak}) - {\cal F}_{P_2}(v_\mathrm{peak})]/{\cal
F}_{P_2}(v_\mathrm{peak})|$ versus the total mass $M$, again evaluated
at the LIGO-I peak-sensitivity GW frequency $f_\mathrm{peak}=153$
Hz.\label{Fig5}}
\end{center}
\end{figure}

A quantitative measure of the difference between the $\mathrm{P}(2,
2.5)$ and $\mathrm{P}(3,3.5)$ waveforms can be seen in Tab.\ \ref{VIa}
in the intersection between the rows and columns labeled P$(\ldots)$.
For three choices of BBH masses, this table shows the maxmax matches
between the search models in the columns and the target models in the
rows, maximized over the search-model parameters $M$ and $\eta$, with
the restriction $0 < \eta \leq 1/4$. These matches are quite high, but
the models are not very faithful to each other.  The same table shows
also the maximized matches (i.e., fitting factors) \emph{between T and
P models}. These matches are low between $\mathrm{T}(2, 2.5)$ and
$\mathrm{P}(2, 2.5)$ (and viceversa), between $\mathrm{T}(2, 2.5)$ and
$\mathrm{P}(3, 3.5)$ (and viceversa), but they are high between
$\mathrm{T}(2, 2)$, $\mathrm{T}(3, 3.5)$ and 3PN P-approximants
(although the estimation of mass parameters is imprecise).  Why this
happens can be understood from Fig.~\ref{Fig5} by noticing that at 3PN
order the percentage difference between the T-approximated and
P-approximated binding energies is rather small ($\leq 0.5\%$), and
that the percentage difference between the T-approximated and
P-approximated fluxes at 3PN order 
(although still $\sim 10\%$) is much smaller than at
2PN order.

\section{Nonadiabatic models}
\label{sec4}

By contrast with the models discussed in Sec.\ \ref{sec3}, in
nonadiabatic models we solve equations of motions that involve
(almost) all the degrees of freedom of the BBH systems. Once again,
all waveforms are computed in the restricted approximation of Eq.\
\eqref{eq:quadapp}, taking the GW phase $\varphi_\mathrm{GW}$ as twice
the orbital phase $\varphi$.

\subsection{Nonadiabatic PN expanded methods: Hamiltonian formalism}
\label{subsec4.1}

Working in the ADM gauge, Damour, Jaranowski and G. Sch\"afer have
derived a PN expanded Hamiltonian for the general-relativistic two-body dynamics~\cite{DJS,DJSd,EOB3PN}:
\beq
\widehat{H}({\mathbf q},{\mathbf p})
= \widehat{H}_{\rm Newt}({\mathbf q},{\mathbf p})
+ \widehat{H}_{\rm 1PN}({\mathbf q},{\mathbf p})
+ \widehat{H}_{\rm 2PN}({\mathbf q},{\mathbf p})
+ \widehat{H}_{\rm 3PN}({\mathbf q},{\mathbf p})\,,
\label{ham}
\eeq
where
\bea
\widehat{H}_{\rm Newt}\left({\bf q},{\bf p}\right) &=& \frac{\pp}{2} -
\frac{1}{q}\,, \label{eq:hfirst}\\
\widehat{H}_{\rm 1PN}\left({\bf q},{\bf p}\right) &=& \frac{1}{8}(3\eta-1)\ppp^2
- \frac{1}{2}\left[(3+\eta)\pp+\eta\np^2\right]\frac{1}{q} + \frac{1}{2q^2}\,,\\
\widehat{H}_{\rm 2PN}\left({\bf q},{\bf p}\right)
&=& \frac{1}{16}\left(1-5\eta+5\eta^2\right)\ppp^3
+ \frac{1}{8} \left[
\left(5-20\eta-3\eta^2\right)\ppp^2-2\eta^2\np^2\pp-3\eta^2\np^4 \right]\frac{1}{q}
\nonumber \\
&& + \frac{1}{2} \left[(5+8\eta)\pp+3\eta\np^2\right]\frac{1}{q^2}
- \frac{1}{4}(1+3\eta)\frac{1}{q^3}\,,\\
\widehat{H}_{\rm 3PN}\left({\bf q},{\bf p}\right)
&=& \frac{1}{128}\left(-5+35\eta-70\eta^2+35\eta^3\right)\ppp^4
\nonumber \\
&& + \frac{1}{16}\left[
\left(-7+42\eta-53\eta^2-5\eta^3\right)\ppp^3
+ (2-3\eta)\eta^2\np^2\ppp^2
+ 3(1-\eta)\eta^2\np^4\pp - 5\eta^3\np^6
\right]\frac{1}{q}
\nonumber \\
&& +\left[ \frac{1}{16}\left(-27+136\eta+109\eta^2\right)\ppp^2
+ \frac{1}{16}(17+30\eta)\eta\np^2\pp + \frac{1}{12}(5+43\eta)\eta\np^4
\right]\frac{1}{q^2} \nonumber \\
&& +\left\{ \left[ -\frac{25}{8} + \left(\frac{1}{64}\pi^2-\frac{335}{48}\right)\eta 
- \frac{23}{8}\eta^2 \right]\pp
+ \left(-\frac{85}{16}-\frac{3}{64}\pi^2-\frac{7}{4}\eta\right)\eta\np^2 
\right\}\frac{1}{q^3}
\nonumber \\
&& + \left[ \frac{1}{8} + \left(\frac{109}{12}-\frac{21}{32}\pi^2\right)\eta 
\right]\frac{1}{q^4}. \label{eq:hlast}
\eea
Here the reduced Non--Relativistic Hamiltonian in the center-of-mass frame,
$\widehat{H} \equiv H^{\rm NR}/\mu$, is written as a
function of the reduced canonical variables ${\bf p}\equiv {\bf
p_1}/\mu=-{\bf p_2}/\mu$, and ${\bf q} \equiv ({\bf x}_1 - {\bf
x}_2)/M$, where ${\bf x_1}$ and ${\bf x_2}$ are the positions of the
BH centers of mass in quasi--Cartesian ADM coordinates (see
Refs.~\cite{DJS,DJSd,EOB3PN}); the scalars $q$ and $p$ are the
(coordinate) lengths of the two vectors; and the vector ${\bf n}$ is
just ${\bf q}/q$.

\subsubsection*{Equations of motion}
\label{subsubsec4.1.1}

We now restrict the motion to a plane, and we introduce
radiation-reaction (RR) effects as in Ref.~\cite{BD2}. The equations
of motion then read (using polar coordinates $r$ and $\vphi$ obtained
from the $\mathbf{q}$ with the usual Cartesian-to-polar
transformation)
\beq
\frac{dr}{d \widehat{t}} = \frac{\pa \widehat{H}}{\pa p_r}(r,p_r,p_\vphi)\,, 
\quad \frac{d \vphi}{d \widehat{t}} \equiv \widehat{\omega} = 
\frac{\pa \widehat{H}}{\pa p_\vphi}(r,p_r,p_\vphi)\,, 
\label{eq:hamone}
\eeq
\beq
\frac{d p_r}{d \widehat{t}} = - \frac{\pa \widehat{H}}{\pa r}(r,p_r,p_\vphi) + \widehat{F}^r(r,p_r,p_{\varphi})\,, 
\quad \frac{d p_\vphi}{d \widehat{t}} = \widehat{F}^\vphi[\widehat{\omega} (r,p_r,p_{\varphi})]\,, 
\label{eq:hamtwo}
\eeq
where $\widehat{t} = t/M$, $\widehat{\omega} = \omega M$; and where
$\widehat{F}^\vphi \equiv {F}^\vphi/\mu $ and $\widehat{F}^r \equiv
{F}^r/\mu$ are the reduced angular and radial components of the RR
force. Assuming ${F}^r \ll {F}^\vphi$~\cite{BD2}, averaging over an
orbit, and using the balance equation (\ref{3.1}), we can express the
angular component of the radiation-reaction force in terms of the GW
flux at infinity~\cite{BD2}. More explicitly, if we use the
P-approximated flux, we have
\beq
\label{fluxP}
\widehat{F}^\vphi \equiv 
{F}_{P_N}[v_{\omega}] = 
- \frac{1}{\eta\,v_\omega^3}\,
{\cal F}_{P_N}[v_\omega]=
- \frac{32}{5}\,\eta\,v_\omega^7\,
\frac{f_{P_N}(v_\omega;\eta)}{1 - v_\omega/v_{\rm pole}^{P_2}(\eta)}\,,
\eeq
while if we use the T-approximated flux we have
\beq
\label{fluxT}
\widehat{F}^\vphi \equiv 
{F}_{T_N}[v_{\omega}] = - \frac{1}{\eta\,v_\omega^3}\,
{\cal F}_{T_N}[v_\omega],
\eeq
where $v_\omega \equiv \widehat{\omega}^{1/3} \equiv (d
\varphi/d\widehat{t})^{1/3}$. This $v_\omega$ is used in Eq.\
\eqref{eq:quadapp} to compute the restricted waveform.  Note that at
each PN order, say $n$PN, we define our Hamiltonian model by evolving
the Eqs.~(\ref{eq:hamone}) and (\ref{eq:hamtwo}) without truncating
the partial derivatives at the $n$PN order (differentiation with
respect to the canonical variables can introduce terms of order higher
than $n$PN).  Because of this choice, and because of the approximation
used to incorporate radiation-reaction effects, these nonadiabatic
models are not, strictly speaking, purely post--Newtonian.
\subsubsection*{Innermost stable circular orbit}

Circular orbits are defined by setting $r=\mathrm{constant}$ while
neglecting radiation-reaction effects. In our PN Hamiltonian models,
this implies $\partial \widehat{H}/\partial p_r = 0$ through
Eq.~\eqref{eq:hamone}; because at all PN orders the Hamiltonian
$\widehat{H}$ [Eqs.\ \eqref{ham}--\eqref{eq:hlast}] is quadratic
in $p_r$, this condition is satisfied for $p_r = 0$; 
in turn, this implies also $\partial \widehat{H}/\partial r = 0$ [through
Eq.~\eqref{eq:hamtwo}], which can be solved for $p_\varphi$. The
orbital frequency is then given by $\hat{\omega} = \partial
\widehat{H}/\partial p_\varphi$.

The stability of circular orbits under radial perturbations depends
on the second derivative of the Hamiltonian:
\bea
\frac{\partial^2
\widehat{H}}{\partial r^2} > 0 \Leftrightarrow \mathrm{stable\,\,orbit}\,;
\quad \quad \frac{\partial^2 \widehat{H}}{\partial r^2} < 0
\Leftrightarrow \mathrm{unstable\,\,orbit}\,.
\eea
For a test particle in Schwarzschild geometry (the $\eta \rightarrow
0$ of a BBH), an Innermost Stable Circular Orbit (ISCO) always exists,
and it is defined by
\beq
{\rm ISCO\,\,(Schwarzschild)}: \quad
{\frac{\partial \widehat{H}^{\rm Schw}}{\partial r}}_{|_{p_r=0}} = 
{\frac{\partial^2 \widehat{H}^{\rm Schw}}{\partial r^2}}_{|_{p_r=0}} =
0,
\eeq
where $\widehat{H}^{\rm Schw}(r,p_r,p_\varphi)$ is the (reduced)
nonrelativistic test-particle Hamiltonian in the Schwarzschild
geometry. Similarly, if such an ISCO exists for the (reduced)
nonrelativistic PN Hamiltonian $\widehat{H}$ [Eq.~\eqref{ham}], it
is defined by
\beq
{\rm ISCO\,\,(Hamiltonian)}: \quad
{\frac{\partial \widehat{H}}{\partial r}}_{|_{p_r=0}} =
{\frac{\partial^2 \widehat{H}}{\partial r^2}}_{|_{p_r=0}} = 0.
\label{eqisco}
\eeq
Any inspiral built as an adiabatic sequence of quasicircular orbits
cannot be extended to orbital separations smaller than the ISCO.  In our
model, we integrate the Hamiltonian equations (\ref{eq:hamone}) and
(\ref{eq:hamtwo}) including terms up to a given PN order, without
re-truncating the equations to exclude terms of higher order that have
been generated by differentiation with respect to the canonical
variables.  Consistently, the value of the ISCO that is relevant to
our model should be derived by solving Eq.~(\ref{eqisco}) without any
further PN truncation.

How is the ISCO related to the Maximum binding Energy for Circular
Orbit (MECO), used above for nonadiabatic models such as T? The PN
expanded energy for circular orbits ${\cal E}_{T_n}(\widehat{\omega})$
at order $n$PN can be recovered by solving the equations
\begin{equation}
\frac{\partial \widehat{H}(r,p_r=0,p_\varphi)}{\partial r} = 0, \quad \quad
\frac{\partial \widehat{H}(r,p_r=0,p_\varphi)}{\partial p_\varphi} = \hat{\omega},
\end{equation}
for $r$ and $p_\varphi$ as functions of $\hat{\omega}$, and by
using the solutions to define
\begin{equation}
\widehat{H}(\hat{\omega}) \equiv \widehat{H}[r(\hat{\omega}),p_r=0,p_\varphi(\hat{\omega})].
\label{CO1}
\end{equation}
Then $\widehat{H}(\hat{\omega} \equiv v^3) = \mathcal{E}_{T_n}(v)$ as
given by Eq.~(\ref{BE}), \emph{if and only if} in this
procedure we are careful to eliminate all terms of order higher than 
$n$PN (see, e.g., Ref.~\cite{LB}).

In the context of nonadiabatic models, the MECO is then defined by
\beq {\rm MECO}: \quad \quad \frac{d \widehat{H}}{d
\widehat{\omega}}=0;
\label{eqMECO}
\eeq  
and it also characterizes the end of adiabatic sequences of circular
orbits. Computing the variation of Eq.~(\ref{CO1}) between nearby circular orbits, 
and setting $p_r=0$, $d p_r=0$, we get
\beq
d \widehat{\omega} = \frac{\partial^2 \widehat{H}}{\partial r \partial p_\varphi}\,
d r + \frac{\partial^2 \widehat{H}}{\partial p_\varphi^2}\,d p_\varphi\,,
\quad \quad 
\frac{\partial^2 \widehat{H}}{\partial r^2}\,
d r + \frac{\partial^2 \widehat{H}}{\partial r \partial p_\varphi}\,dp_\varphi = 0\,;
\eeq
and combining these two equations we get
\beq
\frac{d p_\varphi}{d \widehat{\omega}} = -\frac{\partial^2 \widehat{H}}{\partial r^2}\,
\left [ \left (\frac{\partial^2 \widehat{H}}{\partial r \partial p_\varphi}\right )^2 - 
\frac{\partial^2 \widehat{H}}{\partial p_\varphi^2}\,
\frac{\partial^2 \widehat{H}}{\partial r^2}\right ]^{-1}\,.
\eeq
So finally we can write
\beq
\frac{d \widehat{H}}{d \widehat{\omega}} = 
\frac{\partial \widehat{H}}{\partial p_\varphi}\,
\frac{d p_\varphi}{d \widehat{\omega}} = 
-\frac{\partial^2 \widehat{H}}{\partial r^2}\,
\frac{\partial \widehat{H}}{\partial p_\varphi}\,
\left [ \left (\frac{\partial^2 \widehat{H}}{\partial r \partial p_\varphi}\right )^2 - 
\frac{\partial^2 \widehat{H}}{\partial p_\varphi^2}\,
\frac{\partial^2 \widehat{H}}{\partial r^2}\right ]^{-1}\,. 
\label{rel}
\eeq
Not surprisingly, Eqs.~(\ref{rel}) and \eqref{CO1} together are
formally equivalent to the definition of the ISCO, Eq.~(\ref{eqisco})
[note that the second and third terms on the right-hand side of
Eq.~(\ref{rel}) are never zero.]  Therefore, if we knew the
Hamiltonian $\widehat{H}$ \emph{exactly}, we would find that the MECO
defined by Eq.~(\ref{eqMECO}), is numerically the same as the ISCO
defined by Eq.~(\ref{eqisco}).  Unfortunately, we are working only up
to a finite PN order (say $n$PN); thus, to recover the MECO as given
by Eq.~(\ref{BE}), all three terms on the right-hand side of
Eq.~(\ref{rel}) must be written in terms of $\widehat{\omega}$,
truncated at $n$PN order, then combined and truncated again at $n$PN
order. This value of the MECO, however, will \emph{no longer} be the
same as the ISCO obtained by solving Eq.~(\ref{eqisco}) \emph{exactly
without truncation}.

If the PN expansion was converging rapidly, then the difference between
the ISCO and the MECO would be mild; but for the range of BH masses
that we consider the PN convergence is bad, and the discrepancy is
rather important. The ISCO is present only at 1PN order, with $r_\mathrm{ISCO} = 9.907$ and $\widehat{\omega}_{\rm ISCO} = 0.02833$. The corresponding GW frequencies are given in Tab.\ \ref{MECOT} for a few BBHs with equal masses. At 3PN order we find the formal solution $r^{\rm ISCO} = 1.033$ and $p_\varphi^{\rm ISCO} = 0.355$, but since we do not trust the PN expanded Hamiltonian when the radial coordinate gets so small, we conclude that there is no ISCO at 3PN order.
\begin{table*} 
\begin{tabular}{l|c|rrrr|c|c|rrrr|c}
\hline \hline
\multicolumn{1}{c|}{$N$}
& \multicolumn{6}{c|}{$\left <{\rm HT}_N,{\rm HT}_{N+1} \right>$}
& \multicolumn{6}{c}{$\left <{\rm HP}_N,{\rm HP}_{N+1} \right>$} \\
     & \multicolumn{1}{c}{$(5+20)M_\odot$}
     & \multicolumn{4}{c}{$(10+10)M_\odot$}
     & \multicolumn{1}{c|}{$(15+15)M_\odot$}
     & \multicolumn{1}{c}{$(5+20)M_\odot$}
     & \multicolumn{4}{c}{$(10+10)M_\odot$}
     & \multicolumn{1}{c}{$(15+15)M_\odot$}
     \\
\hline
 0 & 0.118 & 0.191 &(0.553,&13.7,&0.243) & 0.206 						& 0.253 & 0.431 &(0.586,& 16.7,& 0.242) & 0.316 \\
 1 & 0.102 & 0.174 &(0.643,&61.0,&0.240) & 0.170   					& 0.096 & 0.161 &(0.623,& 17.4,& 0.239) & 0.151 \\
 2 ($\hat{\theta} = +2$) & 0.292 & 0.476 &(0.656,&18.6,&0.241)  & 0.377 & 0.266 & 0.369 &(0.618,& 17.6,& 0.240) & 0.325 \\
 2 ($\hat{\theta} = -2$) & 0.287 & 0.431 &(0.671,&19.0,&0.241) & 0.377 & 0.252 & 0.354 &(0.622,& 17.4,& 0.239) & 0.312 \\
\hline \hline
\end{tabular}
\caption{\label{CauchyHT}Test for the Cauchy convergence of the
HT- and HP-approximants. The values quoted are maxmax matches obtained by
maximizing with respect to the extrinsic parameters, but not to the
intrinsic parameters (i.e., the matches are computed for H waveforms
with the same masses, but different PN orders).  Here we define
$\mathrm{HT}_0 = \mathrm{HT}(0, 0)$, $\mathrm{HT}_1 = \mathrm{HT}(1,
1.5)$, $\mathrm{HT}_2 = \mathrm{HT}(2, 2)$ [because the 2.5PN flux
goes to zero before the MECO is reached, so we use the 2PN flux],
$\mathrm{HT}_3 = \mathrm{HT}(3, 3.5, \hat{\theta})$; we also
define $\mathrm{HP}_0 = \mathrm{HP}(0, 0)$, $\mathrm{HP}_1 = \mathrm{HP}(1,
1.5)$, $\mathrm{HP}_2 = \mathrm{HP}(2, 2.5)$, and $\mathrm{HP}_3 =
\mathrm{HP}(3, 3.5, \hat{\theta})$.
The values in
parentheses are the maxmax matches obtained by maximizing with respect
to the extrinsic \emph{and} intrinsic parameters, shown together with
the $\mathrm{H}_{N+1}$ parameters $M$ and $\eta$ where the maxima are
attained. In all cases the integration of the equations is started at
a GW frequency of $20$ Hz.}
\end{table*}

\subsubsection*{Definition of the models}
\label{subsubsec4.1.2}
\begin{figure}[t]
\begin{center}
\epsfig{file = 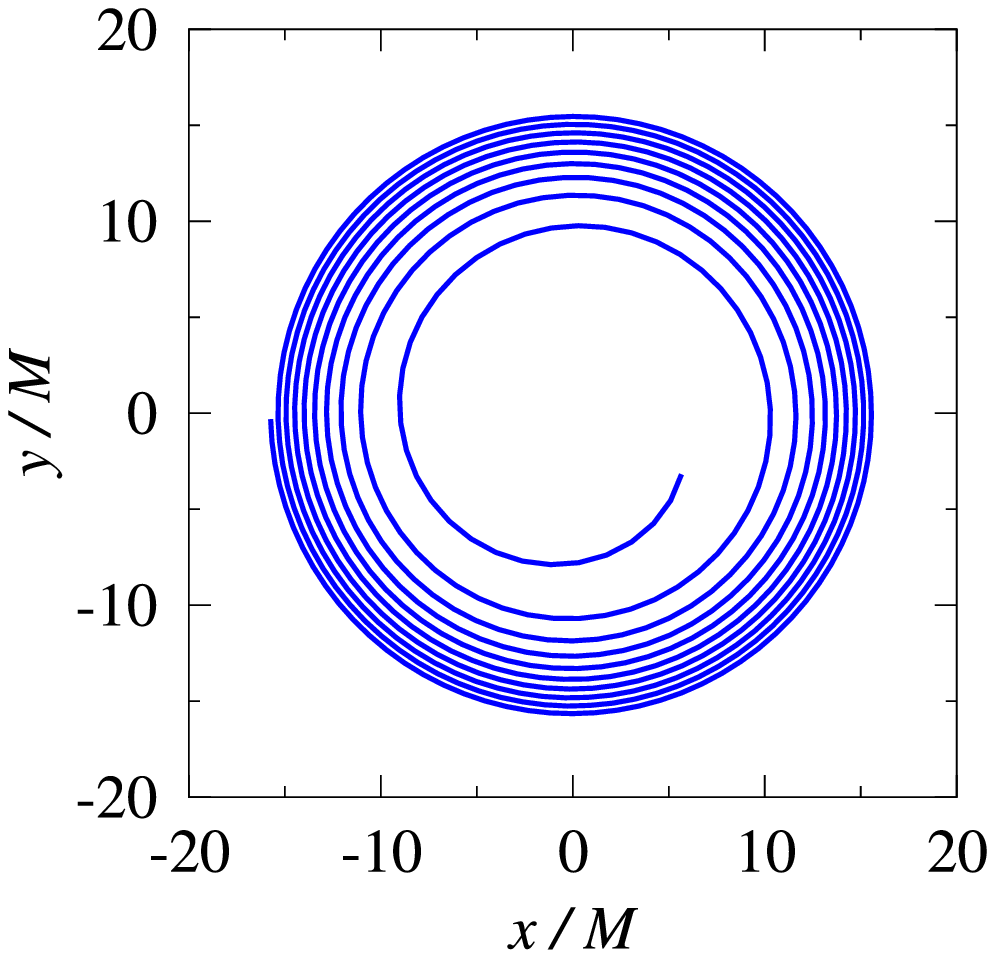,width=\sizetwofig}
\hspace{0.5cm}
\epsfig{file = 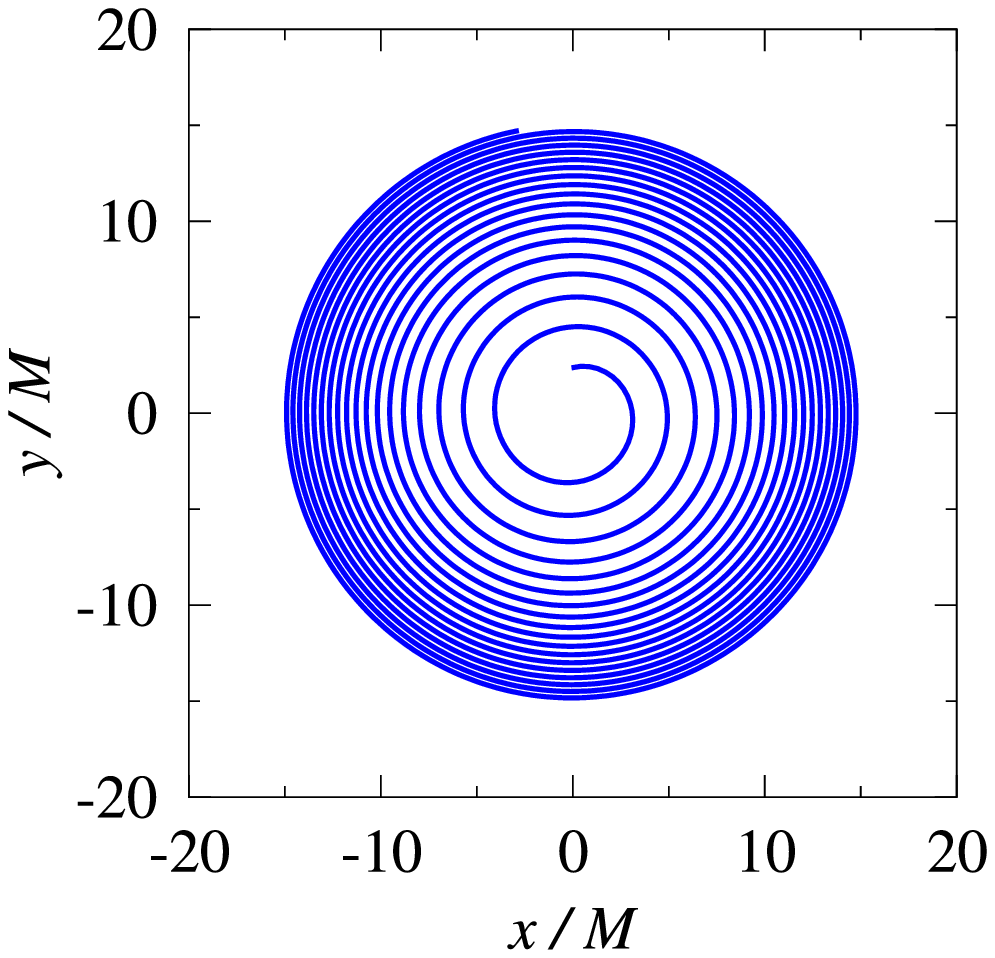,width=\sizetwofig}
\end{center}
\caption{\label{Orb} Inspiraling orbits in $(x,y)$-plane when $\eta=0.25$ 
for HT$(1,1.5)$ (in the left panel) and  HT$(3,3.5,0)$ (in the right panel). For a 
$(15+15)M_\odot$ BBH the evolution starts at $f_{\rm GW}= 34$ Hz and ends at 
$f_{\rm GW}= 97$ Hz for HT$(1,1.5)$
panel and at $f_{\rm GW}= 447$ Hz for  HT$(3,3.5,0)$. The dynamical evolution 
is rather different because at 1PN order there is an ISCO ($r_{\rm ISCO} \simeq 
9.9M$), while at 3PN order it does not exist.}
\end{figure}
In order to build a quasicircular orbit with initial GW frequency
$f_{0}$, our initial conditions $(r_{\rm init},p_{r\,\rm
init},p_{\vphi\,\rm init})$ are set by imposing $\dot{\vphi}_{\rm
init}=\pi f_{0}$, $\dot{p}_{r\,\rm init}=0$ and $d r_{\rm
init}/d\hat{t} = -{\cal F}/(\eta d\hat{H}/dr)_{\rm circ}$, as in Ref.\
\cite{DIS2}. The initial orbital phase $\vphi_{\rm init}$ remains a
free parameter.  For these models, the criterion used to stop the
integration of Eqs.\ \eqref{eq:hamone}, \eqref{eq:hamtwo} is rather
arbitrary.  We decided to push the integration of the dynamical
equations up to the time when we begin to observe unphysical effects
due to the failure of the PN expansion, or when the assumptions that
underlie Eqs.~(\ref{eq:hamtwo}) [such as $\widehat{F}^r \ll
\widehat{F}^\vphi$], cease to be valid. When the 2.5PN flux
is used, we stop the integration when ${\cal F}_{T_N}$ equals 10\% of ${\cal F}_{\rm Newt}$, and we
define the \emph{ending frequency} for these waveforms as the
instantaneous GW frequency at that time. To be consistent with the
assumption of quasicircular motion, we require also that the radial
velocity be always much smaller than the orbital velocity, and we stop
the integration when $|\dot{r}| > 0.3(r\dot{\vphi})$, if this occurs
before ${\cal F}_{T_N}$ equals 10\% of ${\cal F}_{\rm Newt}$. In some cases, during the last stages of
inspiral $\widehat{\omega}$ reaches a maximum and then drops quickly
to zero [see discussion in Sec.~\ref{sec5}].  When this happens, we
stop the evolution at $\dot{\widehat{\omega}}=0$.
\begin{figure}[t]
\begin{center}
\epsfig{file = 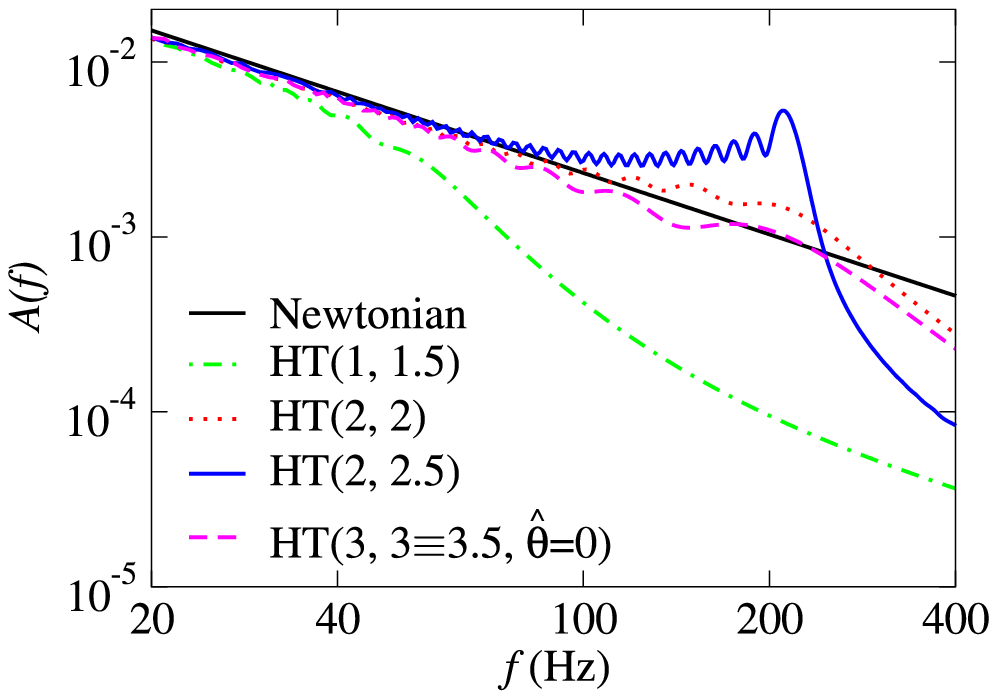,width=\sizetwofig,angle=0}
\hspace{0.5cm}
\epsfig{file = 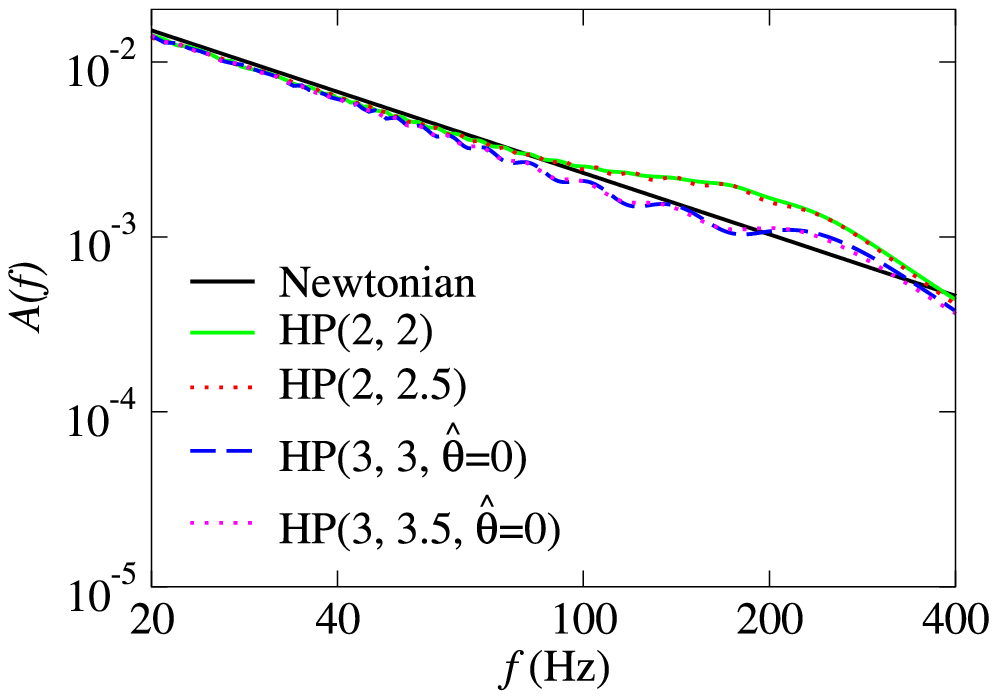,width=\sizetwofig,angle=0}
\end{center}
\caption{\label{AmplH}Frequency-domain amplitude versus frequency for
the HT and HP (restricted) waveforms, at different PN orders, for a
$(15 + 15)M_\odot$ BBH. The HT$(3,3.5,\hat{\theta}=0)$ curve, not plotted, is almost identical to the HT$(3,3,\hat{\theta}=0)$ curve.}
\end{figure}

We shall refer to these models as ${\rm HT}(n{\rm PN},m{\rm PN})$
(when the T-approximant is used for the flux) or ${\rm HP}(n{\rm
PN},m{\rm PN})$ (when the P-approximant is used for the flux), where
$n$PN ($m$PN) denotes the maximum PN order of the terms included in
the Hamiltonian (the flux).  We shall consider $(n\mathrm{PN},
m\mathrm{PN}) = (1, 1.5), (2, 2), (2, 2.5)$, and $(3, 3.5,
\hat{\theta})$ [at 3PN order we need to indicate also a choice of
the arbitrary flux parameter $\hat{\theta}$].

\subsubsection*{Waveforms and matches}
\label{subsubsec4.1.3}
In Tab.~\ref{CauchyHT}, for three typical choices
of BBH masses, we perform a convergence test using Cauchy's
criterion~\cite{DIS1}. The values are very low. For $N=0$ and $N=1$,
the low values are explained by the fact that at 1PN order there is an
ISCO [see the discussion below Eq.\ \eqref{rel}], while at Newtonian and
2PN, 3PN order there is not.  Because of the ISCO, the stopping
criterion [$|\dot{r}| > 0.3 (r \dot{\vphi})$ or
$\dot{\widehat{\omega}}=0$] is satisfied at a much lower frequency,
hence at 1PN order the evolution ends much earlier than in the
Newtonian and 2PN order cases.  In Fig.~\ref{Orb}, we show the
inspiraling orbits in the $(x,y)$ plane for equal-mass BBHs, computed
using the HT$(1, 1.5)$ model (in the left panel) and the HT$(3, 3.5,
0)$ model (in the right panel).  For $N=2$, the low values are due
mainly to differences in the conservative dynamics, that is, to
differences between the 2PN and 3PN Hamiltonians.  Indeed, for a
$(10+10)M_\odot$ BBH we find $\left <{\rm HT}(2, 2),{\rm HT}(3,
2)\right > =0.396$, still low, while $\left <{\rm HT}(2, 2),{\rm
HT}(2, 3.5)\right > =0.662$, considerably higher than the values in
Tab.\ \ref{CauchyHT}.

In Fig.~\ref{AmplH}, we plot the frequency-domain amplitude of the
HT-approximated (restricted) waveforms, at different PN orders, for a
$(15 + 15)M_\odot$ BBH.  The Newtonian amplitude, ${\cal A}_{\rm
Newt}(f) = f^{-7/6}$, is also shown for comparison.  For HT$(1, 1.5)$,
because the ISCO is at $r\simeq 9.9M$, the stopping criterion
$|\dot{r}| > 0.3\, \dot{\vphi}\,r$ is reached at a very low frequency
and the amplitude deviates from the Newtonian prediction already at $f
\sim 50$Hz. For HT$(2, 2.5)$, the integration of the dynamical
equation is stopped as the flux function goes to zero; just before
this happens, the RR effects become weaker and weaker, and in the
absence of an ISCO the two BHs do not plunge, but continue on a
quasicircular orbit until $\mathcal{F}_T(v)$ equals 10\% of ${\cal F}_{\rm Newt}$. 
So the binary spends many cycles at high frequencies, skewing the amplitude with respect to the Newtonian result, and producing the oscillations seen in Fig.~\ref{AmplH}.  We consider this behaviour rather unphysical, and in the following we shall no longer take into account the HT$(2, 2.5)$ model, but at 2PN order we shall use HT$(2, 2)$.

The situation is similar for the HP models. Except at 1PN order, the
HT and HP models do not end their evolution with a plunge.  As a
result, the frequency-domain amplitude of the HT and HP waveforms does
not decrease markedly at high frequencies, as seen in
Fig.~\ref{AmplH}, and in fact it does not deviate much from the
Newtonian result (especially at 3PN order).

Quantitative measures of the difference between HT and HP models at
2PN and 3PN orders, and of the difference between the Hamiltonian
models and the adiabatic models, can be seen in Tables~\ref{VIa},
\ref{VIb}. For some choices of BBH masses, these tables show the
maxmax matches between the search models in the columns and the target
models in the rows, maximized over the search-model parameters $M$ and
$\eta$, with the restriction $0 < \eta \leq 1/4$. The matches between
the H$(2, 2)$ and the H$(3, 3.5)$ waveforms are surprisingly low.
More generally, the H$(2, 2)$ models have low matches with all the
other PN models. We consider these facts as an indication of the
unreliability of the H models. In the following we shall not give much
credit to the H$(2, 2)$ models, and when we discuss the construction
of detection template families we shall consider only the H$(3, 3.5)$
models. [We will however comment on the projection of the H$(2, 2)$
models onto the detection template space.]

As for the H$(3, 3.5)$ models, their matches with the 2PN adiabatic
models are low; but their matches with the 3PN adiabatic models are
high, at least for $M \leq 30 M_\odot$. For $M = 40 M_\odot$ (as shown
in Tables~\ref{VIa} and \ref{VIb}), the matches can be quite low, as
the differences in the late dynamical evolution become significant.

\subsection{Nonadiabatic PN expanded methods: Lagrangian formalism}
\label{subsec4.2}

\subsubsection*{Equations of motion}
\label{subsubsec4.2.1}

In the harmonic gauge, the equations of motion for the
general-relativistic two-body dynamics in the Lagrangian formalism
read~\cite{DD,KWW,IW}:
\beq {\bf \ddot{x}} = {\bf a}_{\rm N} + {\bf a}_{\rm PN} + {\bf
a}_{\rm 2PN} + {\bf a}_{\rm 2.5RR} + {\bf a}_{\rm 3.5RR}\,, 
\label{eqL}
\eeq
where
\beq {\bf a}_{\rm N} = - {\frac{M}{r^2}} {\bf \hat n}\,, \eeq
\beq {\bf a}_{\rm PN}= - {\frac{M}{r^2}} \left\{ {\bf \hat n} \left[
(1+3\eta)v^2 - 2(2+\eta){\frac{M}{r}} - {\frac{3}{2}} \eta \dot r^2
\right] -2(2-\eta) \dot r {\bf v} \right\}\,, \eeq
\bea {\bf a}_{\rm 2PN} &=& - {\frac{M}{r^2}} \biggl\{ {\bf \hat n}
\biggl[ {\frac{3}{4}} (12+29\eta) \left ( {\frac{M}{r}} \right )^2 +
\eta(3-4\eta)v^4 + {\frac{15}{8}} \eta(1-3\eta)\dot r^4 \nonumber \\ && -
{\frac{3}{2}} \eta(3-4\eta)v^2 \dot r^2 - {\frac{1}{2}} \eta(13-4\eta)
{\frac{M}{r}} v^2 - (2+25\eta+2\eta^2) {\frac{M}{r}} \dot r^2 \biggr]
\nonumber \\ && - {\frac{1}{2}} \dot r {\bf v} \left[ \eta(15+4\eta)v^2
- (4+41\eta+8\eta^2){\frac{M}{r}} -3\eta(3+2\eta) \dot r^2 \right]
\biggr\} , \eea
\beq {\bf a}_{\rm 2.5RR} = {\frac{8}{5}} \eta {\frac{M^2}{r^3}} \left\{
\dot r {\bf \hat n} \left[ 18v^2 + {\frac{2}{3}} {\frac{M}{r}} -25
\dot r^2 \right] - {\bf v} \left[ 6v^2 - 2 {\frac{M}{r}} -15 \dot r^2
\right] \right\} , \eeq
\bea {\bf a}_{\rm 3.5RR} &=& {\frac{8}{5}} \eta {\frac{M^2}{r^3}}
\biggl \{ \dot r {\bf \hat n} \biggl [ \left (\frac{87}{14}-48\eta
\right )v^4 - \left (\frac{5379}{28}+\frac{136}{3}\eta \right )
v^2\,\frac{M}{r}+ \frac{25}{2}(1+5\eta)v^2\dot{r}^2+ \left
(\frac{1353}{4}+133\eta \right )\dot{r}^2\frac{M}{r} \nonumber \\ &&
-\frac{35}{2}(1-\eta)\dot{r}^4 + \left ( \frac{160}{7}+\frac{55}{3}\eta
\right ) \left ( \frac{M}{r} \right )^2 \biggr ] - {\bf v} \biggl [
-\frac{27}{14}v^4 - \left ( \frac{4861}{84}+\frac{58}{3}\eta \right
)v^2 \frac{M}{r} + \frac{3}{2}(13-37 \eta)v^2\,\dot{r}^2 \nonumber \\
&& + \left ( \frac{2591}{12}+97\eta \right ) \dot{r}^2\, \frac{M}{r} -
\frac{25}{2}(1-7\eta)\dot{r}^4 + \frac{1}{3}\left ( \frac{776}{7}+55\eta
\right )\, \left ( \frac{M}{r}\right )^2 \biggr ] \biggr \}\,.  \eea
For the sake of convenience, in this section we are using same symbols
of Sec.~\ref{subsec4.1} to denote different physical quantities (such
as coordinates in different gauges).
Here the vector ${\bf x} \equiv {\bf x_1}-{\bf x_2}$ is the
difference, in pseudo--Cartesian harmonic coordinates \cite{DD}, between
the positions of the BH centers of mass; the vector ${\bf v}={d{\bf
x}/dt}$ is the corresponding velocity; the scalar $r$ is the
(coordinate) length of ${\bf x}$; the vector ${\bf \hat n}\equiv{{\bf
x}/r}$; and overdots denote time derivatives with respect to the
post--Newtonian time. We have included neither the 3PN order corrections $a_{\rm
3PN}$ derived in Ref.~\cite{DBF}, nor the 4.5PN order term $a_{\rm
4.5PN}$ for the radiation-reaction force computed in
Ref.~\cite{IW4.5}. Unlike the Hamiltonian models, where the
radiation-reaction effects were averaged over circular orbits but 
were present up to 3PN order, here radiation-reaction effects are
instantaneous, and can be used to compute generic orbits, but are given
only up to 1PN order beyond the leading quadrupole term.

We compute waveforms in the quadrupole approximation
of Eq.\ \eqref{eq:quadapp}, defining the orbital phase $\varphi$ as
the angle between ${\bf x}$ and a fixed direction in the orbital
plane, and the invariantly defined velocity $v$ as $(M
\dot{\varphi})^{1/3}$.
\begin{figure}[t]
\begin{center}
\epsfig{file = 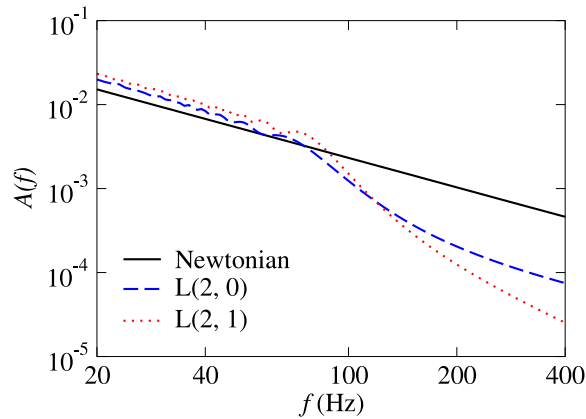,width=\sizeonefig}
\end{center}
\caption{\label{amplL}  
Frequency-domain amplitude versus frequency for the L-approximated 
(restricted) waveforms, at different PN orders, for a $(15 + 15)M_\odot$ BBH.}
\end{figure}

\subsubsection*{Definition of the models}
\label{subsubsec4.2.2}

For these models, just as for the HT and HP models, the choice of the
endpoint of evolution is rather arbitrary. We decided to stop the
integration of the dynamical equations when we begin to observe
unphysical effects due to the failure of the PN expansion.  For many
(if not all) configurations, the PN-expanded center-of-mass binding
energy (given by Eqs.~(2.7a)--(2.7e) of Ref.~\cite{kidder}) begins to
increase during the late inspiral, instead of continuing to decrease.
When this happens, we stop the integration.  The instantaneous GW
frequency at that time will then be the \emph{ending frequency} for
these waveforms. We shall refer to these models as ${\rm L}(n{\rm
PN},m{\rm PN})$, where $n{\rm PN}$ ($n{\rm PN}$) denotes the maximum
PN order of the terms included in the Hamiltonian (the
radiation-reaction force). We shall consider $(n{\rm PN},m{\rm
PN})=(2, 0),(2, 1)$.

\subsubsection*{Waveforms and matches}
\label{subsubsec4.2.3}

In Fig.~\ref{amplL}, we plot the frequency-domain amplitude versus
frequency for the L-approximated (restricted) waveforms, at different
PN orders, for a $(15 + 15)M_\odot$ BBH. The amplitude deviates from
the Newtonian prediction slightly before $100$ Hz. Indeed, the GW
ending frequencies are $116$ Hz and $107$ Hz for the L$(2, 0)$ and
L$(2, 1)$ models, respectively.  These frequencies are quite low,
because the unphysical behavior of the PN-expanded center-of-mass
binding energy appears quite early [at $r_{\rm end} = 6.6$ and $r_{\rm
end} = 7.0$ for the L$(2, 0)$ and L$(2, 1)$ models, respectively]. So
the L models do not provide waveforms for the last stage of inspirals
and plunge.

Table~\ref{Lapprox} shows the maxmax matches between the
L-approximants and a few other selected PN models. The overlaps 
are quite high, except with the EP$(2,2.5)$ and EP$(3,3.5,0)$
at high masses, but extremely unfaithful.
Moreover, we could expect the L$(2, 0)$ and L$(2, 1)$ models to have high fitting factors with the
adiabatic models T$(2,0)$ and T$(2,1)$. However, this is not the
case. As Table~\ref{tableTL} shows, the T models are neither effectual
nor faithful in matching the L models, and vice versa. This might be
due to one of the following factors: (i) the PN-expanded conservative
dynamics in the adiabatic limit (T models) and in the nonadiabatic
case (L models) are rather different; (ii) there is an important
effect due to the different criteria used to end the evolution in the
two models, which make the ending frequencies rather different. 
All in all, the L models do not seem very reliable, so we shall not
give them much credit when we discuss detection template families. However, 
we shall investigate where they lie in the detection template space.
\newcolumntype{d}{D{.}{.}{1.3}}
\newcolumntype{e}{D{.}{.}{2.2}}
\newcolumntype{f}{D{.}{.}{1.2}}

\begingroup
\squeezetable
\begin{table*} 
\begin{tabular}{cr|def|def|def|def|def|def|def}
\input tableL.tex
\end{tabular}
\caption{\label{Lapprox}Fitting factors [see Eq.\ \eqref{eq:ffd}] for
the projection of the L$(2,1)$ (target) waveforms onto the T, P, EP
and HP (search) models at 2PN and 3PN order. The values quoted are
obtained by maximizing the maxmax (mm) match over the search-model
parameters $M$ and $\eta$.}
\end{table*}
\endgroup
\begin {table*} 
\begin {tabular}{lr|ccc|ccc|ccc|ccc}
\input tableTL.tex
\end{tabular}
\caption{\label{tableTL} Fitting factors [see Eq.\
\eqref{eq:ffd}] for the projection of the L$(2,1)$ and L$(2,0)$ (target) waveforms
onto the T$(2,0)$ and T$(2,1)$ (search)
models. The values quoted are obtained by maximizing the maxmax (mm)
match over the search-model parameters $M$ and $\eta$.}
\end{table*}
\subsection{Nonadiabatic PN resummed methods: the Effective-One-Body approach}
\label{subsec4.3}

The basic idea of the effective-one-body (EOB) approach~\cite{BD1} is to map
the {\em real} two-body conservative dynamics, generated by the
Hamiltonian (\ref{ham}) and specified up to 3PN order, onto an {\em
effective} one-body problem where a test particle of mass $\mu=m_1
m_2/M$ (with $m_1$ and $m_2$ the BH masses, and $M=m_1+m_2$) moves in
an effective background metric $g_{\mu \nu}^{\rm eff}$ given by
\beq 
ds_{\rm eff}^2 \equiv g_{\mu \nu}^{\rm eff}\,dx^\mu\, dx^\nu =
-A(R)\,c^2dt^2 + \frac{D(R)}{A(R)}\,dR^2+
R^2\,(d\theta^2+\sin^2\theta\,d\varphi^2) \,, \eeq
where
\bea A(R) &=& 1 + a_1\,\frac{GM}{c^2 R} +
a_2\left(\frac{GM}{c^2R}\right)^2 + a_3\left(\frac{GM}{c^2 R}\right)^3
+ a_4\left(\frac{GM}{c^2R}\right)^4 + \cdots \,,\\ 
D(R) &=& 1 +
d_1\,\frac{GM}{c^2 R} + d_2\left(\frac{GM}{c^2 R}\right)^2 +
d_3\left(\frac{GM}{c^2 R}\right)^3 + \cdots\,. \eea
The motion of the particle is described by the action 
\beq
\label{action}
S_{\rm eff} = - \mu c \int ds_{\rm eff}\,.
\eeq
For the sake of convenience, in this section we shall use the same symbols
of Secs.~\ref{subsec4.1} and ~\ref{subsubsec4.2.2} to denote different
physical quantities (such as coordinates in different gauges).  The
mapping between the real and the effective dynamics is worked out
within the Hamilton--Jacobi formalism, by imposing that the action
variables of the real and effective description coincide (i.e.,
${J_{\rm real}} = {J_{\rm eff}}$, ${{\cal I}_{\rm real}}= {{\cal
I}_{\rm eff}}$, where $J$ denotes the total angular momentum, and
${\cal I}$ the radial action variable~\cite{BD1}), while allowing the
energy to change,
\beq
\frac{{\cal E}_{\rm eff}^{\rm NR}}{\mu c^2} = \frac{{\cal E}_{\rm real}^{\rm 
NR}}{\mu c^2} \left[ 1 + \alpha_1 \, \frac{{\cal E}_{\rm real}^{\rm 
NR}}{\mu c^2} + \alpha_2 \left( \frac{{\cal E}_{\rm real}^{\rm NR}}{\mu c^2} 
\right)^2 + \alpha_3 \left( \frac{{\cal E}_{\rm real}^{\rm NR}}{\mu c^2} 
\right)^3 + \cdots \right] \,,
\label{mapE}
\eeq
here ${\cal E}^{\rm NR}_{\rm eff}$ is the Non--Relativistic
\emph{effective} energy, while is related to the relativistic
effective energy ${\cal E}_{\rm eff}$ by the equation ${\cal E}^{\rm
NR}_{\rm eff} = {\cal E}_{\rm eff}-\mu\,c^2$; ${\cal E}_{\rm eff}$ is
itself defined uniquely by the action (\ref{action}). The
Non--Relativistic \emph{real} energy ${\cal E}_{\rm real}^{\rm NR}
\equiv H({\bf q},{\bf p})$, where $H({\bf q},{\bf p})$ is given by
Eq.~(\ref{ham}) with $H({\bf q},{\bf p})= \mu \widehat{H}({\bf
q},{\bf p})$. From now on, we shall relax our notation and set $G = c
= 1$.

\subsubsection*{Equations of motion}
\label{subsubsec4.3.1}

Damour, Jaranowski and Sch\"afer~\cite{EOB3PN} found that, at 3PN
order, this matching procedure contains more equations to satisfy than
free parameters to solve for ($a_1, a_2, a_3, d_1, d_2, d_3$, and
$\alpha_1, \alpha_2, \alpha_3$). These authors suggested the following
two solutions to this conundrum.  At the price of modifying the
energy map and the coefficients of the effective metric at the 1PN and
2PN levels, it is still possible at 3PN order to map uniquely the
real two-body dynamics onto the dynamics of a test mass moving on a
geodesic (for details, see App.\ A of Ref.~\cite{EOB3PN}). However,
this solution appears very complicated; more importantly, it seems
awkward to have to compute the 3PN Hamiltonian as a foundation for
deriving the matching at the 1PN and 2PN levels.  The second solution
is to abandon the hypothesis that the effective test mass moves along
a geodesic, and to augment the Hamilton--Jacobi equation with
(arbitrary) higher-derivative terms that provide enough coefficients
to complete the matching. With this procedure, the Hamilton-Jacobi
equation reads
\beq 0 = \mu^2 + g_{\rm eff}^{\mu \eta} (x) \, p_{\mu} \, p_{\eta} +
A^{\mu \eta \rho \sigma} (x) \, p_{\mu} \, p_{\eta} \, p_{\rho} \,
p_{\sigma} + \cdots\,.  
\eeq
Because of the quartic terms $A^{\alpha \beta \gamma \delta}$, the
effective 3PN relativistic Hamiltonian is not uniquely fixed by the
matching rules defined above; the general expression is~\cite{EOB3PN}:
\beq \mathcal{E}^\mathrm{NR}_\mathrm{eff} \equiv \widehat{H}_{\rm
eff}({\mathbf q},{\mathbf p}) = \sqrt{A (q) \left[ 1 + {\mathbf p}^2 +
\left( \frac{A(q)}{D(q)} - 1 \right) ({\mathbf n} \cdot {\mathbf p})^2
+ \frac{1}{q^2} \left( z_1 ({\mathbf p}^2)^2 + z_2 \, {\mathbf p}^2
({\mathbf n} \cdot {\mathbf p})^2 + z_3 ({\mathbf n} \cdot {\mathbf
p})^4 \right) \right]} \,,
\label{eq:genexp}
\eeq
here we use the reduced relativistic effective Hamiltonian
$\widehat{H}_{\rm eff} = {H}_{\rm eff}/\mu$, and ${\bf q}$ and ${\bf p}$
are the reduced canonical variables, obtained by rescaling the
canonical variables by $M$ and $\mu$, respectively. The coefficients
$z_1,z_2$ and $z_3$ are arbitrary, subject to the constraint
\beq 8z_1 + 4z_2 +3z_3 = 6(4-3\eta)\,\eta\,.  
\eeq
Moreover, we slightly modify the EOB model at 3PN order of Ref.~\cite{EOB3PN} 
by requiring that in the test mass limit the 3PN EOB 
Hamiltonian equal the Schwarzschild Hamiltonian. Indeed, one of the original rationales of the PN
resummation methods was to recover known exact results in the
test-mass limit.  To achieve this, $z_1, z_2$ and $z_3$ must go to
zero as $\eta \rightarrow 0$.  A simple way to enforce this limit is
to set $z_1 = \eta \tilde{z}_1, z_2 =\eta \tilde{z}_2$ and $z_3
=\eta \tilde{z}_3$. With this choice the coefficients $A(r)$ and $D(r)$ 
in Eq.~(\ref{eq:genexp}) read:
\bea
\label{coeffA}
A(r) &=& 1 - \frac{2}{r}+\frac{2\eta}{r^3}+ \left [ \left (
\frac{94}{3}-\frac{41}{32}\pi^2\right ) -\tilde{z}_1\right ]\,\frac{\eta}{r^4}\,,\\ 
\label{coeffD}
D(r) &=& 1 -\frac{6\eta}{r^2}+\left [7\tilde{z}_1 +\tilde{z}_2+ 2(3\eta-26)\right ]\,\frac{\eta}{r^3}\,, 
\eea
where we set $r = |{\bf q}|$. The authors of Ref.~\cite{EOB3PN} restricted 
themselves to the case ${z}_1={z}_2=0$ ($\tilde{z}_1=\tilde{z}_2=0$). 
Indeed, they observed that for quasicircular orbits the terms
proportional to $z_2$ and $z_3$ in Eq.~(\ref{eq:genexp}) are very small, 
while for circular orbits the term proportional
to $z_1$ contributes to the coefficient $A(r)$, as seen in Eq.~(\ref{coeffA}). 
So, if the coefficient $z_1=\eta\tilde{z}_1 \neq 0$, its value could be chosen such as  
to cancel the 3PN contribution in $A(r)$. To avoid this fact, which can be also thought as 
a gauge effect due to the choice of the coordinate system in the effective 
description, the authors of Ref.~\cite{EOB3PN} decided to pose ${z}_1=0$ 
($\tilde{z}_1=0$). By contrast, in this paper we prefer to explore the
effect of having $z_{1,2} \neq 0$.  So we shall depart from the
general philosophy followed by the authors in Ref.~\cite{EOB3PN},
pushing (or expanding) the EOB approach to more extreme regimes.

Now, the reduction to the one-body dynamics
fixes the arbitrary coefficients in Eq.~(\ref{mapE}) uniquely to
$\alpha_1 = \eta/2$, $\alpha_2 =0$, and $\alpha_3=0$, and provides the
\emph{resummed} (improved) Hamiltonian [obtained by solving for
${\cal E}_{\rm real}^{\rm NR}$ in Eq.~(\ref{mapE}) and imposing
$H^{\rm improved} \equiv {\cal E}_{\rm real}^{\rm NR}$]:
\beq 
\label{himpr}
H^{\rm improved} = M\,\sqrt{1 + 2\eta\,\left ( \frac{H_{\rm eff}-
\mu}{\mu}\right )}\,.  
\eeq
Including radiation-reaction effects, we can then write the Hamilton
equations in terms of the reduced quantities $\widehat{H}^{\rm
improved}= H^{\rm improved}/\mu$, $\widehat{t} = t/M$,
$\widehat{\omega} = \omega\,M$ \cite{BD2},
\bea 
\frac{dr}{d \widehat{t}} &=& \frac{\pa \widehat{H}^{\rm
improved}}{\pa p_r}(r,p_r,p_\vphi)\,, \label{eq:eobhamone} \\
\frac{d \vphi}{d \widehat{t}} &\equiv& \widehat{\omega} = \frac{\pa \widehat{H}^{\rm improved}}
{\pa p_\vphi}(r,p_r,p_\vphi)\,, \\
\frac{d p_r}{d t} &=& - \frac{\pa \widehat{H}^{\rm improved}}{\pa r}(r,p_r,p_\vphi)\,, \\
\frac{d p_\vphi}{d \widehat{t}} &=&
\widehat{F}^\vphi[\widehat{\omega} (r,p_r,p_{\varphi})]\,, \label{eq:eobhamfour}
\eea
where for the $\varphi$ component of the radiation-reaction force we
use the T- and P-approximants to the flux function [see Eqs.~(\ref{fluxP}), (\ref{fluxT})].  
Note that at each PN order, say $n$PN, we integrate the
Eqs.~(\ref{eq:eobhamone})--(\ref{eq:eobhamfour}) without further
truncating the partial derivatives of the Hamiltonian at $n$PN order
(differentiation with respect to the canonical variables can introduce
terms of order higher than $n$PN).

Following the discussion around Eq.~(\ref{eqisco}), the ISCO of these
models is determined by setting $\partial H_0^{\rm improved}/\partial r =
\partial^2 H_0^{\rm improved}/\partial r^2 = 0$, where $H_0^{\rm
improved}(r,p_r,p_\vphi) = H^{\rm improved}(r,0,p_\vphi)$. If we define
\beq
\widehat{H}_{\rm eff}^2(r,0,p_\vphi) \equiv W_{p_\vphi}= A(r)\,\left (1+\frac{p_\vphi^2}{r^2} 
+ \eta\,\tilde{z}_1\,\frac{p_{\vphi}^4}{r^6}\right )\,,
\eeq
we extract the ISCO by imposing $\partial W_{p_\vphi}(r)/\partial r =0
=\partial^2 W_{p_\vphi}(r)/\partial^2 r$.
Damour, Jaranowski and Sch\"afer~\cite{EOB3PN} noticed that at 3PN
order, for $\tilde{z}_1=\tilde{z}_2=0$, and using the PN expanded form
for $A(r)$ given by Eq.~(\ref{coeffA}), there is no ISCO.  To improve
the behavior of the PN expansion of $A(r)$ and introduce an ISCO, they
proposed replacing $A(r)$ with the Pad\'e approximants
\beq
A_{P_2}(r) = \frac{r(-4+2r+\eta)}{2r^2+2\eta+r\eta}\,,
\eeq
and
\beq
\label{coeffPA}
A_{P_3}(r) = \frac{r^2[(a_4(\eta,0)+8\eta-16) + r(8-2\eta)]}{
r^3\,(8-2\eta)+r^2\,(a_4(\eta,0)+4\eta)+r\,(2a_4(\eta,0)+8\eta)+4(\eta^2+a_4(\eta,0))}\,,
\eeq
where
\beq
a_4(\eta,\tilde{z}_1) = \left [\frac{94}{3}-\frac{41}{32}\pi^2
-\tilde{z}_1\right ]\,\eta\,.
\eeq
In Table~\ref{MECOT}, we show the GW frequency at the ISCO for some
typical choices of BBH masses, computed using the above expressions
for $A(r)$ in the improved Hamiltonian (\ref{himpr}) with
$\tilde{z}_1=\tilde{z}_2=0$.

We use the Pad\'e resummation for $A(r)$ of Ref.~\cite{EOB3PN} also
for the general case $\tilde{z}_1 \neq 0$, because for the PN expanded
form of $A(r)$ the ISCO does not exist for a wide range of values of
$\tilde{z}_1$.  [However, when we discuss Fourier-domain detection
template families in Sec.~\ref{sec6}, we shall investigate also EOB
models with PN-expanded $A(r)$.]

\begin{figure}[t]
\begin{center}
\epsfig{file = 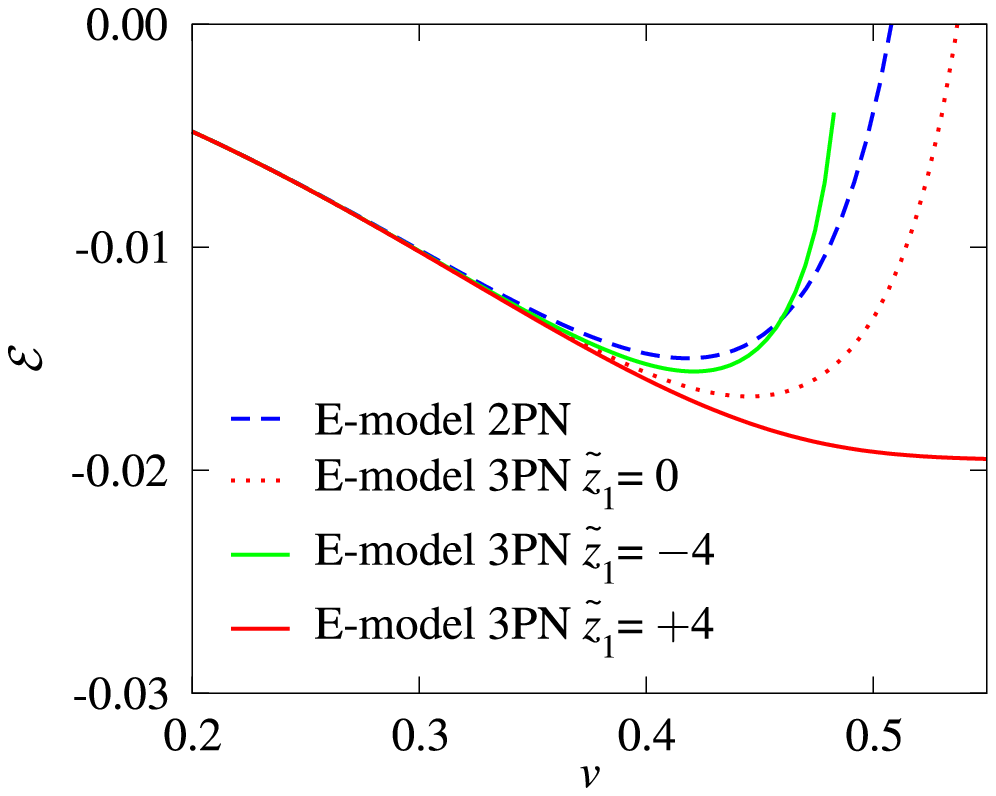,width=\sizeonefig}
\hspace{0.5cm}
\epsfig{file = 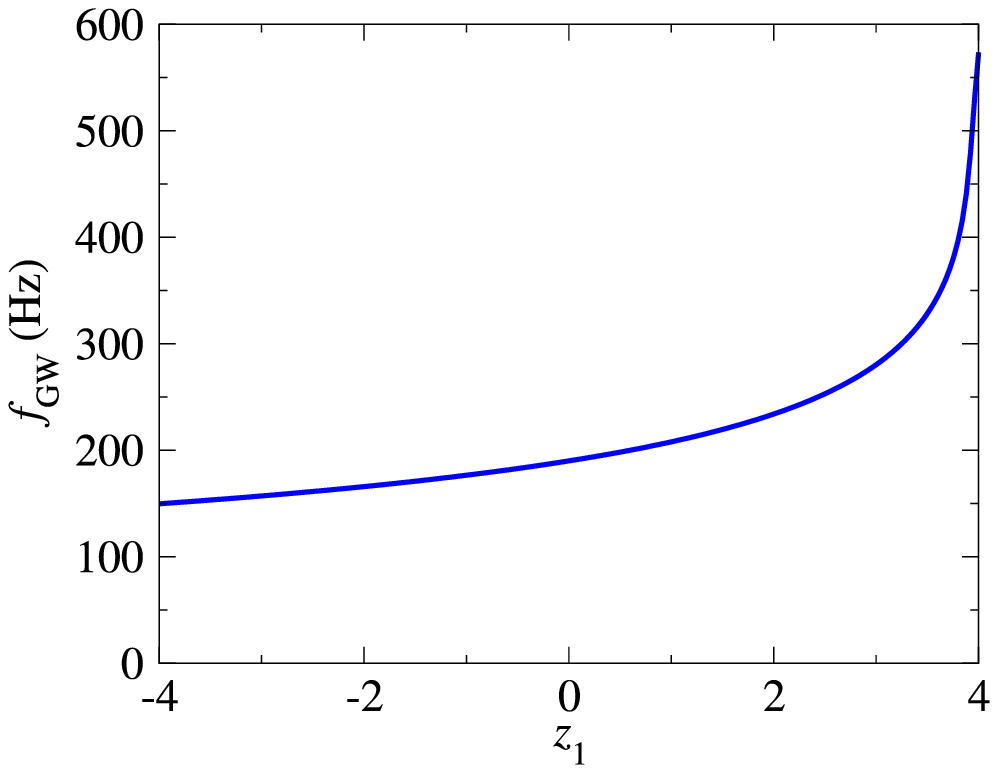,width=\sizeonefig}
\caption{\label{enerE} In the left panel, we plot the binding energy
evaluated using the improved Hamiltonian (\protect\ref{himpr}), as a
function of the velocity parameter $v$, for equal-mass BBHs, $\eta =
0.25$. We plot different PN orders for the E-model varying also the
parameter $\tilde{z}_1$. In the right panel, we plot the GW frequency
at the ISCO at 3PN order as a function of the parameter $\tilde{z}_1$
for (15+15)$M_\odot$ BBH.}
\end{center}
\end{figure}

\begin{figure}[t]
\begin{center}
\epsfig{file = 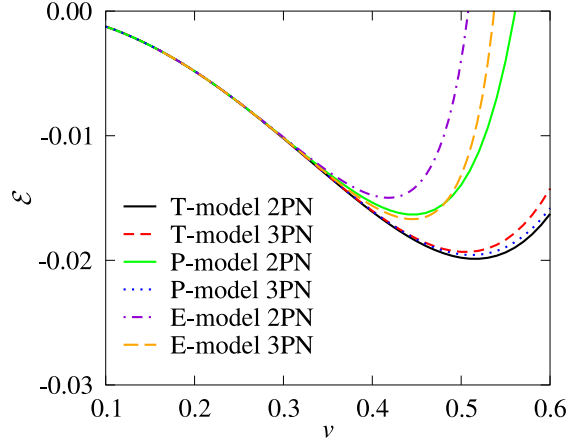,width=\sizeonefig}
\caption{\label{Fig7}Binding energy as a function of the
velocity parameter $v$, for equal-mass BBHs. We plot different PN
orders for selected PN models. For the E model at 3PN order we 
fix $\tilde{z}_1=0=\tilde{z}_2$.}
\end{center}
\end{figure}

In Fig.~\ref{enerE}, we plot the binding energy as evaluated using the
improved Hamiltonian (\ref{himpr}), at different PN orders, for
equal-mass BBHs. At 3PN order, we use as typical values $\tilde{z}_1
= 0, \pm 4$. [For $\tilde{z}_1 > 4$ the location of
the ISCO is no longer a monotonic function of $\tilde{z}_1$. So we set
$\tilde{z}_1 \leq 4$.] In the right panel of Fig.~\ref{enerE}, we show
the variation in the GW frequency at the ISCO as a function of
$\tilde{z}_1$ for a (15+15)$M_\odot$ BBH.
Finally, in Fig.~\ref{Fig7}, we compare the binding energy for a few selected PN
models, where for the E models we fix $\tilde{z}_1=\tilde{z}_2=0$ [see
the left panel of Fig.~\ref{enerE} for the dependence of the binding
energy on the coefficient $\tilde{z}_1$]. Notice, in the left panel,
that the 2PN and 3PN T energies are much closer to each other than the
2PN and 3PN P energies are, and than the 2PN and 3PN E energies are;
notice also that the 3PN T and P energies are very close. The
closeness of the binding energies (and of the MECOs and ISCOs)
predicted by PN expanded and resummed models at 3PN order (with $\tilde{z}_1 =0$), 
and of the binding energy predicted by the numerical quasiequilibrium BBH models
of Ref.~\cite{GGB} was recently pointed out in Refs.~\cite{LB,DGG}.
However, the EOB results are very close to the numerical results of
Ref.~\cite{GGB} only if the range of variation of $\tilde{z}_1$ is
restricted.

\subsubsection*{Definition of the models}
\label{subsubsec4.3.2}
For these models, we use the initial conditions laid down in Ref.\
\cite{DIS2}, and also adopted in this paper for the HT and HP models
(see Sec.\ \ref{subsec4.1}). At 2PN order, we stop the
integration of the Hamilton equations at the light ring given by the
solution of the equation $r^3-3r^2+5\eta=0$~\cite{BD2}. At 3PN order,
the light ring is defined by the solution of
\begin{equation}
\frac{d}{du} \left[ u^2\,A_{P_3}(u) \right] = 0,
\end{equation}
with $u=1/r$ and $A_{P_3}$ is given by Eq.~(\ref{coeffPA}).  For some
configurations, the orbital frequency and the binding energy start to
decrease before the binary can reach the 3PN light ring, so we stop
the evolution when $\dot{\hat{\omega}} = 0$ [see discussion in
Sec.~\ref{subsec5.1} below].  For other configurations, it happens
that the radial velocity becomes comparable to the angular velocity
before the binary reaches the light ring; in this case, the
approximation used to introduce the RR effects into the conservative
dynamics is no longer valid, and we stop the integration of the
Hamilton equations when $|\dot{r}/(r\dot{\vphi})|$ reaches $0.3$. 
For some models, usually those with $\tilde{z}_{1,2}\neq 0$,
the quantity $|\dot{r}/(r\dot{\vphi})|$ reaches a maximum 
during the last stages of evolution, then it starts 
decreasing, and $\dot{r}$ becomes positive. In such cases, we 
choose to stop at the maximum of $|\dot{r}/(r\dot{\vphi})|$.

In any of these cases, the instantaneous GW frequency at the time when
the integration is stopped defines the \emph{ending frequency} for
these waveforms.

We shall refer to the EOB models (E-approximants) as ${\rm ET}(n{\rm
PN},m{\rm PN})$ (when the T-approximant is used for the flux) or ${\rm
EP}(n{\rm PN},m{\rm PN})$ (when the P-approximant is used for the
flux), where $n{\rm PN}$ ($m{\rm PN}$) denotes the maximum PN order of
the terms included in the Hamiltonian (flux). We shall consider
$(n{\rm PN},m{\rm PN}) = (1, 1.5)$, $(2, 2.5)$, and
$(3,3.5,\hat{\theta})$ [at 3PN order we need to indicate also a choice
of the arbitrary flux parameter $\hat{\theta}$].

\subsubsection*{Waveforms and matches}
\label{subsubsec4.3.3}

In Table~\ref{IX}, we investigate the dependence of the E waveforms on
the values of the unknown parameters $\tilde{z}_1$ and $\tilde{z}_2$
that appear in the EOB Hamiltonian at 3PN order.  The coefficients
$\tilde{z}_1$ and $\tilde{z}_2$ are in principle completely arbitrary.
When $\tilde{z}_1 \neq 0$, the location of the ISCO changes, as shown
in Fig.~\ref{enerE}.  Moreover, because in Eq.~(\ref{eq:genexp})
$\tilde{z}_1$ multiplies a term that is not zero on circular orbits,
the motion tends to become noncircular much earlier, and the criteria
for ending the integration of the Hamilton equations are satisfied
earlier. [See the discussion of the ending frequency in the previous section.]  This
effect is much stronger in equal-mass BBHs with high $M$.  For
example, for (15+15)$M_\odot$ BBHs and for $\tilde{z}_2=0$, the
fitting factor (the maxmax match, maximized over $M$ and $\eta$)
between an EP target waveform with $\tilde{z}_1=0$ and EP search
waveforms with $-40 \,\laq\, \tilde{z}_1 < -4$ can well be $\leq\,
0.9$. However, if we restrict $\tilde{z}_1$ to the range $[-4,4]$, we
get very high fitting factors, as shown in Table~\ref{IX}.

In Eq.~(\ref{eq:genexp}), the coefficients $\tilde{z}_2$ and
$\tilde{z}_3$ multiply terms that are zero on circular orbits.  [The
coefficient $\tilde{z}_2$ appears also in $D(r)$, given by
Eq.~(\ref{coeffD}).]  So their effect on the dynamics is not very
important, as confirmed by the very high matches obtained in
Table~\ref{IX} between EP waveforms with $\tilde{z}_2 = 0$ and EP
waveforms with $\tilde{z}_2 = \pm 4$. It seems that the effect of
changing $\tilde{z}_2$ is nearly the same as a remapping of the BBH
mass parameters.

We investigated also the case in which we use the PN expanded form for
$A(r)$ given by Eq.~(\ref{coeffA}). For example, for (15+15)$M_\odot$
BBHs and $\tilde{z}_2=0$, the fitting factors between EP target
waveforms with $\tilde{z}_1=-40, -4, 4, 40$ and EP search waveforms
with $\tilde{z}_1=0$ are $({\rm maxmax},M,\eta) = (0.767, 39.55,
0.240)$, $(0.993, 30.83, 0.241)$, $(0.970, 30.03, 0.241)$, and
$(0.915, 28.23, 0.242)$, respectively. So the overlaps can be quite
low.

\begin{figure}[t]
\begin{center}
\epsfig{file = 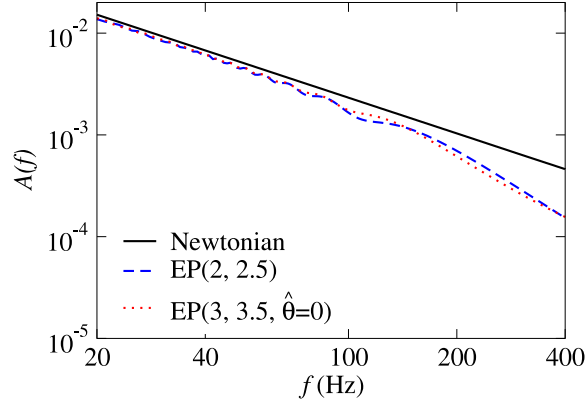,width=\sizeonefig,angle=0}
\end{center}
\caption{\label{AmplE}Frequency-domain amplitude versus frequency for
the EP-approximated (restricted) waveform, at different PN orders, for
a $(15 + 15)M_\odot$ BBH.}
\end{figure}

In Table~\ref{CauchyE}, for three typical choices of BBH masses, we
perform a convergence test using Cauchy's criterion. The values are
quite high. However, as for the P-approximants, we have no way to test
whether the E-approximants are converging to the true limit. In
Fig.~\ref{AmplE}, we plot the frequency-domain amplitude of the
EP-approximated (restricted) waveforms, at different PN orders, for a
(15 +15)$M_\odot$ BBH. The evolution of the EOB models contains a
plunge characterized by quasicircular motion~\cite{BD2}. This plunge
causes the amplitude to deviate from the Newtonian amplitude,
$\mathcal{A}_\mathrm{Newt} = f^{-7/6}$ around $200$ Hz, which is a
higher frequency than we found for the adiabatic models [see
Figs.~\ref{AmplT}, \ref{AmplP}].

In Table~\ref{VIII}, for some typical choices of the masses, we
evaluate the fitting factors between the ET$(2, 2.5)$ and ET$(3, 3.5)$
waveforms (with $\tilde{z}_1=\tilde{z}_2=0$) and the $\mathrm{T}(2,
2.5)$ and $\mathrm{T}(3, 3.5)$ waveforms. This comparison should
emphasize the effect of moving from the adiabatic orbital evolution,
ruled by the energy-balance equation, to the (almost) full Hamiltonian
dynamics, ruled by the Hamilton equations. More specifically, we see
the effect of the differences \emph{in the conservative dynamics}
between the PN expanded T-model and the PN resummed E-model (the
radiation-reaction effects are introduced in the same way in both
models). While the matches are quite low at 2PN order, they are high
($\geq 0.95$) at 3PN order, at least for $M \leq 30 M_\odot$, but the
estimation of $m_1$ and $m_2$ is poor.  This result suggests that, for
the purpose of signal detection as opposed to parameter estimation,
the conservative dynamics predicted by the EOB resummation and by the
PN expansion are very close at 3PN order, at least for $M \leq 30
M_\odot$.  Moreover, the results of Table~\ref{VIII} suggest also that
the effect of the unknown parameter $\hat{\theta}$ is rather small, at
least if $\hat{\theta}$ is of order unity, so in the following we
shall always set $\hat{\theta}=0$.

In Tables~\ref{VIa} and \ref{VIb} we study the difference between the
EP$(2, 2.5)$ and EP$(3, 3.5)$ models (with
$\tilde{z}_1=\tilde{z}_2$=0), and all the other adiabatic and
nonadiabatic models. For some choices of BBH masses, these tables show
the maxmax matches between the search models in the columns and the
target models in the rows, maximized over the search-model parameters
$M$ and $\eta$, with the restriction $0 < \eta \leq 1/4$. At 2PN
order, the matches with the T$(2, 2.5)$, HT$(2, 2)$ and HP$(2, 2.5)$
models are low, while with the matches with the T$(2, 2)$ and P$(2,
2.5)$ models are high, at least for $M \leq 30 M_\odot$ (but the
estimation of the BH masses is poor). At 3PN order, the matches with
T$(3, 3.5, \hat{\theta})$, P$(3, 3.5, \hat{\theta})$, HP$(3, 3.5,
\hat{\theta})$ and HT$(3, 3.5, \hat{\theta})$ are quite high if $M
\leq 30M_\odot$. However, for $M=40 M_\odot$, the matches can be quite
low.  We expect that this happens because in this latter case the
differences in the late dynamical evolution become crucial.

\begin {table*} 
\begin {tabular}{l|c|rrrr|c}
\hline \hline
\multicolumn{1}{c|}{$N$} & \multicolumn{6}{c}{$\left <{\rm EP}_N,{\rm
EP}_{N+1} \right > $} \\
     & \multicolumn{1}{c}{$(5+20)M_\odot$}
     & \multicolumn{4}{c}{$(10+10)M_\odot$}
     & \multicolumn{1}{c}{$(15+15)M_\odot$}\\
\hline
 0 & 0.677 & 0.584 &(0.769, &17.4, &0.246) & 0.811 \\
 1 & 0.766 & 0.771 &(0.999, &21.8, &0.218) & 0.871 \\
 2 ($\hat{\theta} =+2$) & 0.862 & 0.858 &(0.999, &21.3, &0.222) & 0.898 \\
 2 ($\hat{\theta} =-2$) & 0.912 & 0.928 &(0.999, &21.9, &0.211) & 0.949 \\
\hline \hline
\end{tabular}
\caption{\label{CauchyE}Test for the Cauchy convergence of the
EP-approximants.  The values quoted assume optimization on the
extrinsic parameters but the same intrinsic parameters (i.e., they
assume the same masses).  Here we define $\mathrm{EP}_0 =
\mathrm{EP}(0, 0)$, $\mathrm{EP}_1 = \mathrm{EP}(1, 1.5)$,
$\mathrm{EP}_2 = \mathrm{EP}(2, 2.5)$, and $\mathrm{EP}_3 =
\mathrm{EP}(3, 3.5, \hat{\theta}, \tilde{z}_1 = \tilde{z}_2 = 0)$.
The values in parentheses are the maxmax matches obtained by
maximizing with respect to the extrinsic \emph{and} intrinsic
parameters, shown together with the $\mathrm{EP}_{N+1}$ parameters $M$
and $\eta$ where the maxima are attained. In all cases the integration
of the equations is started at a GW frequency of $20$ Hz.}
\end{table*}
\subsection{Features of the late dynamical evolution in nonadiabatic models}
\label{subsec5.1}
\begin{figure}
\includegraphics[width=\textwidth]{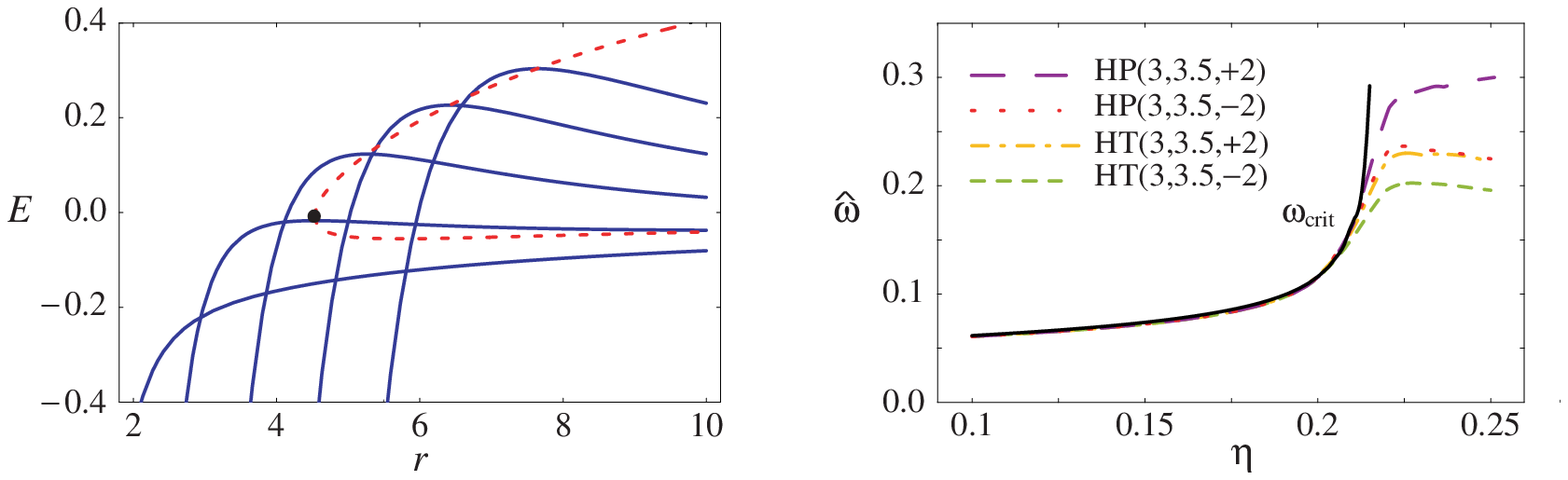}
\caption{Ending points of the H models at 3PN order for low values of
$\eta$. In the left panel, we plot as a function of $r$ the
Hamiltonian $\widehat{H}(r,p_r=0,p_{\varphi})$ [given by
Eq.~(\ref{ham})], evaluated at $\eta=0.16$ for a (5+20)$M_\odot$ BBH,
for various values of the (reduced) angular momentum
$p_{\varphi}$. The circular-orbit solutions are found at the values of
$r$ and $\hat{H}$ joined by the dashed line. At $r_{\rm crit} = 4.524$
there is a critical radius, below which there is no circular orbit. In
the right panel, we plot as a function of $\eta$ the orbital angular frequency
$\widehat{\omega}_{\rm crit}(\eta)$ corresponding to the critical radius,
for $0.1 < \eta < 0.21$ (solid line). This curve agrees well
with the ending frequencies of the HT and HP models at 3PN order,
which are shown as dotted and dashed lines in the figure.\label{HT3PN}}
\end{figure}

While studying the numerical evolution of nonadiabatic models, we
encounter two kinds of dynamical behavior that are inconsistent with
the assumption of quasicircular motion used to include the
radiation-reaction effects, so when one of these two behaviors occurs,
we immediately stop the integration of the equations of motion. First,
in the late stage of evolution $\widehat{\omega}$ can reach a maximum,
and then drop quickly to zero; so we stop the integration if
$\dot{\widehat{\omega}}=0$. Second, the radial velocity $\dot{r}$ can
become a significant portion of the total speed, so we stop the
integration if $\dot{r} = 0.3(r\widehat{\omega})$.

The first behavior is found mainly in the H models at 3PN order, when
$\eta$ is relatively small ($\stackrel{<}{_\sim}0.21$). As we shall
see below, it is \emph{not} characteristic of either the Schwarzschild
Hamiltonian or the EOB Hamiltonian.  In the left panel of
Fig.~\ref{HT3PN}, we plot the binding energy evaluated from 
$\widehat{H}(r,p_r=0,p_{\varphi})$ [given by
Eq.~(\ref{ham})] as a function of $r$ at $\eta=0.16$, for various
values of the (reduced) angular momentum $p_{\varphi}$.  As this plot
shows, there exists a \emph{critical radius}, $r_{\rm crit}$, below
which no circular orbits exist. This $r_\mathrm{crit}$ can be derived
as follows. From Fig.~\ref{HT3PN} (left), we deduce that
\beq
\label{Hcritical}
\left. {{\frac{d \widehat{H}}{dr}}} \right|_{\rm circ} \rightarrow \infty\,,\quad
\quad r \rightarrow r_{\rm crit}\,.
\eeq
Because circular orbits satisfy the conditions
\beq
p_r=0\,,\quad \quad \frac{\partial \widehat{H}}{\partial r}=0\,,
\eeq
and 
\beq
\left. {{\frac{d p_{\varphi}}{dr}}} \right|_{\rm circ} = -\frac{\partial^2\widehat{H}}{\partial r^2} 
\,\left(\frac{\partial^2\widehat{H}}{\partial r \partial p_{\varphi}}\right)^{-1}\,,
\eeq 
we get 
\beq
\left. {{\frac{d \widehat{H}}{dr}}} \right|_{\rm circ}=
\left. \frac{\partial
\widehat{H}}{\partial r}+\frac{\partial \widehat{H}}{\partial
p_{\varphi}}\,{{\frac{d p_{\varphi}}{dr}}} \right|_{\rm circ}
=-\frac{\partial \widehat{H}}{\partial p_{\varphi}}\,
\frac{\partial^2\widehat{H}}{\partial r^2}\,
\left(\frac{\partial^2\widehat{H}}{\partial r \partial p_{\varphi}}\right)^{-1}\,.
\eeq
Combining these equations we obtain two conditions that define
$r_\mathrm{crit}$:
\beq
\left. \frac{\partial \widehat{H}}{\partial r} \right|_{r_\mathrm{crit}}=0\,,\quad \quad 
\left. \frac{\partial^2\widehat{H}}{\partial r \partial p_{\varphi}}\right|_{r_\mathrm{crit}}=0\,.
\eeq
In the right panel of Fig.~\ref{HT3PN}, we plot the critical orbital
frequency $\widehat{\omega}_{\rm crit}$ as a function of $\eta$ in the
range [0.1, 0.21]. In the same figure, we show also the ending
frequencies for the HT$(3,3.5,\pm 2)$ and HP$(3,3.5,\pm 2)$
models. For $0.1 < \eta < 0.21$, these ending frequencies are in good
agreement with the critical frequencies $\hat{\omega}_\mathrm{crit}$;
for $\eta > 0.21$, the ending condition $\dot{r} = 0.3(r
\widehat{\omega})$ is satisfied before $\dot{\hat{\omega}}=0$.  For
$0.1<\eta<0.21$, this good agreement can be explained as follows: for
the H models at 3PN order with $\eta\stackrel{<}{_\sim}0.21$, the
orbital evolution is almost quasicircular (i.e., $\dot{r}$ remains
small and $\widehat{\omega}$ keeps increasing) until the critical
point is reached; beyond this point, there is no way to keep the orbit
quasicircular, as the angular motion is converted significantly into
radial motion, and $\widehat{\omega}$ begins to decrease.  This
behavior ($\dot{\widehat{\omega}} \rightarrow 0$) is also present in
the E model in the vicinity of the light ring, because the light ring
is also a minimal radius for circular orbits [the conditions
(\ref{Hcritical}) are satisfied also in this case].  However, the
behavior of the energy is qualitatively different for the H and E
models: in the E models (just as for a test particle in Schwarzchild
spacetime) the circular-orbit energy goes to infinity, while this is not the case for the H models.

The second behavior is usually caused by radiation-reaction effects,
and accelerated by the presence of an ISCO (and therefore of a
\emph{plunge}).  However, it is worth to mention another interesting
way in which the criterion $\dot{r} = 0.3(r \hat{\omega})$ can be
satisfied for some E evolutions at 3PN order.  During the late stages
of evolution, $\dot{r}$ sometimes increases suddenly and drastically,
and the equations of motion become singular.  This behavior is quite
different from a plunge due to the presence of an ISCO (in that case
the equations of motion do \emph{not} become singular). The cause of
this behavior is that at 3PN order the coefficient $D(r)$ [see
Eq.~(\ref{coeffD})] can go to zero and become negative for a
sufficiently small $r$. For $\tilde{z}_1=\tilde{z}_2=0$, this
occurs at the radius $r_D$ given by
\beq
\label{Dzero}
r_{\rm D}^3-6\eta r_{\rm D}+ 2(3\eta-26)\eta=0;
\eeq
$r_D$ can fall outside the light ring. For example, for $\eta=0.25$
we have $r_{\rm D}=2.54$, while the light rings sits at $r=2.31$.
On the transition from $D(r)>0$ to $D(r)<0$, the effective
EOB metric unphysical, and the E model then becomes invalid. Using the
Hamiltonian equation of motion (\ref{eq:eobhamone}), it is straightforward to 
prove that a negative $D(r)$ causes the radial velocity to blow up: 
\beq
\dot{r}=\frac{\partial \widehat{H}}{\partial p_r} \propto \frac{p_r}{D(r)} \rightarrow\infty 
\quad \quad {\rm as} \quad r \rightarrow r_{\rm D}\,.
\eeq

\section{Signal-to-noise ratio for the two-body models}
\label{sec5}
\begin{figure}[t]
\begin{center}
\epsfig{file = 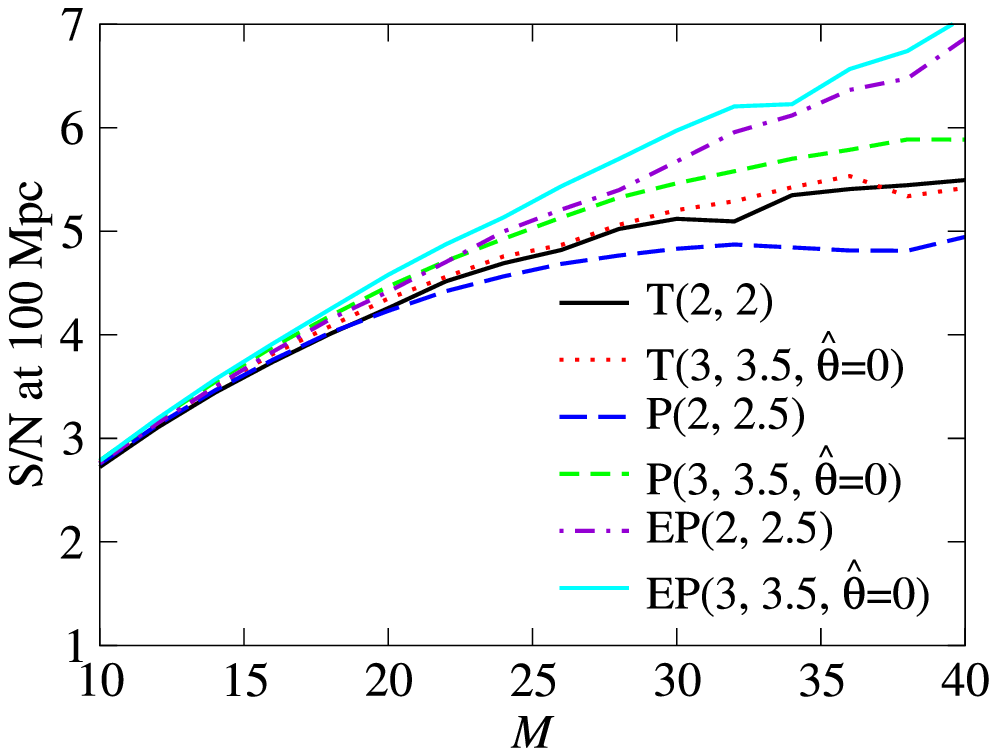,width=\sizetwofig,angle=0}
\hspace{0.5cm}
\epsfig{file = 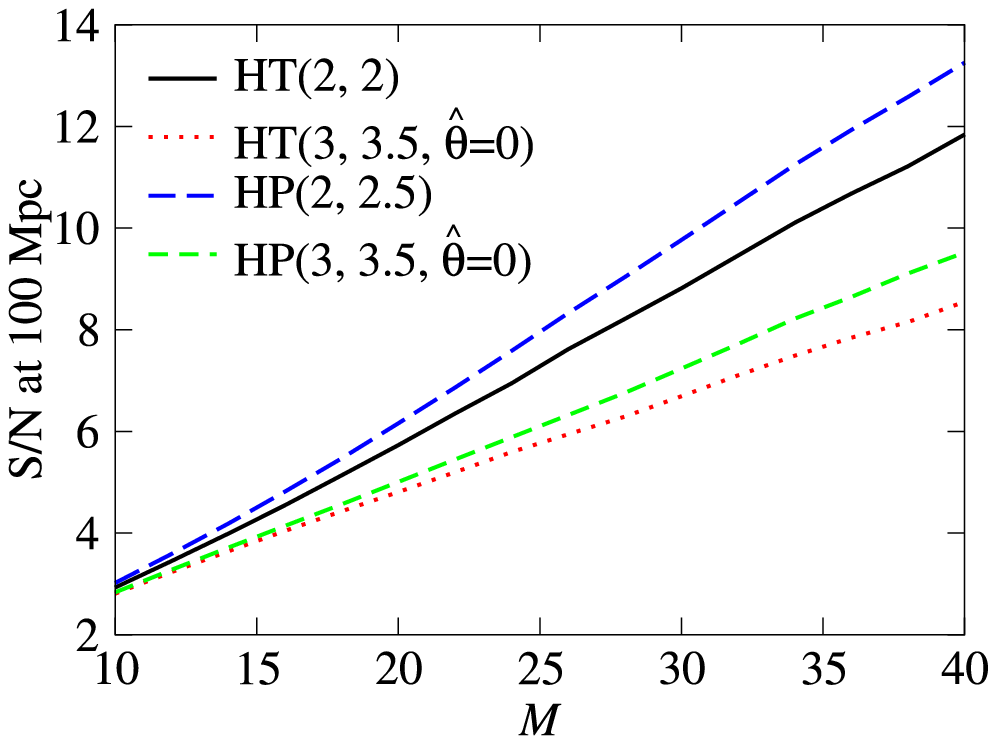,width=\sizetwofig,angle=0}
\caption{\label{sn}Signal-to-noise ratio at 100 Mpc versus total mass
$M$, for selected PN models. The S/N is computed for equal-mass BBHs
using the LIGO-I noise curve \eqref{ligoI} and the waveform expression
\eqref{eq:quadapp} with the rms $\Theta = 8/5$; for the E model at 3PN
we set $\tilde{z}_1=\tilde{z}_2=0$.}
\end{center}
\end{figure}
\begin{figure}[t]
\begin{center}
\epsfig{file = 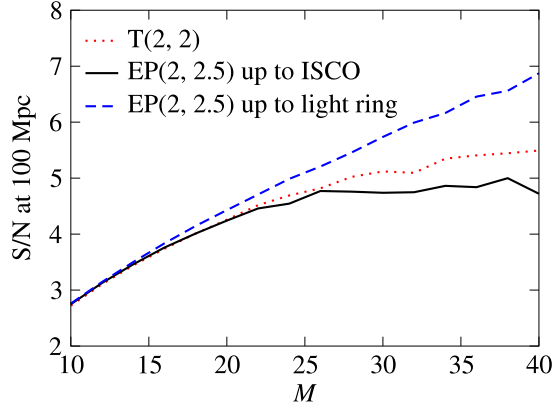,width=\sizeonefig}
\caption{\label{snr}Effect of the plunge on the signal-to-noise
ratio. The S/N is computed at 100 Mpc for equal-mass BBHs, as a
function of the total mass, for the T$(2, 2)$ adiabatic model (for
comparison), and for the EP$(2, 2.5)$ model with ending frequency at
the ISCO, and at the light ring (in this latter case the signal
includes a plunge). Here we use the LIGO-I noise curve \eqref{ligoI}
and the waveform expression \eqref{eq:quadapp} with the rms $\Theta =
8/5$.}
\end{center}
\end{figure}

In Fig.~\ref{sn}, we plot the optimal signal-to-noise ratio
$\rho_\mathrm{opt}$ for a few selected PN models. The value of
$\rho_\mathrm{opt}$ is computed using Eqs.\ \eqref{eq:innerproduct}
and \eqref{eq:optsn} with the waveform given by Eq.\
\eqref{eq:quadapp}, for a luminosity distance of 100 Mpc and the rms
$\Theta = 8/5$ [see discussion around Eq.\ \eqref{eq:quadapp}]; for
the EP model we set $\tilde{z}_1=\tilde{z}_2=0$.  Notice that, because
the E models have a plunge, their signal-to-noise ratios are much
higher (at least for $M \geq 30 M_\odot$) than those for the adiabatic
models, which we cut off at the MECO. See also Fig.~\ref{snr}, which
compares the S/N for EP$(2, 2.5)$ waveforms with and without the
plunge; for $M = 20 M_\odot$, excluding the plunge decreases the
S/N by $\sim$ 4\% (which corresponds to a decrease in detection rate
of 12\% for a fixed detection threshold); while for $M = 30 M_\odot$,
excluding the plunge decreases the S/N by $\sim$ 22\% (which
corresponds to a decrease in detection rate of 54\%). This result
confirms the similar conclusion drawn in Ref.~\cite{DIS3}.

Because at 2PN and 3PN order the H models do not have a plunge, but
the two BHs continue to move on quasicircular orbits even at close separations, the number of total GW cycles is increased, and so is the
signal-to-noise ratio, as shown in the right panel of Fig.~\ref{sn}.
However, we do not trust the H models much, because they show a
very different behavior at different PN orders, as already
emphasized in Sec.~\ref{subsec4.1}.

\section{Performance of Fourier-domain detection templates, 
and construction of a Fourier-domain detection-template bank}
\label{sec6}

In the previous sections we have seen [for instance, in Table~\ref{VIa}] 
that the overlaps between the various PN waveforms are not very high, 
and that there could be an important loss in event rate if, for the purpose of detection, we restricted ourselves to \emph{only one} of the 
two-body models [see Figs.~\ref{sn}, \ref{snr}]. 
To cope with this problem we propose the following strategy.
We \emph{guess} that the conjunction of the waveforms from all the 
PN models spans a region in signal space that includes (or
almost includes) the true signals, and we build a \emph{detection} 
template family that embeds all the PN models in a higher-dimensional 
space. The PN models that we 
have considered (expanded and resummed, adiabatic and nonadiabatic) rely on a wide variety of very different dynamical
equations, so the task of consolidating them under a single set of generic equations seems
arduous. On the other hand, we have reason to suspect, from the values
of the matches, and from direct investigations, that the
frequency-domain amplitude and phasing (the very ingredients that
enter the determination of the matches) are, qualitatively, rather similar functions for all the
PN models. We shall therefore create a family of templates that model
\emph{directly} the Fourier transform of the GW signals, by writing
the amplitude and phasing as simple polynomials in the GW frequency
$f_\mathrm{GW}$. We shall build these polynomials with the specific powers of $f_\mathrm{GW}$ that appear in the Fourier transform of PN
expanded adiabatic waveforms, as computed in the stationary-phase
approximation. However, we shall not constrain the coefficients of these powers to have the same functional dependence on the physical parameters that they have in that scheme. 
More specifically, we define our generic family of Fourier-domain effective templates as 
\beq
\label{fourier}
h_{\rm eff}(f) = {\cal A}_{\rm eff}(f)\,e^{i \psi_{\rm eff}(f)}\,,
\eeq
where 
\bea
{\cal A}_{\rm eff}(f) &=& f^{-7/6}\,\left (1 - \alpha\,f^{2/3}\right )\,
\theta(f_{\rm cut}-f)\,, \\
{\psi}_{\rm eff}(f) &=& 2\pi f t_0+\phi_0+ f^{-5/3}\,\left ( \psi_0 + \psi_{1/2}\,f^{1/3}+
\psi_{1}\,f^{2/3}+ \psi_{3/2}\,f + \psi_{2}\,f^{4/3} + \cdots \right )\,,
\label{phasing}
\eea
where $t_0$ and $\phi_0$ are the time of arrival and the
frequency-domain phase offset, and where $\theta(\ldots)$ is the
Heaviside step function. This detection template family is similar in some respects to the template banks implicitly used in Fast Chirp Transform techniques \cite{JP}.  However,
because we consider BBHs with masses $10\mbox{--}40 M_\odot$, the physical GW signal can end within the LIGO frequency band; and the predictions for the ending frequency given by different PN models can be 
quite different. Thus, we modify also the Newtonian
formula for the amplitude, by introducing the cutoff frequency $f_{\rm cut}$ and the shape parameter $\alpha$. 
  
The significance of $f_{\rm cut}$ with  respect to true physical signals
deserves some discussion. If the best match for the physical signal
$g$ is the template $h_{f_\mathrm{cut}}$, which ends at the instantaneous
GW frequency $f_\mathrm{cut}$ (so that $h_{f_\mathrm{cut}}(f) \simeq
g(f)$ for $f < f_\mathrm{cut}$ and $h_{f_\mathrm{cut}}(f) = 0$ for $f
> f_\mathrm{cut}$), then we can be certain to lose a
fraction of the optimal $\rho$ that is given approximately by
\begin{equation}
\frac{\rho_\mathrm{cut}}{\rho_\mathrm{opt}} \leq 
\frac{
\sqrt{\int_0^{f_\mathrm{cut}} \frac{|\tilde{g}(f)|^2}{S_n(f)} df }
}{
\sqrt{\int_0^{\infty} \frac{|\tilde{g}(f)|^2}{S_n(f)} df }
} \simeq
1 - \frac{1}{2} \frac{
\int_{f_\mathrm{cut}}^{\infty} \frac{|\tilde{g}(f)|^2}{S_n(f)} df
}{
\int_0^{\infty} \frac{|\tilde{g}(f)|^2}{S_n(f)} df
}.
\end{equation}
On the other hand, if we try to match $g$ with the same template
family \emph{without cuts} (and if indeed the $h$'s are completely
inadequate at modeling the amplitude and phasing of $g$ above
$f_\mathrm{cut}$), then even the best-match template
$h_\mathrm{no\,cut}$ (defined by $h_\mathrm{no\,cut}(f) \simeq g(f)$
for $f < f_\mathrm{cut}$, and by zero correlation,
$\overline{h_\mathrm{no\,cut}(f) g^*(f)} \simeq 0$ for $f >
f_\mathrm{cut}$) will yield an additional loss in $\rho$ caused by the
fact that we are spreading the power of the template beyond the range
where it can successfully match $g$. Mathematically, this loss comes
from the different normalization factor for the templates
$h_{f_\mathrm{cut}}$ and $h_\mathrm{no\,cut}$, and it is given by
\begin{equation}
\frac{\rho_\mathrm{no\,cut}}{\rho_\mathrm{cut}} \leq \frac{
\sqrt{\int_0^{f_\mathrm{cut}} \frac{|\tilde{h}(f)|^2}{S_n(f)} df } }{
\sqrt{\int_0^{\infty} \frac{|\tilde{h}(f)|^2}{S_n(f)} df } } \simeq 1
- \frac{1}{2} \frac{ \int_{f_\mathrm{cut}}^{\infty}
\frac{|\tilde{h}(f)|^2}{S_n(f)} df }{ \int_0^{\infty}
\frac{|\tilde{h}(f)|^2}{S_n(f)} df }.
\end{equation}
If we assume that $g$ and $h_\mathrm{no\,cut}$ have roughly the same amplitude
distribution, the two losses are similar.

In the end, we might be better off cutting templates if we cannot be
sure that their amplitude and phasing, beyond a certain frequency, are
faithful representations of the true signal. Doing so, we
approximately halve the \emph{worst-case} loss of $\rho$, because
instead of losing a factor
\begin{equation}
\frac{\rho_\mathrm{no\,cut}}{\rho_\mathrm{cut}}
\frac{\rho_\mathrm{cut}}{\rho_\mathrm{opt}} \simeq
1 - \frac{1}{2} \frac{ \int_{f_\mathrm{cut}}^{\infty}
\frac{|\tilde{h}(f)|^2}{S_n(f)} df }{ \int_0^{\infty}
\frac{|\tilde{h}(f)|^2}{S_n(f)} df } - 
\frac{1}{2} \frac{ \int_{f_\mathrm{cut}}^{\infty}
\frac{|\tilde{g}(f)|^2}{S_n(f)} df }{ \int_0^{\infty}
\frac{|\tilde{g}(f)|^2}{S_n(f)} df } \simeq 
1- \frac{ \int_{f_\mathrm{cut}}^{\infty}
\frac{|\tilde{g}(f)|^2}{S_n(f)} df }{ \int_0^{\infty}
\frac{|\tilde{g}(f)|^2}{S_n(f)} df },
\end{equation}
we lose only the factor $\rho_\mathrm{cut}/\rho_\mathrm{opt}$.
On the other hand, we do not want to lose the signal-to-noise ratio
that is accumulated at high frequencies if our templates have a
fighting chance of matching the true signal there; so it makes sense
to include in the detection bank the \emph{same} template with several
different values of $f_\mathrm{cut}$.

It turns out that using only the two parameters $\psi_0$ and $\psi_{3/2}$ in the phasing (and setting all other $\psi$ 
coefficients to zero) and the two amplitude parameters, $f_{\rm cut}$ and $\alpha$, we obtain a family that can already match all the PN models of Secs.~\ref{sec3},
\ref{sec4} with high fitting factors FF. This is possible largely
because we restrict our focus to BBHs with relatively high masses,
where the number of GW cycles in the LIGO range (and thus the total
range of the phasing $\psi(f)$ that we need to consider) is small.

In Tab.\ \ref{tableETB1} we list
the minmax (see Sec.~\ref{sec2}) fitting factor for the projection of
the PN models onto our frequency-domain effective templates, for a set
of BBH masses ranging from $(5+5)M_{\odot}$ to $(20+20)M_{\odot}$. In
computing the fitting factors, we used the simplicial search algorithm
\texttt{amoeba}~\cite{nrc} to search for the optimal set of parameters
$(\psi_0,\,\psi_{3/2},\,f_{\rm cut},\,\alpha)$ (as always, the
time of arrival and initial phase of the templates were automatically
optimized as described in Sec.~\ref{sec2}). From Tab.\ \ref{tableETB1} we draw the
following conclusions:
\begin{enumerate}
\item All the adiabatic models (T and P) are matched with fitting
factors $\mathrm{FF} > 0.97$. Lower-mass BBHs are matched better than
higher-mass BBHs, presumably because for the latter the inspiral ends
at lower frequencies within the LIGO band, producing stronger edge
effects, which the effective templates cannot capture fully. 3PN models
are matched better than 2PN models.
\item The Effective-One-Body models (ET and EP) are matched even
better than the adiabatic models, presumably because they have longer
inspirals and less severe edge effects at the end of inspiral. Unlike
the adiabatic models, however, ET and EP are matched better for
higher-mass BBHs. In fact, all the FFs are $> 0.99$ except for 
$(5+5)M_\odot$ BBHs, where $\mathrm{FF} \gtrsim 0.979$. The reason for this
is probably that this low-mass BBH has more GW cycles in the LIGO
frequency band than any other one, and the two phasing parameters of
our effective templates cannot quite model the evolution of the
phasing. [In the adiabatic models, these effects may be 
overshadowed by the loss in signal to noise ratio due to the edge effects 
at high frequencies.] 
When the parameters $\tilde{z}_{1,2}$ are allowed to be nonzero, the
matches get worse, but not by much. For all the plausible values of
$\tilde{z}_1$, the worst situation seems to happen at
$\tilde{z}_1 = - 40$, where the overlaps are still higher than $\sim
0.95$ [with minimum 0.947.]
\item The Hamiltonian models (HT and HP) at 3PN order are not 
matched as precisely, but the detection template family 
still works reasonably well. We usually have  
$\mathrm{FF} > 0.96$, but there are several
exceptions, with FF as low as 0.948. For these models, the overlaps
are lower in the equal-mass cases, where the ending frequencies of the
waveforms are
much higher than for the other models; it seems that the effective
templates are not able to reproduce this late portion of the waveforms
(this might not be so bad, because it does not seem likely that this
part of the signal reflects the true behavior of BBH waveforms).
\item The Lagrangian models (L) are matched a bit worse than the
Hamiltonian models (HT and HP) at 3PN, but they still have $\mathrm{FF}$ higher
than $0.95$ in most cases, with several exceptions [at either
$(20+20)M_{\odot}$ or $(5+5)M_{\odot}$], which can be as low as $0.93$.
\item HT and HP models at 2PN are matched the worst, with typical
values lower than  $0.95$ and higher than $0.85$.
\end{enumerate}

Finally, we note that our amplitude function ${\cal A}_{\rm eff}(f)$
is a linear combination of two terms, so we can search automatically
over the correction coefficient $\alpha$, in essentially the same way
as discussed in Sec.~\ref{sec2} for the orbital phase. In other words,
$\alpha$ is an \emph{extrinsic parameter}. [Although we do
search over $\alpha$, it is only to show the required range, which
will be a useful piece of information when one is deciding how to lay
down a mesh of discrete templates on the continuous detection-template space.]

\subsection{Internal match and metric}
\label{subsec6.1}

To understand the matches between the Fourier-domain templates and the
PN models, and to prepare to compute the number of templates needed to
achieve a given (internal) MM, we need to derive an expression
for the match between two Fourier-domain effective templates.

We shall first restrict our consideration to effective templates with
the same amplitude function (i.e., the same $\alpha$ and $f_{\rm
cutoff}$). The overlap $\langle h(\psi_0,\psi_{3/2}),h(\psi_0+\Delta
\psi_0,\psi_{3/2}+\Delta \psi_{3/2})\rangle$ between templates with
close values of $\psi_0$ and $\psi_{3/2}$ can be described (to second
order in $\Delta \psi_0$ and $\Delta \psi_{3/2}$) by the mismatch metric
$g_{ij}$ \cite{O}:
\beq
\label{metric}
\langle
h(\psi_0,\psi_{3/2}),h(\psi_0+\Delta \psi_0,\psi_{3/2}+\Delta
\psi_{3/2})\rangle = 1-\sum_{i,j=0,3/2}g_{\rm ij}\,\Delta \psi_i \Delta \psi_j.
\eeq
The metric coefficients $g_{ij}$ can be evaluated analytically from the overlap
\bea
&&
\langle
h(\psi_0,\psi_{3/2}),h(\psi_0+\Delta \psi_0,\psi_{3/2}+\Delta
\psi_{3/2})\rangle \simeq \nonumber \\
&&\left[\max_{\Delta \phi_0,\Delta t_0}\int df\frac{|{\cal
A}{(f)}|^2}{S_h(f)}\cos\left(\sum_{i}\frac{\Delta
\psi_i}{f^{n_i}}+\Delta\phi_0+2\pi f\Delta
t_0\right)\right]\Bigg/\left[\int df\frac{|{\cal A}{(f)}|^2}{S_h(f)}\right] \nonumber \simeq \\
&& 1-\frac{1}{2}\left[\max_{\Delta \phi_0,\Delta t_0}\int
df\frac{|{\cal A}{(f)}|^2}{S_h(f)}\left(\sum_{i}\frac{\Delta
\psi_i}{f^{n_i}}+\Delta\phi_0+2\pi f\Delta t_0\right)^2\right]\Bigg/\left[\int df\frac{|{\cal
A}{(f)}|^2}{S_h(f)}\right].
\eea
where $n_0\equiv5/3$ and $n_{3/2}\equiv 2/3$. Comparison with Eq.~(\ref{metric}) then gives 
\bea
\label{gijmin}
\sum_{i,j}g_{\rm ij}\,\Delta \psi_i \Delta \psi_j =
\frac{1}{2} \min_{\Delta \phi_0,\Delta t_0}
&& \left\{ 
\left(\begin{array}{cc} \Delta \psi_0 & \Delta \psi_{3/2} \end{array}\right) {\bf M}_{(1)} 
\left(\begin{array}{c} \Delta \psi_0 \\ \Delta \psi_{3/2} \end{array}\right) \right.
+ \nonumber \\
&& \left.2\left(\begin{array}{cc} \Delta\phi_0 & 2\pi  \Delta t_0  \end{array}\right) {\bf M}_{(2)} 
\left(\begin{array}{c} \Delta \psi_0 \\ \Delta \psi_{3/2} \end{array}\right)
+
\left(\begin{array}{cc}  \Delta \phi_0 & 2\pi  \Delta t_0  \end{array}\right) {\bf M}_{(3)} 
\left(\begin{array}{c}  \Delta \phi_0 \\ 2\pi  \Delta t_0 \end{array}\right)
\right\} 
\eea
where the ${\bf M}_{(1)\ldots(3)}$ are the matrices
\bea
{\bf M}_{(1)}&=&\left[\begin{array}{cc} J(2n_0) & J(n_0+n_{3/2}) \\ J(n_0+n_{3/2}) &
J(2n_{3/2}) \end{array}\right]\,,\\
{\bf M}_{(2)}&=&\left[\begin{array}{cc} J(n_0) & J(n_{3/2}) \\ J(n_{0}-1) &
J(n_{3/2}-1) \end{array}\right]\,,\\
{\bf M}_{(3)}&=&\left[\begin{array}{cc} J(0) & J(-1) \\ J(-1) & J(-2)
\end{array} \right]\,,
\eea
and  where
\beq
J(n)\equiv\left[\int df \frac{|{\cal A}(f)|^2}{S_h(f)}\frac{1}{f^n}\right]\bigg/\left[\int df \frac{|{\cal A}(f)|^2}{S_h(f)}\right]\,.
\eeq

Since ${\bf M}_{(3)}$ describes the mismatch caused by $(\Delta
\phi_0,\Delta t_0)$, it must be positive definite; because the
right-hand side of (\ref{gijmin}) reaches its minimum with respect to variations of $\Delta \phi_0$ and $\Delta t_0$ when
\beq
2{\bf M}_{(2)} \left(\begin{array}{c} \Delta \psi_0 \\ \Delta \psi_{3/2}
\end{array}\right) + 2{\bf M}_{(3)} \left(\begin{array}{c} \Delta
\phi_0 \\ 2\pi \Delta t_0 \end{array}\right)=0\,,
\eeq
we obtain
\beq
\label{gij}
g_{\rm ij}=\frac{1}{2}\left[{\bf M}_{(1)}-{\bf M}_{(2)}^{\rm T}{\bf M}_{(3)}^{-1}{\bf M}_{(2)}\right]_{ij}\,.
\eeq
We note also that the mismatch $\langle
h(\psi_0,\psi_{3/2}),h(\psi_0+\Delta \psi_0,\psi_{3/2}+\Delta
\psi_{3/2})\rangle$ is translationally invariant in the
$(\psi_0,\psi_{3/2})$ plane, so the metric $g_{ij}$ is constant everywhere.
\begin{figure}[t]
\begin{center}
\includegraphics[width=\textwidth]{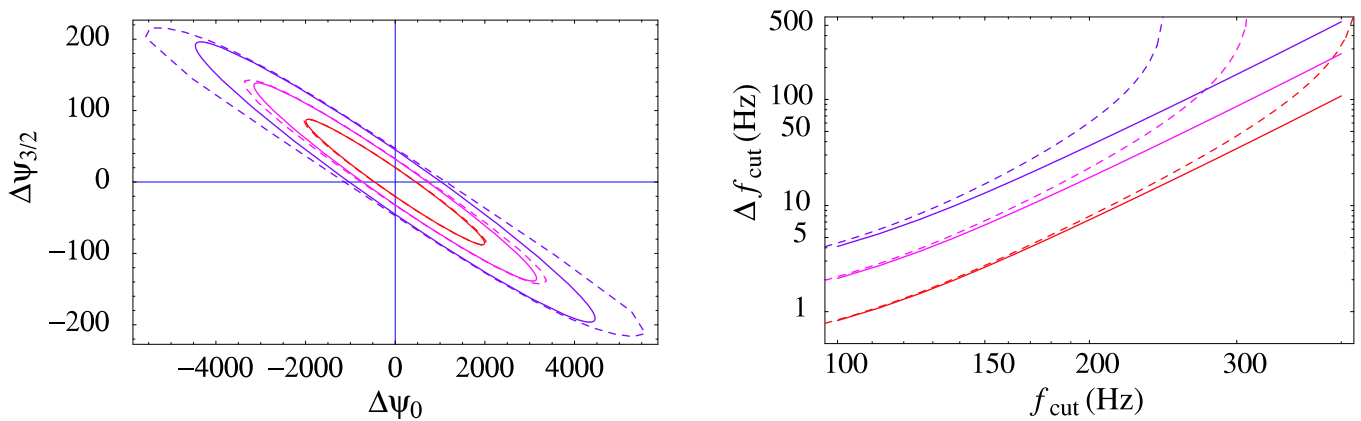}
\caption{\label{ADfmismatch} In the left panel, we plot
the iso-match contours for the function $\langle
h(\psi_0,\psi_{3/2}),h(\psi_0+\Delta \psi_0,\psi_{3/2}+\Delta
\psi_{3/2})\rangle$; contours are given at matches of 0.99, 0.975 and
0.95. Solid lines give the indications of the mismatch metric; dashed
lines give actual values. Here we use a Newtonian amplitude function
${\cal A}(f) = f^{-7/6}$ [we set $\alpha = 0$ and we do not cut the
template in the frequency domain. In fact $f_\mathrm{cut}=400\,\mathrm{Hz}$].
In the right panel we plot the values of $\Delta f_\mathrm{cut}$
(versus $f_{\rm cut}$) required to obtain matches $\langle
h(f_\mathrm{cut}),h(f_\mathrm{cut}+\Delta f_\mathrm{cut}) \rangle$ of
0.95 (uppermost curve), 0.975 and 0.99 (lowermost). In the region
below each contour the match is larger than the value quoted for the
contour. Again, here we use a Newtonian amplitude function 
${\cal A}(f) = f^{-7/6}$ [we set $\alpha = 0$].}
\end{center}
\end{figure}
In the left panel of Fig.~\ref{ADfmismatch}, we plot the iso-match
contours (at matches of 0.99, 0.975 and 0.95) in the $(\Delta
\psi_0,\Delta \psi_{3/2})$ plane, as given by the metric (\ref{gij})
[solid ellipses], compared with the actual values obtained from the
numerical computation of the matches [dashed lines].  For our
purposes, the second-order approximation given by the metric is quite
acceptable. In this computation we use a Newtonian amplitude function
${\cal A}(f)=f^{-7/6}$ [i.e., we set $\alpha = 0$ and we set 
our cutoff frequency at $400\,\mathrm{Hz}$].

We move now to the mismatch induced by different cutoff frequencies
$f_\mathrm{cut}$. Unlike the case of the $\psi_0$, $\psi_{3/2}$ parameters,
this mismatch is first order in $\Delta f_{\rm cut}$, so it cannot be
described by a metric. Suppose that we have two effective templates
$h(f_\mathrm{cut})$ and $h(f_\mathrm{cut} + \Delta f_\mathrm{cut})$
with the same phasing and amplitude $\Delta f>0$, but different cutoff
frequencies. The match is then given by
\bea \langle h(f_{\rm cut}) , h(f_{\rm cut}+\Delta f_\mathrm{cut})
\rangle &=& \frac{\left[\int_0^{f_{\rm cut}} df\frac{|{\cal
A}(f)|^2}{S_h(f)}\right]}{ \left[\int_0^{f_{\rm cut}} df\frac{|{\cal
A}(f)|^2}{S_h(f)}\right]^{1/2} \left[\int_0^{f_{\rm cut}+\Delta
f_\mathrm{cut}} df\frac{|{\cal A}(f)|^2}{S_h(f)}\right]^{1/2}} \\
\label{dfexact}
&=&
\left[\frac{
\int_0^{f_{\rm cut}} df\frac{|{\cal A}(f)|^2}{S_h(f)}}{
\int_0^{f_{\rm cut}+\Delta f_\mathrm{cut}} df\frac{|{\cal A}(f)|^2}{S_h(f)}
}\right]^{1/2}\, \simeq 
1 - \left[\frac{\Delta f_\mathrm{cut}}{2}\frac{|{\cal A}(f_\mathrm{cut})|^2
}{S_h(f_\mathrm{cut})}\right]\bigg/\left[\int_0^{f_{\rm cut}} df\frac{|{\cal
A}(f)|^2}{S_h(f)}\right]^{1/2}\,.
\eea
This result depends strongly on $f_\mathrm{cut}$. In the right panel
of Fig.~\ref{ADfmismatch} we plot the values of $\Delta
f_\mathrm{cut}$ that correspond to matches of 0.95, 0.975 and 0.99,
according to the first order approximation [solid lines], and to the
exact numerical calculations [dashed lines], both of which are given
in the second line of Eq.~(\ref{dfexact}). In the region below each
contour the match is larger than the value that characterizes the
contour. As we can see from the graph, the linear approximation 
is not very accurate, thus in the following we shall use the exact formula.

\subsection{Construction of the effective template bank: parameter range}
\label{subsec6.3}
\begin{figure}[t]
\begin{center}
\includegraphics[width=\sizeonefig]{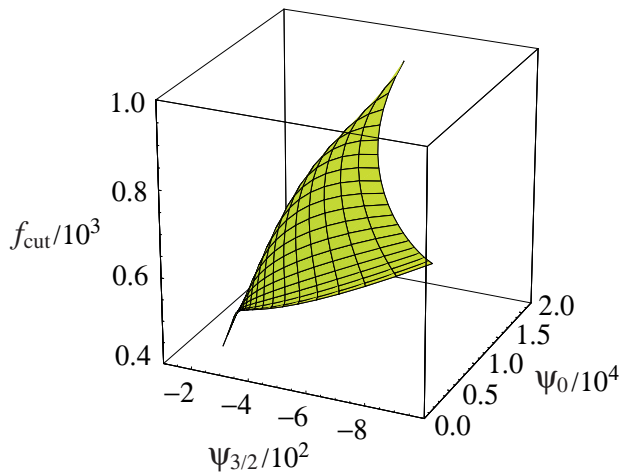}
\caption{\label{3dproj}Projection of
the ET$(2,2.5)$ waveforms onto the frequency-domain effective template
space. For $\alpha$ we choose the optimal value found by the 
search. The $(\psi_0, \psi_{3/2}, f_{\rm cut})$ surface is interpolated from the then mass
pairs shown in Tab.\ \ref{tableETB1}.}
\end{center}
\end{figure}

All the PN target models are parametrized by two independent numbers
(e.g., the two masses or the total mass and the mass ratio); if we
select a range of interest for these parameters, the resulting set
of PN signals can be seen as a two-dimensional region in the
$(m_1,m_2)$ or $(M,\eta)$ plane. Under the mapping that takes each PN
signal into the Fourier-domain effective template that matches it
best, this two-dimensional region is \emph{projected} into a
two-dimensional surface in the $(\psi_0,\psi_{3/2},f_\mathrm{cut})$
parameter space (with the fourth parameter $\alpha = 0$).
As an example, we show in Fig.~\ref{3dproj} the projection
of the ET$(2, 2.5)$ waveforms with (single-BH) masses $5$--$20\,{\rm
M}_{\odot}$. The 26 models tested in Secs.~\ref{sec3}, \ref{sec4}
would be projected into 26 similar surfaces. 
In constructing the detection template families, we shall first  
focus on  17 of the 26 models, namely the adiabatic T and P models at
2PN and 3PN, the E models at 2PN and at 3PN but with $\tilde{z}_{1,2}=0$, and
the H models at 3PN.  We will comment on the E models with 
$\tilde{z}_{1,2}\neq0$, on the L models, and on the HT and HP models at 2PN order 
at the end of this section.

\begin{figure*}[t]
\begin{center}
\includegraphics[width=\textwidth]{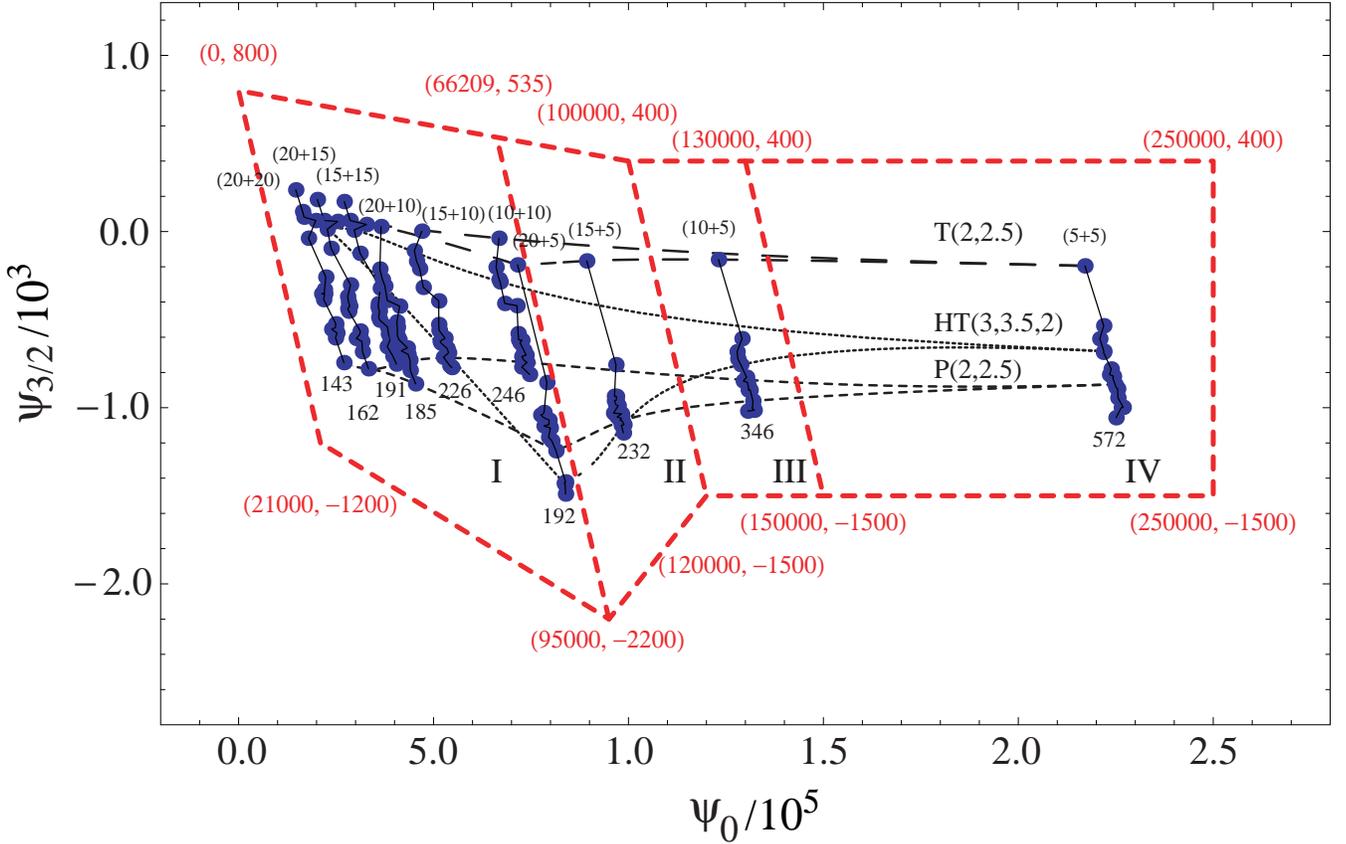}
\end{center}
\caption{Projection of the PN waveforms onto the
($\psi_0$,$\psi_{3/2}$) plane, for BBHs with masses $(5+5)M_{\odot}$,
$(10+5)M_{\odot}$, \ldots, $(20+20)M_{\odot}$ (see Tab.\ \ref{tableETB1}). The projection was computed by maximizing the maxmax
match over the parameters $\psi_0$, $\psi_{3/2}$ and $f_\mathrm{cut}$;
the correction coefficient $\alpha$ was set to zero. The thin dotted
and dashed lines show the boundaries of the projected images for the
models (from the top) T$(2, 2.5)$, HT$(3, 3.5, \hat{\theta}=2)$ and
P$(2, 2.5)$. Solid lines (the \emph{BH mass lines}) link the images of the
same BBH for different PN models. The ends of the BH mass lines are
marked with the BBH masses and with the minimum value 
$\min\{f_\mathrm{end},f_\mathrm{cut}\}$ across all the PN models.
The thick dashed lines delimit the region that
will be covered by the effective template bank; the
$(\psi_0,\psi_{3/2})$ coordinates are marked on the vertices. The
region is further subdivided into four subregions I--IV that group the
BH mass lines with very similar ending frequencies
$f_\mathrm{end\,min}$.\label{adall}}
\end{figure*}

It is hard to visualize all three parameters at once, so we shall start
with the phasing parameters $\psi_0$ and $\psi_{3/2}$.  In
Fig.~\ref{adall}, we plot the $(\psi_0,\psi_{3/2})$ section of the
PN-model projections into the ($\psi_0,\psi_{3/2},f_{\rm cutoff})$
space, with solid diamonds showing the projected points corresponding to BBHs with the same set of ten mass pairs as in Tab.\ \ref{tableETB1}. Each PN model is projected to a
curved-triangular region, with boundaries given by the sequences of
BBHs with masses $(m+m)$ (equal mass), $(20+m)$ and $(m+5)$.
In Fig.~\ref{adall}, these boundaries are plotted using
thin dashed lines, for the models T$(2, 2.5)$ (the uppermost in the plot),
HT$(3, 3.5, \hat{\theta}=2)$ (in the middle), and P$(2, 2.5)$ (lowest).

As we can see, different PN models can occupy regions with very
different areas, and thus require a very different number of effective
templates to match them with a given $\mathrm{MM}_\mathrm{T}$. Among
these three models, T$(2, 2.5)$ requires the least number of templates,
P$(2, 2.5)$ requires a few times more, and HT$(3, 3.5, \hat{\theta}=2)$
requires many more.
This is consistent with the result by Porter~\cite{P02} who found that, 
for the same range of physical parameters, T waveforms are more closely spaced
than P waveforms, so fewer are needed to achieve a certain MM.
In this plot we have also linked the points
that correspond to the same BBH parameters in different PN models. In
Fig.~\ref{adall}, these lines (we shall call them \emph{BH mass lines})
lie all roughly along one direction.
\begin{table}
\begin{center}
\begin{tabular}{r||c|c||c|c|c}
\hline \hline
\multicolumn{1}{c||}{$M$} & end-to-end match & ${\cal N}_{\rm end\,to\,end}$
& $f_{\rm cut\,min}$ 
&  $\langle h(f_{\rm cut\,min}),h(+\infty)\rangle$ &  ${\cal N}_{\rm mass\,line}^{\rm cut}$ \\
\hline
$(5+5)M_\odot$  & 0.478  & 37 & 572    & 1.00  & 0.2 \\
$(10+5)M_\odot$ & 0.434  & 41 & 346    & 0.98  & 0.9 \\
$(15+5)M_\odot$ & 0.398  & 46 & 232    & 0.94  & 3.1 \\
$(10+10)M_\odot$ & 0.449  & 40 & 246    & 0.95  & 2.6 \\
$(20+5)M_\odot$ & 0.347  & 52 & 192    & 0.90  & 5.3 \\
$(15+10)M_\odot$ & 0.443  & 40 & 226    & 0.94  & 3.3 \\
$(20+10)M_\odot$ & 0.428  & 42 & 185    & 0.89  & 5.9 \\
$(15+15)M_\odot$ & 0.482  & 36 & 191    & 0.90  & 5.4 \\
$(20+15)M_\odot$ & 0.464  & 38 & 162    & 0.84  & 8.5 \\
$(20+20)M_\odot$ & 0.438  & 41 & 143    & 0.79  & 11.9 \\
\hline \hline
\end{tabular}
\end{center}
\caption{\label{tab:endtoend}End-to-end matches and ending frequencies along the BH mass lines of Fig.\ \ref{adall}. The first three columns show the end-to-end matches and the corresponding number of templates (for MM $\simeq 0.98$) along the BH mass lines; the remaining columns
show the minimum ending frequencies of PN waveforms along the BH mass lines, the match between the two effective templates at the ends of the range, and the number of templates needed to step along the range while always maintaining a match $\simeq 0.98$ between neighboring templates.  When computing these matches, we use a Newtonian amplitude function ${\cal A}(f) = f^{-7/6}$ [we set $\alpha = 0$], and we maximize over the parameters $\psi_0$ and $\psi_{3/2}$ (which is equivalent to assuming perfect phasing synchronization).
}
\end{table}

A simple way to characterize the difference between the PN target
models is to evaluate the maxmax \emph{end-to-end match} between
effective templates at the two ends of the BH mass lines (i.e., the match
between the effective templates with the largest and smallest
$\psi_{3/2}$ among the projections of PN waveforms with the same mass
parameters $m_1$, $m_2$); we wish to focus first on the effects of the
phasing parameters, so we do not cut the templates in the frequency
domain and we set $\alpha=0$.  We compute also a naive end-to-end
number of templates, ${\cal N}_{\mathrm{end\,to\,end}}$, by counting
the templates required to step all along the BH mass line while
maintaining at each step a match $\simeq 0.98$ between neighboring
templates. A simple computation yields ${\cal
N}_{\mathrm{end\,to\,end}} = \log(\mbox{end-to-end
match})/\log(0.98)$. The results of this procedure are listed in
Table~\ref{tab:endtoend}.  Notice that, as opposed to the fitting
factors between template families computed elsewhere in this
paper (which are maximized over the BBH mass parameters of one
of the families), these matches give a measure of the dissimilarity
between different PN models \emph{for the same values of the BBH
parameters}; thus, they provide a crude estimate of how much the
effective template bank must be enlarged to embed all the various PN
models.

We expect that the projection of a true BBH waveform onto the
$(\psi_0,\psi_{3/2})$ plane will lie near the BH mass line with the true
BBH parameters, or perhaps near the extension of the BH mass line in either
direction. For this reason we shall lay down our effective
templates in the region traced out by the thick dashed lines in
Fig.~\ref{adall}, which was determined by extending the BH mass lines in both directions by half of their length.

We move on to specifying the required range of $f_{\rm cut}$ for each
$(\psi_0,\psi_{3/2})$.  For a given PN model and BBH mass parameters,
we have defined the \emph{ending frequency} $f_\mathrm{end}$ as the
instantaneous GW frequency at which we stop the integration of the PN
orbital equations. We find that usually the $f_{\rm cut}$ of the
optimally-matched projection of a PN template is larger than the
$f_\mathrm{end}$ of the PN template.  This is because the abrupt
termination of the PN waveforms in the time domain creates a tail in
the spectrum for frequencies higher than $f_\mathrm{end}$.  With
$f_\mathrm{cut} > f_\mathrm{end}$ and $\alpha > 0$, the effective
templates can mimic this tail and gain a higher match with the PN
models. In some cases, however, the optimal $f_{\rm cut}$ can be smaller 
than $f_{\rm end}$ [for example, P$(2, 2.5)$ with $(10+5)M_{\odot}$, 
$(15+5)M_{\odot}$ and $(10+10)M_{\odot}$] suggesting that the match of 
the phasing in the entire frequency band up to $f_{\rm end}$ is not very 
good and we have to shorten the Fourier-domain template. 
Now, since we do not know the details of the plunge for true
BBH inspiral, it is hard to estimate where the optimal
$f_\mathrm{cut}$ might lie, except perhaps imposing that it should be
larger than ${\rm min}(f_\mathrm{end}, f_\mathrm{cut})$. 
A possibility is to set the range of $f_\mathrm{cut}$ to
be  above  $f_{\rm cut\,min}\equiv \min\{f_\mathrm{cut},f_\mathrm{end}\}$, with 
the minimum evaluated among all the PN models.

In Table \ref{tab:endtoend} we show the $f_\mathrm{cut\,min}$ found across the PN models for given BBH mass parameters. We have also marked this minimum frequency in Fig.~\ref{adall} under the corresponding BH 
mass lines. In the table we also show the match of the two detection  templates
$h(f_\mathrm{cut}=f_\mathrm{cut\,min})$ and
$h(f_\mathrm{cut}=+\infty)$, and the number ${\cal 
N}^\mathrm{cut}_{\mathrm{mass\,line}}$ of intermediate templates with
different $f_\mathrm{cut}$ needed to move from
$h(f_\mathrm{cut\,min})$ to $h(+\infty)$ while maintaining
at each step a match $\simeq 0.98$ between neighboring templates.  
It is easy to see that this number is ${\cal N}^\mathrm{cut}_
{\mathrm{mass\,line}}=\log \langle h(f_{\rm
cut\,min}), h(+\infty)\rangle/\log(0.98)$.  The match was
computing using a Newtonian amplitude function ${\cal A}(f) =
f^{-7/6}$ [we set $\alpha = 0$], and maximizing over the parameters
$\psi_0$ and $\psi_{3/2}$.
Under our previous hypothesis that the projection of a true BBH
waveform would lie near the corresponding BH mass line, we can use the
numbers in Table~\ref{tab:endtoend} to provide a rough estimate of the
range of $f_\mathrm{cut}$ that should be taken at each point
$(\psi_0,\psi_{3/2})$ within the dashed contour of
Fig.~\ref{adall}. We trace out four subregions I, II, III, IV, such
that the BH mass lines of each subregion have approximately the same
values of $f_{\rm cut\,min}$; we then use these minimum ending
frequencies to set a lower limit for the values of $f_\mathrm{cut}$
required in each subregion: $f_\mathrm{cut\,min}(\mathrm{I})=143$,
$f_\mathrm{cut\,min}(\mathrm{II})=192$,
$f_\mathrm{cut\,min}(\mathrm{III})=232$,
$f_\mathrm{cut\,min}(\mathrm{IV})=346$.  The maximum $f_\mathrm{cut}$
is effectively set by the detector noise curve, which limits the
highest frequency at which signal-to-noise can be still accumulated.

Moving on to the last parameter, $\alpha$, we note that it is
probably only meaningful to have  $\alpha f_\mathrm{cut}^{2/3} \leq
1$, so that $\mathcal{A}_\mathrm{eff}(f)$ cannot become negative for $f <
f_\mathrm{cut}$. [A negative amplitude in the detection template will
usually give a negative contribution to the overlap, unless the
phasing mismatch is larger than $\pi/2$, which does not seem
plausible in our cases.] Indeed, the optimized values found for $\alpha$
in Tab.\ \ref{tableETB1} seem to follow this rule, except for a few slight violations that are probably due to 
numerical error (since we had performed a search to find the optimal 
value of $\alpha$). 
For the 17 models considered here, the optimal $\alpha$ is 
always positive [Tab.~\ref{tableETB1}]
which means that, due to cutoff effects, the amplitude at high frequencies 
becomes always lower than the $f^{-7/6}$ power law. So for the 17 
models considered in this section $0\le\alpha f_\mathrm{cut}^{2/3} \le 1$.
[Note that this range will have to be extended to include negative $\alpha$'s if we want to 
incorporate the models discussed in Sec.~\ref{subsec6.6}.]

\subsection{Construction of the effective templates bank: parameter density}
\label{subsec6.4}

At this stage, we have completed the specification of the region in
the $(\psi_0,\psi_{3/2},f_{\rm cut},\alpha)$ parameter space where we
shall lay down our bank of templates. We expect that the FF for the
projection of the true physical signals (emitted by
\emph{nonspinning} BBHs with $M = 10 \mbox{--} 40 M_\odot$) onto this
template bank should be very good. We now wish to evaluate the total
number of templates $\mathcal{N}$ needed to achieve a certain MM.

We shall find it convenient to separate the mismatch due to the
phasing from the mismatch due to the frequency cuts by introducing two
minimum match parameters $\mathrm{MM}_\psi$ and
$\mathrm{MM}_\mathrm{cut}$, with $\mathrm{MM} = \mathrm{MM}_\psi \cdot
\mathrm{MM}_\mathrm{cut} \simeq \mathrm{MM}_\psi +
\mathrm{MM}_\mathrm{cut} - 1$.  As mentioned at the beginning of this
section, the correction coefficient $\alpha$ is essentially an
\emph{extrinsic} parameter [see Sec.\ \ref{subsec2.2}]: we do not need
to discretize the template bank with respect to $\alpha$, and there
is no corresponding $\mathrm{MM}$ parameter.

We evaluate $\mathcal{N}$ in three refinement steps:
\begin{enumerate}
\item We start by considering only the phasing parameters, and we
compute the parameter area $S_i$ [in the $(\psi_0,\psi_{3/2})$ plane]
for each of the subregions $i = $ I, II, III, IV of
Fig.~\ref{adall}. We then multiply by the determinant $\sqrt{g}$ of
the constant metric, and divide by 2(1--MM$_\psi$), according to
Eq.~\eqref{eq:metric}, to get
\begin{equation}
\mathcal{N} = \sum_i \frac{S_i \sqrt{g}}{2(1-\mathrm{MM}_\psi)}.
\end{equation}
This expression is for the moment only formal, because we cannot
compute $\sqrt{g}$ without considering the amplitude
parameters $\alpha$ and $f_\mathrm{cut}$.
\item Next, we include the effect of $f_\mathrm{cut}$. In the previous
section, we have set $f_\mathrm{min\,cut}$ for each of the subregions
by considering the range swept by $f_\mathrm{end}$ along the mass
lines. Recalling our discussion of
$\mathcal{N}^\mathrm{cut}_\mathrm{mass\,line}$, we approximate the
number of distinct values of $f_\mathrm{cut}$ that we need to include
for each parameter pair $(\psi_0,\psi_{3/2})$ as
\begin{equation}
n^\mathrm{cut}_i(\psi_0,\psi_{3/2},\alpha) \simeq 1 + \frac{\log
  \left\langle h(\psi_0,\psi_{3/2},\alpha,f_\mathrm{min\,cut}),
  h(\psi_0,\psi_{3/2},\alpha,\mathrm{no\,cut}) \right\rangle}{\log 
\mathrm{MM}_\mathrm{cut}}.
\end{equation}
For $\alpha$ in the physical range $0 \leq \alpha \leq
f^{-2/3}_\mathrm{cut}$ this match is minimized for $\alpha = 0$, so
this is the value that we use to evaluate the $n^\mathrm{cut}_i$'s. 
Note that the choice of cutoff frequencies does not depend on the
values of the phasing parameters. This allows us to have a single set of
cutoff frequencies for all points in one subregion. For subregion $i$,
we denote this set by $F_i$.
\item The final step is to include the effect of $\alpha$ and
$f_\mathrm{cut}$ on the computation of $\sqrt{g}$. For simplicity, we
shoot for an upper limit by maximizing $\sqrt{g}$ with respect to
$\alpha$. [Because $\alpha$ is essentially an extrinsic
parameter, we do not multiply $\mathcal{N}$ by the number of its
discrete values: the matches are automatically maximized on the
continuous range $0 \leq \alpha \leq f_\mathrm{cut}^{-2/3}$.] Our
final estimate for the total number of templates is
\begin{equation}
\mathcal{N} = \frac{1}{2(1-\mathrm{MM}_\psi)}\sum_i S_i
\sum_{f_{\mathrm{cut}} \in F_i} \max_{\alpha} [\sqrt{g}]
\end{equation}
\end{enumerate}

We have evaluated this $\mathcal{N}$ numerically. \emph{We find that
the contributions to the total number of templates from the four
subregions, for $\mathrm{MM} = 0.96$ (taking $\mathrm{MM}_\psi =
\mathrm{MM}_\mathrm{cut} = 0.98$), are $\mathcal{N}(\mathrm{I}) \simeq
6,410$, $\mathcal{N}(\mathrm{II}) \simeq 2,170$,
$\mathcal{N}(\mathrm{III}) \simeq 1,380$, $\mathcal{N}(\mathrm{IV})
\simeq 1,230$, for a total of $\mathcal{N} = 11,190$. This number
scales approximately as $[0.04/(1-\mathrm{MM})]^2$.}  Notice that
subregion I, which contains all the BBHs with total mass above
$25M_{\odot}$, requires by far the largest number of templates. This
is mostly because these waveforms end in the LIGO band, and many
values of $f_{\rm cut}$ are needed to match different ending
frequencies.

Remember that the optimal signal-to-noise ratio $\rho$ for filtering
the true GW signals by a template bank is approximately degraded (in
the worst case) by the factor $\mathrm{MM}_\mathrm{T} = \mathrm{FF} +
\mathrm{MM} - 1$ \footnote{This is true only when the waveform and the
neighboring detection templates are all sufficiently close so that the
metric formalism is still valid. As we have seen in Fig.~\ref{ADfmismatch}, 
by imposing $\mathrm{MM}_{\psi}=0.98$, the overlaps between the neighboring detection 
templates are well described by the metric. However, due to the
fact we do not know the true waveforms, and thus the true FF, it is not
quite certain how exact this formula will eventually be. In some
sense, this formula could be regarded an additional assumption.}.
 
While $\mathrm{MM}$ depends on the geometry of the
template bank, we can only guess at the fitting factor $\mathrm{FF}$
for the projection of the true signal onto the template space. In this
section we have seen that all PN models can be projected onto the
effective frequency-domain templates with a good FF: for a vast
majority of the waveforms $\mathrm{FF} \gtrsim 0.96$ (and the few
exceptions can be explained). \emph{It is therefore reasonable to hope
that the FF for the true GW signals is $\sim 0.96$, so the total
degradation from the optimal $\rho$ will be $\mathrm{MM}_\mathrm{T}
\gtrsim 0.92$, corresponding to a loss of $\lesssim$ 22\% in event
rate}. This number can be improved by scaling up the number of
templates, but of course the actual FF represents an upper limit for
$\mathrm{MM}_\mathrm{T}$. For instance, about 47,600 templates should
get us $\mathrm{MM}_\mathrm{T} \gtrsim 0.94$, corresponding to a loss
of $\lesssim$ 17\% in event rate.

\subsection{Parameter estimation with the detection  template family}
\label{subsec6.5}

Although our family of effective templates was built for the main
purpose of detecting BBHs, we can still use it (once a detection is
made) to extract partial information about the BH masses.  It is
obvious from Fig.~\ref{adall} that the masses cannot in general be
determined unambiguously from the best-match parameters [i.e., the
projection of the true waveform onto the $(\psi_0,\psi_{3/2})$ plane],
because the images of different PN models in the plane have
overlaps. Therefore different PN models will have different ideas, as
it were, about the true masses. Another way of saying this is that the BH
mass lines can cross.

However, it still seems possible to extract at least one mass
parameter, the chirp mass ${\cal M}=M\eta^{3/5}$,
with some accuracy. Since the phasing is dominated by the term
$\psi_0 f^{-5/3}$ at low frequencies, we can use the leading Newtonian
term $\psi_{\rm N}(f)= \frac{3}{128}(\pi {\cal M} f)^{-5/3}$ obtained
for a PN expanded adiabatic model in the stationary-phase
approximation to infer
\beq
\label{chirpA}
\psi_0 \sim \frac{3}{128}\left(\frac{1}{\pi {\cal M}}\right)^{5/3} \quad
\Longrightarrow \quad
{\cal M}^{\rm approx}=\frac{1}{\pi}\left(\frac{3}{128\,\psi_0}\right)^{3/5}.
\eeq
If this correspondence was exact, the BH mass lines in
Fig.~\ref{adall} would all be vertical. They are not, so this
estimation has an error that gets larger for smaller $\psi_0$ (i.e.,
for binaries with higher masses). In Table~\ref{tab:estimate} we show
the range of chirp-mass estimates obtained from Eq.\ (\ref{chirpA})
for the values of $\psi_0$ at the projections of the PN models in
Fig.\ \ref{adall}, together with their percentage error $\epsilon
\equiv ({\cal M}^{\rm approx}_{\rm max}-{\cal M}^{\rm approx}_{\rm
min})/{\cal M}$.  In this table, ${\cal M}_{\rm max}$ and ${\cal
M}_{\rm min}$ correspond to the endpoints of the BH mass lines. If we
take into account the extension of the BH mass lines by a factor of
two in the effective template bank, we should double the $\epsilon$ of
the table.
\begin{table}
\begin{center}
\begin{tabular}{r||r|r|r|r}
\hline \hline
\multicolumn{1}{c||}{$M$} & ${\cal M}$ & ${\cal M}^{\rm approx}_{\rm min}$ &
${\cal M}^{\rm approx}_{\rm max}$ & $\epsilon \%$ \\
\hline
$(5+5)M_\odot$ &  4.35 &  4.16 &  4.27 & 2.6\\
$(10+5)M_\odot$ &  6.08 &  5.75 &  6.00 & 4.2\\
$(15+5)M_\odot$ &  7.33 &  6.85 &  7.28 & 5.9\\
$(10+10)M_\odot$ & 8.71 &  8.10 &  8.72 & 7.1\\
$(20+5)M_\odot$ &  8.33 &  7.55 &  8.31 & 9.1\\
$(15+10)M_\odot$ & 10.62 & 9.76 & 10.96 & 11.3\\
$(20+10)M_\odot$ & 12.17 & 10.92 & 12.50 & 13.0\\
$(15+15)M_\odot$ & 13.06 & 11.69 & 14.88 & 24.4 \\
$(20+15)M_\odot$ & 15.05 & 13.15 & 17.74 & 30.6\\
$(20+20)M_\odot$ & 17.41 & 14.91 & 21.52 & 38.0\\
\hline \hline
\end {tabular}
\end{center}
\caption{\label{tab:estimate}Estimation of the chirp masses
$\mathcal{M}$ from the projections of the PN target models onto the
Fourier-domain effective template space.  The numbers in the second
column (labeled ``$\mathcal{M}$'') give the values of the chirp mass
corresponding to the BH masses to their left; the numbers in the
third and fourth columns give the range of estimates obtained from
Eq.\ \eqref{chirpA} for the values of $\psi_0$ at the projections of
the target models shown in Fig.\ \ref{adall}.
The last column shows the percentage error $\epsilon \equiv ({\cal
M}^{\rm approx}_{\rm max}-{\cal M}^{\rm approx}_{\rm min})/{\cal M}$.}
\end{table}

It seems quite possible that a more detailed investigations
of the geometry of the projections into the effective template space
(and especially of the BH mass lines) could produce better algorithms to
estimate binary parameters. But again, probably only one 
parameter can be estimated with certain accuracy.

\subsection{Extension of the two-dimensional Fourier-domain detection template}
\label{subsec6.6}

\begin{figure}
\includegraphics[width=\sizeonefig]{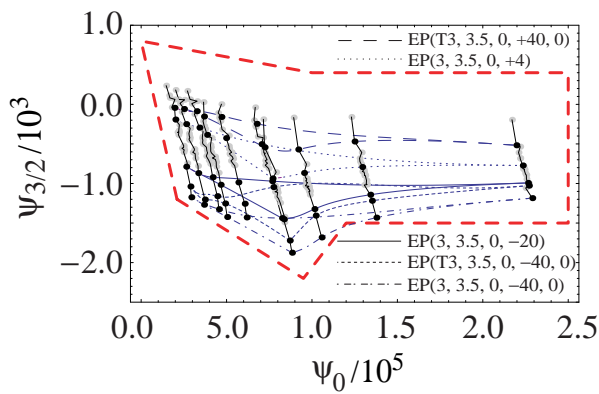}
\caption{Projection of the E models with nonzero $\tilde{z}_{1}$ 
into the $(\psi_0,\psi_{3/2})$ plane (shown in black dots.) The new points 
sit quite well along the BH mass lines of the 17 models investigated in 
Secs.~\protect\ref{subsec6.3}, \protect\ref{subsec6.4} and 
\protect\ref{subsec6.5}. We use the notation EP$(3,3.5,\hat{\theta},\tilde{z}_1,\tilde{z}_2)$ 
and denote by EP(T3, ...) the two-body model in which the 
coefficient $A(r)$ is PN expanded [see Eq.~(\protect\ref{coeffA})].
\label{Ezproj}}
\end{figure}
\begin{figure}
\begin{center}
\includegraphics[width=\sizeonefig]{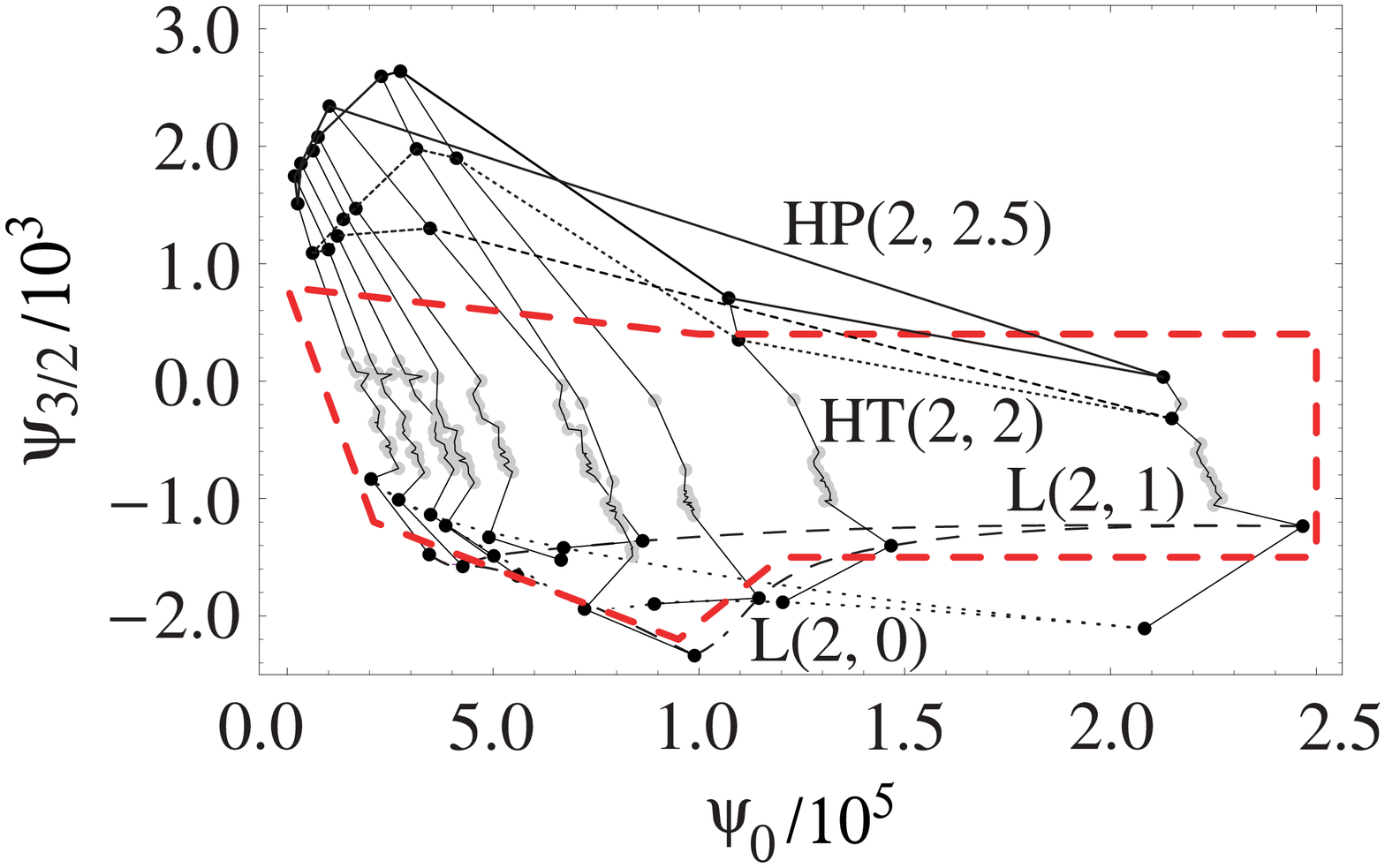}
\end{center}
\caption{Projections of HT and HP models at 2PN and L models into the $(\psi_0,\psi_{3/2})$ plane (shown in black dots.) The projections of the previous 17 models are shown in gray dots. \label{HLproj}} 
\end{figure}
In our construction of the effective template bank, we have been focusing 
until now on a subset of 17 models. The models we left out are:  
E models at 3PN with $\tilde{z}_{1,2}$ nonzero, 
HT and HP models at 2PN, and L models.

As we can see from Fig.~\ref{Ezproj}, E models with $\tilde{z}_{1,2}$ 
nonzero have a very similar behavior to the 17 models investigated  above. 
Indeed: (i) the projection of the PN waveforms from
the same model occupy regions that are triangular, and (ii) the
projections of PN waveforms of a given mass lies on the BH mass line
spanned by the previous 17 models. In addition,  their projections lie
roughly in the region we have already defined in Secs.~\ref{subsec6.3}, 
\ref{subsec6.4} and \ref{subsec6.5}. However, the ending
frequencies of these models can be much lower than the values we have set
for the detection templates: the detection templates (in all four
subregions) should be extended to lower cutoff frequencies if we decide 
to match these models, up to FF $\sim 0.95$. A rough estimate shows that  
this increases the number of templates to about twice the original value.

In Fig.~\ref{HLproj}, we plot the projections of the L$(2, 0)$, L$(2, 1)$,
HT$(2, 2)$
and HP$(2, 2.5)$ waveforms into the $(\psi_0, \psi_{3/2})$ plane. As we 
already know, these models are not matched by the detection 
templates as well as the other 17 models. Here we can see 
that their projections onto the $(\psi_{0},\psi_{3/2})$ plane are also 
quite dissimilar from those models.  For L models, although different
masses project into a triangular region, the projection of each mass
configuration does not align along the corresponding BH mass line
generated by the 17 models. In order to cover the $L$ models up to FF$\sim 0.93$, we need to expand the
$(\psi_0,\psi_{3/2})$ region only slightly. However, as we read from 
Tab.\ \ref{tableETB1}, the cutoff frequencies need to be extended to even lower values than for the E models with nonzero $\tilde{z}_{1,2}$. Luckily, this expansion will not cost much. In the end the total number of templates needed should be about three times the original value.

For HT and HP models at 2PN, the projections almost lie along the BH mass lines, but the regions 
occupied by these projections have weird shapes. We have to
extend the $(\psi_0,\psi_{3/2})$ region by a factor $\sim 2$ in order to
cover the phasings. [The ending/cutoff frequencies for these models 
are higher than for the previous two types of models.] An additional
subtlety in this case is that, as we can read from
Tab.\ \ref{tableETB1}, the optimal values of $\alpha$ are often negative, since the amplitude
becomes higher than the $f^{-7/6}$ power law at higher
frequencies. This expansion of the range of  $\alpha$ affects both
the choice of the discrete cutoff frequencies and the placement 
of $(\psi_0,\psi_{3/2})$ lattices. This effect is yet to be estimated. 

Finally, we notice that if these extensions are made, then the 
estimation of the chirp mass from the coefficient $\psi_0$ 
becomes less accurate than the one given in Table~\ref{tab:estimate}. 

\subsection{Extension of the Fourier-domain detection template family to more than two phasing parameters}
\label{subsec6.7}

\begin{figure}[ht]
\begin{center}
\includegraphics[width=\sizeonefig]{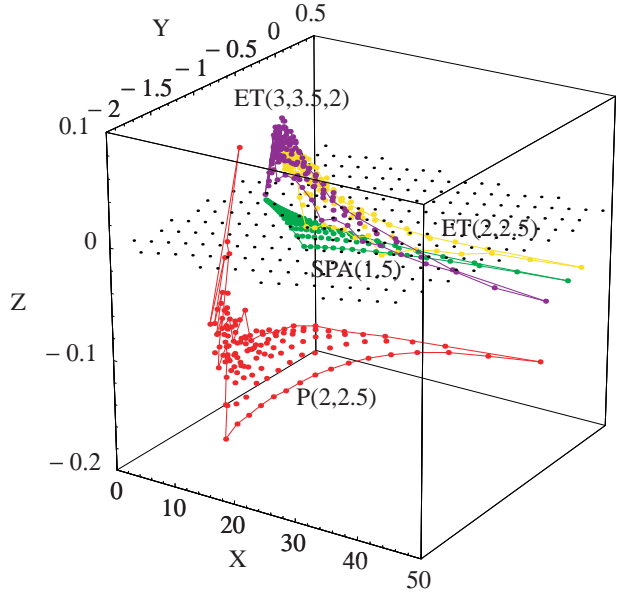}
\end{center}
\caption{Projection of the models P$(2, 2.5)$, ET$(2, 2.5)$, ET$(3,
3.5, 0)$, and SPA(1.5) onto the three-parameter Fourier-domain
detection template, for many BBH masses that lie within the same
ranges taken in Fig.~\ref{adall}. The variables $(X,Y,Z)$ are related
to $(\psi_0,\psi_{1},\psi_{3/2})$ by a linear transformation,
constructed so that the mismatch metric is just $\delta_{ij}$ and that
the $(\psi_0,0,\psi_{3/2})$ plane is mapped to the $(X,Y,0)$
plane. The dots show the value of the
parameters $(X,Y,Z)$ where the match with
one of the PN waveforms is maximum.\label{acdprojrot}}
\end{figure}
\begin{figure}
\begin{center}
\includegraphics[width=\sizeonefig]{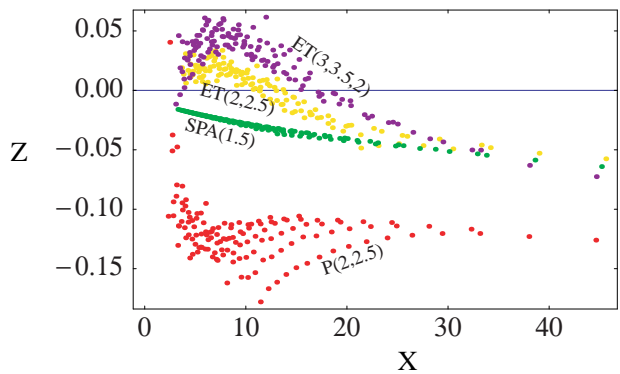}
\end{center}
\caption{$(X,Z)$ section of Fig.~\protect\ref{acdprojrot}. Comparison with Fig.\ \ref{acdprojrot} shows that all the projections lie near the $(X,0,Z)$ plane. \label{acddev}}
\end{figure}

It might seem an accident that by using only two phasing
parameters, $\psi_0$ and $\psi_{3/2}$, we are able to match very
precisely the wide variety of PN waveforms that we have considered.
Indeed, since the waveforms predicted by each PN model span a
two-dimensional manifold (generated by varying the two BH masses $m_1$
and $m_2$ or equivalently the mass parameters $M$ and $\eta$), we
could naturally expect that a \emph{third parameter} is required to
incorporate all the PN models in a more general family, and to add
even more signal shapes that extrapolate beyond the phasings and
amplitudes seen in the PN models.

In particular, because the accumulation of signal-to-noise ratio is
more sensitive to how well we can match \emph{the phasing} (rather
than the amplitude) of PN templates, such a third parameter should
probably interpolate between phasings predicted by different PN
models. As a consequence, the amplitude parameters $f_{\rm cut}$ and
${\cal A}$ do not generate a real dimensional extension of our
detection template family. In this section, we present a qualitative
study of the extension of our detection template family obtained by
adding one phasing parameter, the parameter $\psi_1$ of
Eq.\ \eqref{phasing}.

We use the $(\psi_0,\psi_1,\psi_{3/2})$ Fourier-domain detection
templates to match the PN waveforms from the models P$(2, 2.5)$,
ET$(2, 2.5)$, and ET$(3, 3.5, 0)$; these models were chosen because their projections onto the $(\psi_0,\psi_{3/2})$ detection templates were rather distant in the $(\psi_0,\psi_{3/2})$ parameter space.
Throughout this section (and unlike the rest of this paper), we use an
approximated search procedure whereby we essentially replace the
amplitude of the target models with the Newtonian amplitude ${\cal
A}(f) = f^{-7/6}$ with a cutoff frequency $f_\mathrm{cut}$ [we always
assumed ${\cal A}=0$ and $f_{\rm cut}=400$ Hz].  As expected, the
matches increase, and indeed they are almost perfect: always higher
than $0.994$ (it should be remembered however that these should be
considered as matches \emph{of the {\rm PN} phasings} rather than as matches of the PN waveforms; especially for high masses, the frequency
dependence of the amplitude is likely to change these values).

If we plot the projections of the PN waveforms in the $(\psi_0,\psi_1,\psi_{3/2})$ space, we find that the clusters of points corresponding to each PN target model look quite different from the projections [onto the $(\psi_0,\psi_{3/2})$ template space] shown in Fig.~\ref{adall}; but this is just an artifact of the parametrization. We can perform a linear transformation $(\psi_0,\psi_1,\psi_{3/2}) \rightarrow (X,Y,Z)$, defined in such a
way that (i) in the $(X,Y,Z)$ parameters, the mismatch metric is just
$\delta_{ij}$, and that (ii) the $(\psi_0,0,\psi_{3/2})$ plane is
mapped to the $(X,Y,0)$ plane. These conditions define the linear
transformation up to a translation and a rotation along the $Z$ axis;
to specify the transformation completely we require also that all the
projections of the PN models lie near the origin, and be concentrated
around the $X$ axis.
Figure~\ref{acdprojrot} shows the projection of the PN models P$(2,
2.5)$, ET$(2, 2.5)$, and ET$(3, 3.5, 0)$ onto the
$(\psi_0,\psi_1,\psi_{3/2})$ detection template family, as parametrized by the $(X,Y,Z)$ coordinate system, for many BBH
masses that lie within the same ranges of Fig.~\ref{adall}.  Each
dot marks the parameters $(X,Y,Z)$ that best match
the phasing of one of the PN waveforms.  We include also the
projection of a further PN model, SPA(1.5), obtained by solving the
frequency-domain version of the balance equation, obtained in the
stationary-phase approximation from our T model. The expression of the
SPA(1.5) phasing as a function of $f$ coincides with our
Eq.\ \eqref{phasing}, but the coefficients that correspond to
$(\psi_0,\psi_{1},\psi_{3/2})$ \emph{are functions of the two mass
parameters} $M$ and $\eta$.

By construction, the match between nearby detection templates 
is related to their Euclidian distance in the $(X,Y,Z)$ by 
\beq
1-{\rm overlap}=\Delta X^2+\Delta Y^2+\Delta Z^2\,.
\eeq
We see immediately that all the PN models are not very distant from the
$(X,Y,0)$ plane [also shown in the figure], which coincides with the
$(\psi_0,\psi_{3/2})$ plane. The farthest model is P$(2, 2.5)$, with a
maximum distance $\sim 0.18$. It is important to notice that, since
this number is obtained by assuming $f_{\rm cut}=400$ Hz and ${\cal
A}=0$, it tends to underestimate the true overlaps for models that end
below $400$ Hz, such as the P models at higher masses.  See also
Fig.~\ref{acddev} for an $(X,Z)$ section of Fig.~\ref{acdprojrot}.

\begin{figure}
\begin{center}
\begin{tabular}{ccc}
\includegraphics[width=0.45\textwidth]{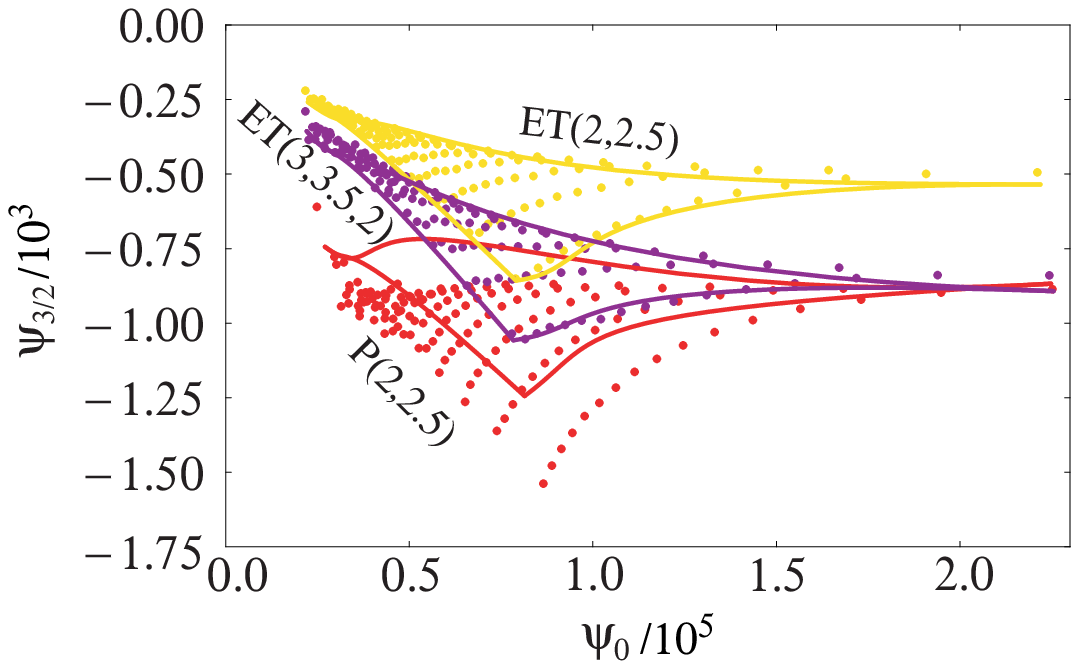} &
\hspace{0.03\textwidth} &
\includegraphics[width=0.45\textwidth]{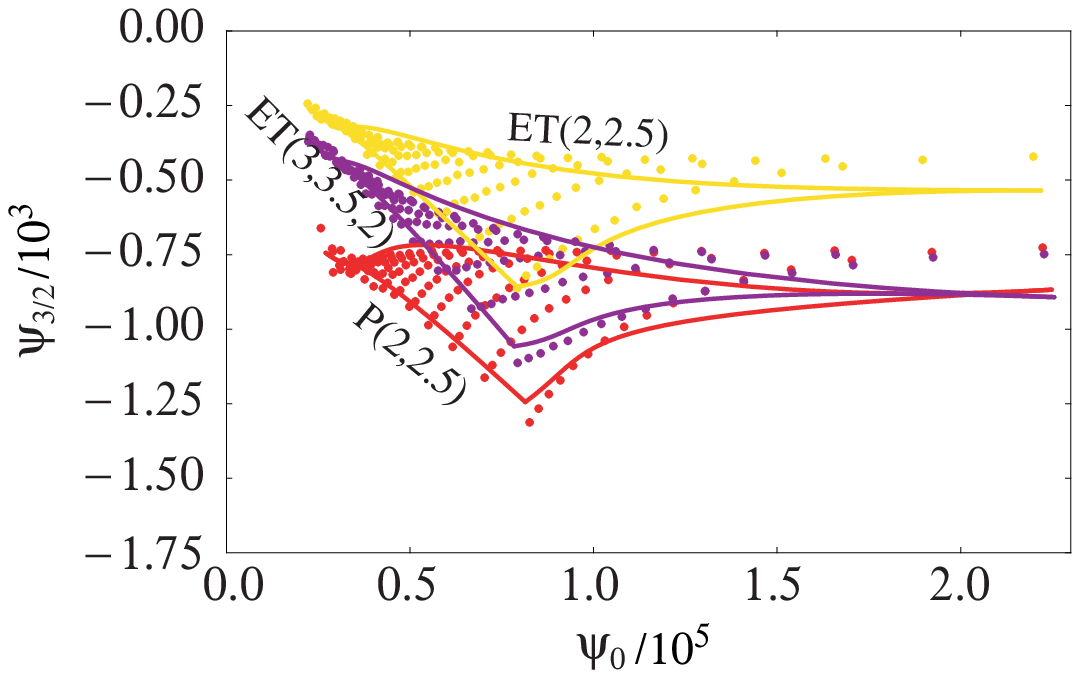}
\end{tabular}
\caption{In this figure, we compare the projection of the PN models
onto the three-dimensional $(\psi_0,\psi_{1},\psi_{3/2})$
Fourier-domain detection template family [shown by the dots as a
two-dimensional section in the $(\psi_0,\psi_{3/2})$ submanifold] with
the projection of the PN models in the two-dimensional
$(\psi_0,\psi_{3/2})$ template family [shown by the lines].  In the
left panel, we use ${\cal A}=0$ and $f_{\rm cut}=400$ Hz to maximize
the matches; in the right panel we use ${\cal A}=0$ and $f_{\rm
cut}=200$ Hz.\label{acdtoad}}
\end{center}
\end{figure}

We can study the relation between this three-dimensional family of
templates and the two-dimensional family considered earlier by
projecting the points of Fig.~\ref{acdprojrot} onto the $(X,Y,0)$
plane [which corresponds to the $(\psi_0,0,\psi_{3/2})$ plane]. The
resulting images resemble closely the projections of the PN models
onto the $(\psi_0,\psi_{3/2})$ parameter space of the two-dimensional
family, as seen in the left panel of Fig.~\ref{acdtoad}.  However, the
agreement is poor for P$(2, 2.5)$ because of the relatively high cut
frequency $f_{\rm cut}=400$ Hz. The right panel of Fig.~\ref{acdtoad}
was obtained by taking $f_{\rm cut}=200$ Hz. The agreement is much
better.  This result goes some way toward explaining why using
only two phasing parameters was enough to match most PN models in a
satisfactory way.

As stated at the beginning of this section, the parameter $Z$ can
indeed be used to expand the dimensionality of our detection template
family, because it appears to interpolate between different PN
models. It is possible that the number of $Z$ values needed when
laying down a discrete template family might not be too large, because
the PN models do not seem to lie very far from the $Z=0$ plane
[remember that distances in the $(X,Y,Z)$ parameter space are
approximately mismatch distances].

The good performance that we find for the two- and three-dimensional
Fourier-domain families confirms the results obtained in
Refs.~\cite{DIS3}, \cite{CA} and \cite{EPT}.  In Ref.~\cite{DIS3}, the
authors point out that the waveforms obtained from the stationary
phase approximation at 2PN and 2.5PN order are able to approximate the
E models, throughout most of the LIGO band, 
by maximizing over the mass parameters [see Ref.~\cite{DIS3}, 
in particular the discussion of 
their model ``Tf2,'' and the discussion around their Fig.~2].

In Ref.~\cite{CA}, Chronopolous and Apostolatos show that what would
be in our notation the SPA$(2)$ model (where the phasing is described
by a fourth-order polynomial in the variable $f^{1/3}$) can be
approximated very well, at least for the purpose of signal detection,
by the SPA$(1.5)$ model, with the advantage of having a much lower
number of templates.  In Ref.~\cite{EPT}, the authors go even further,
investigating the possibility of approximating the SPA$(2)$ phasing
with a polynomial of third, second and even first degree obtained
using Chebyshev approximants.

It is important to underline that in all of these analyses the
coefficients that appear in the expression of the phasing
[corresponding to our $\psi_0, \psi_1, \ldots$ in Eq.~\eqref{phasing}]
depend on only two BBH mass parameters, either directly~\cite{DIS3,CA},
or indirectly~\cite{EPT} through specific PN relations at each PN
order.  As a consequence, the phasings assumed in these references are
confined to a two-dimensional submanifold analog to the surface
labeled ``SPA$(1.5)$'' in Fig.\ \ref{acdprojrot}.

In this paper we follow a more general approach, because the phasing
coefficients $\psi_i$ are initially left completely arbitrary.  Only
after studying systematically the projection of the PN models onto the
template bank we have determined the region where a possible detection
template bank would be laid down.  The high matches that we find between
detection templates and the various PN models depend crucially on this
assumption. As a consequence, our parameters $\psi_i$ do not have a
direct physical meaning, and they cannot easily be traced back to specific
functions of the BBH masses, except for the chirp mass, as seen in
Sec.~\ref{subsec6.5}. This is natural, because our detection templates
are built to interpolate between different PN models, each of
which has, as it were, a different idea of what the waveform for a BBH
of given masses should be.

\section{Performance of the time-domain detection templates and construction of 
the detection bank in time domain}
\label{sec7}
Another possibility of building a detection template family 
is to adopt one or more of the physical models
discussed in Secs.~\ref{sec4} as the effective template bank used for
detection. Under the general hypothesis that underlies this work (that
is, that the \emph{target} models span the region in signal space
where the true physical signals reside), if we find that one of the
target models matches all the others very well, we can use it as the
effective model; and we can estimate its effectualness in matching the
true physical signal from its effectualness in matching all the other
models.

As shown in Tables~\ref{VIa}, \ref{VIb} and discussed in Sec.~\ref{sec5}, the
fitting factors FF for the projection of the PN models onto each
other are low (at least for PN order $n \leq 2.5$ or for high masses); in other
words, the models appear to be quite distant in signal space.
This conclusion is overturned, however, if we let the dimensionless
mass ratio $\eta$ move beyond its physical range $0 \leq \eta \leq
1/4$. For instance, the P$(2,2.5)$ and EP$(3,3.5,0)$ models can be
extended formally to the range $0 \leq \eta \leq 1$. Beyond those
ranges, either the equations (of energy-balance, or motion) become
singular, or the determination of the MECO or light ring (the
evolutionary endpoint of the inspiral for the P$(2,2.5)$ model and the
EP$(3,3.5,0)$ model, respectively) fails.

\begin{figure*}
\begin{center}
\includegraphics[width=\sizeonefig]{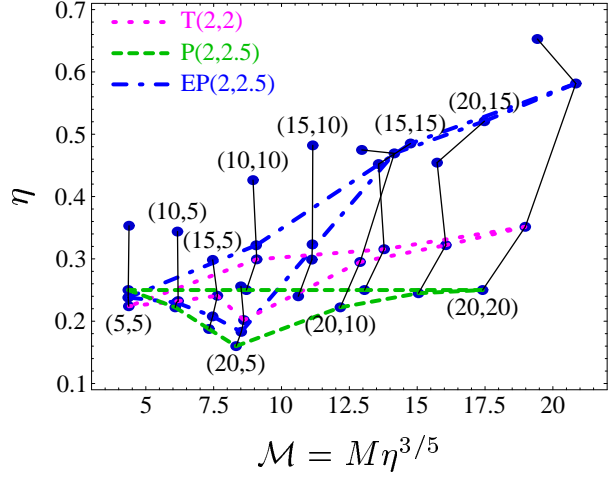}
\end{center}
\caption{Projection of 2PN waveforms onto the P$(2,2.5)$ effective
template space.  Dots are shown for the same BBH masses of
Tab.\ \ref{tableETB1}, and for PN
models T$(2,2.5)$, P$(2,2.5)$, ET$(2,2.5)$, and EP$(2,2.5)$. The thin
solid lines show the \emph{BH mass lines} (introduced in
Sec.~\ref{subsec6.3}), while the dashed and dotted lines show the
contours of the projections of selected PN models.
\label{fig:padeprojecta}}
\end{figure*}

\begin{figure*}
\begin{center}
\includegraphics[width=\sizeonefig]{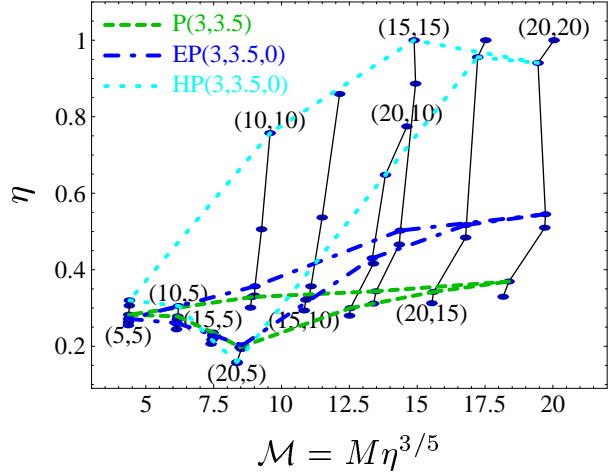}
\end{center}
\caption{Projection of 3PN waveforms onto the P$(2,2.5)$ effective
template space.  Dots are shown for the same BBH masses of
Tab.\ \ref{tableETB1}, and for PN
models T$(3,3.5,+2)$, P$(3,3.5,+2)$, ET$(3,3.5,+2)$, EP$(3,3.5,+2)$,
HT$(3,3.5,+2)$, and HP$(3,3.5,0)$. The dots for $\hat{\theta} = -2$
are only slightly displaced, and they are not shown.  The thin solid
lines show the \emph{BH mass lines} (introduced in Sec.~\ref{subsec6.3}),
while the dashed and dotted lines show the contours of the projections of
selected PN models.
\label{fig:padeprojectb}}
\end{figure*}

\begin{figure*}
\begin{center}
\includegraphics[width=\sizeonefig]{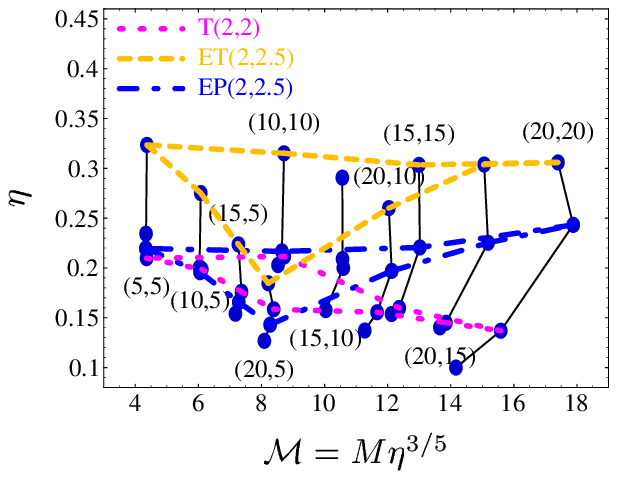}
\end{center}
\caption{Projection of 2PN waveforms onto the EP$(3,3.5)$ effective
template space. This projection includes the effect of the frequency
cut.  Dots are shown for the same BBH masses of
Tab.\ \ref{tableETB1}, and for PN
models T$(2,2.5)$, P$(2,2.5)$, ET$(2,2.5)$, and EP$(2,2.5)$. The thin
solid lines show the \emph{BH mass lines} (introduced in
Sec.~\ref{subsec6.3}), while the dashed and dotted lines show the
contours of the projections of selected PN models.
\label{fig:eobprojecta}}
\end{figure*}

\begin{figure*}
\begin{center}
\includegraphics[width=\sizeonefig]{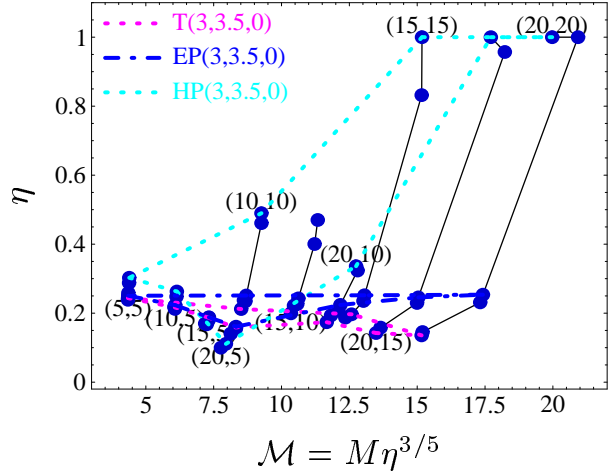}
\end{center}
\caption{Projection of 3PN waveforms onto the EP$(3,3.5)$ effective
template space. This projection includes the effect of the frequency
cut.  Dots are shown for the same BBH masses of
Tab.\ \ref{tableETB1}, and for PN
models T$(3,3.5,+2)$, P$(3,3.5,+2)$, ET$(3,3.5,+2)$, EP$(3,3.5,+2)$,
HT$(3,3.5,+2)$, and HP$(3,3.5,+2)$. The dots for $\hat{\theta} = -2$
are only slightly displaced, and they are not shown.  The thin solid
lines show the \emph{BH mass lines} (introduced in Sec.~\ref{subsec6.3}),
while the dashed and dotted lines show the contours of the projections of
selected PN models.
\label{fig:eobprojectb}}
\end{figure*}

When the models are extended to $0 < \eta \leq 1$, they appear to lie
much closer to each other in signal space. In particular, the
P$(2,2.5)$ and EP$(3,3.5,0)$ models are able to match all the other
models, with minmax $\mathrm{FF} > 0.95$, for almost all the masses in
our range, and in any case with much improved $\mathrm{FF}$ for most
masses; see Tables~\ref{tableXV} and \ref{tableXVI}. Apparently, part
of the effect of the different resummation and approximation schemes
is just to modulate the strength of the PN effects in a way that can
be simulated by changing $\eta$ to nonphysical values in any one
model. This fact can be appreciated by looking at
Figs.~\ref{fig:padeprojecta}, \ref{fig:padeprojectb} and \ref{fig:eobprojecta}, \ref{fig:eobprojectb}, which show the
projection of several models onto the P$(2,2.5)$ and EP$(3,3.5,0)$
effective template spaces, respectively. For instance, in comparison
with T$(2,2.5)$, the model P$(2,2.5)$ seems to underestimate
systematically the effect of $\eta$, so a satisfactory FF for $\eta_T =
0.25$ can be obtained only if we let $\eta_P > 0.25$ (quite
consistently, in the comparison of Tables~\ref{VIa}, \ref{VIb}, where $\eta$ was
confined to its physical range, T$(2, 2.5)$ could match P$(2,2.5)$
effectively, but the reverse was not true).

The other (and perhaps crucial) effect of raising $\eta$ is to change
the location of the MECO for the P-approximant model (or the light
ring, for the EP model), where orbital evolution ends. (Remember that
one of the differences between the Pad\'e and the EOB models is that
the latter includes a plunge part between the ISCO and the light
ring.)  More specifically, for P$(2,2.5)$ [EP$(3,3.5,0)$] the position
of the MECO [light ring] is pushed to smaller radii as $\eta$ is
increased. This effect can increase the FF for target models that have
very different ending frequencies from those of P$(2,2.5)$ and
EP$(3,3.5)$ at comparable $\eta$'s.

Because for the EP model the frequency at the light ring is already
quite high, we cannot simply operate on $\eta$ to improve the match
between the EP model and other models that end at much lower
frequencies [see the values of minmax matches in
Table~\ref{tableXVI}]. Thus, we shall enhance the effectualness of EP
by adding an arbitrary \emph{cut} parameter that modifies the radius
$r$ (usually the light-ring radius) at which we stop the integration
of the Hamilton equations \eqref{eq:eobhamone}--\eqref{eq:eobhamfour};
the effect is to modify the final instantaneous GW frequency of the
waveform. This is therefore a \emph{time-domain cut}, as opposed to
the frequency-domain cuts of the frequency-domain effective templates
examined in the previous section.

We can then compute the FF by searching over $f_\mathrm{cut}$ in
addition to $M$ and $\eta$, and we shall correspondingly account for the
required number of distinct $f_\mathrm{cut}$ when we estimate the
number of templates required to give a certain
$\mathrm{MM}_\mathrm{tot}$.  Even so, if we are unsure whether we can
model successfully a given source over a certain range of frequencies
that falls within LIGO range (as it is the case for the heavy BBHs
with MECOs at frequencies $<$ 200 Hz), the correct way to estimate the
optimal $\rho$ (and therefore the expected detection rate) is to
include only the signal power in the frequency range that we know
well.

The best matches shown in Tables~\ref{tableXV} and \ref{tableXVI}, and
in Figs.~\ref{fig:padeprojecta}--\ref{fig:eobprojectb} were obtained
by searching over the target model parameter space with the simplicial
\texttt{amoeba} algorithm~\cite{nrc}. We found (empirically) that it
was expedient to conduct the searches on the parameters $\beta \equiv
M \eta^{2/5}$ and $\eta$ rather than on $M$ and $\eta$. This is because
iso-match surfaces tend to look like thin ellipses clustered around
the best match parameter pair, with principal axes along the $\beta$
and $\eta$ directions. As shown in Table~\ref{tableXV}, the values of
the maxmax and minmax FFs are very close to each other for the
P$(2,2.5)$ model; the same is true for the EP$(3,3.5)$ model (so in
Table~\ref{tableXVI} we do not show both). For EP$(3,3.5)$, the search
over the three parameters $(\beta,\eta,f_\mathrm{cut})$ was performed
as a refinement step after a first search on $(\beta,\eta)$.

We have evaluated the mismatch metric~\cite{O} $g_{ij}$ (see
Sec.~\ref{sec2}) with respect to the parameters $(\beta,\eta)$ for the
models P$(2,2.5)$ and EP$(3,3.5,0)$ (while evaluating $g_{ij}$, the EP
waveforms were not cut). The metric components at the point
$(\beta_0,\eta_0)$ were obtained by first determining the ranges
$(\beta_\mathrm{min},\beta_\mathrm{max})$,
$(\eta_\mathrm{min},\eta_\mathrm{max})$ for which
\begin{eqnarray}
\braket{u(\beta_0,\eta_0)}{u(\beta_\mathrm{min},\eta_0)} &=&
\braket{u(\beta_0,\eta_0)}{u(\beta_\mathrm{max},\eta_0)} = 1 - 0.05 \\
\braket{u(\beta_0,\eta_0)}{u(\beta_0,\eta_\mathrm{min})} &=&
\braket{u(\beta_0,\eta_0)}{u(\beta_0,\eta_\mathrm{max})} = 1 - 0.05;
\end{eqnarray}
then a quadratic form was least-squares--fit to 16 values of the match
along the ellipse $\Gamma_1$ with axes given by
$(\beta_\mathrm{min},\beta_\mathrm{max})$ and
$(\eta_\mathrm{min},\eta_\mathrm{max})$. The first quadratic form was
used only to determine the principal axes of two further ellipses
$\Gamma_2$ and $\Gamma_3$, at projected matches of $1 - 0.025$ and $1
- 0.0125$. Another quadratic form (giving the final result for the
metric) was then fit at the same time to 16 points along $\Gamma_2$
and to 16 points along $\Gamma_3$, but the two ellipses were given
different fitting weights to cancel the quartic correction terms in
the Taylor expansion of the match around $(\beta_0,\eta_0)$ [the cubic
terms were canceled automatically by taking symmetric points along the
ellipses]. The rms error of the fit was in all cases very good,
establishing that the quadratic approximation held in the close
vicinity (matches $\sim 0.95$) of each point.

We estimate that the numerical error $\sim 20 \%$ is in any case less 
than the error associated with using Eq.~(\ref{eq:metric}) to evaluate 
the required number of templates, instead of laying down a lattice 
of templates more accurately.

\begin{figure}
\begin{center}
\includegraphics[width=\textwidth]{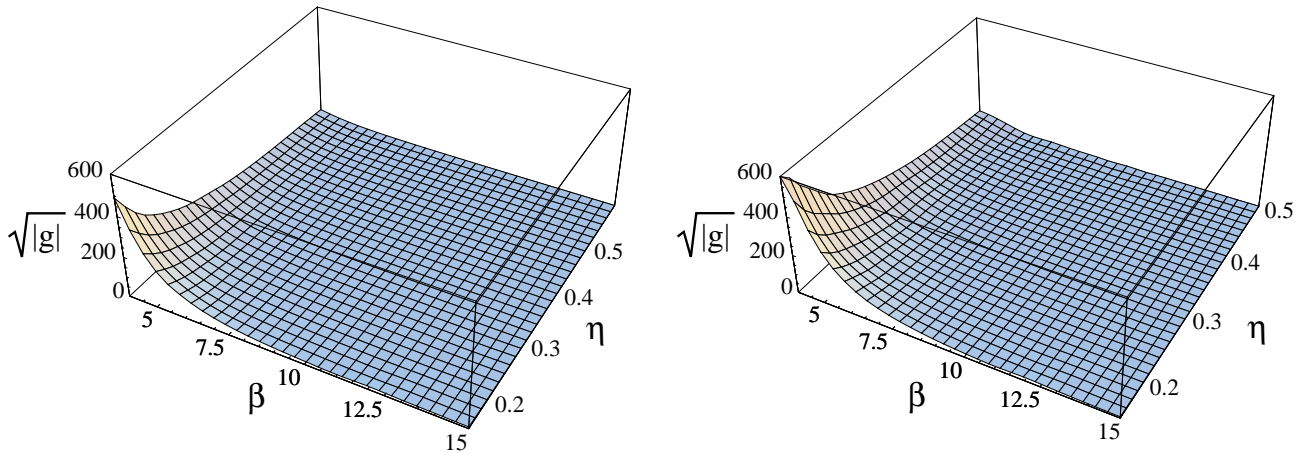}
\end{center}
\caption{Determinant of the mismatch metric for the P$(2,2.5)$
models [left panel], and for the EP$(3,3.5,0)$ models [right
panel]. The determinant $\sqrt{|g|}$ is shown as a function of $\eta$
and $\beta = M\eta^{2/5}$.\label{fig:pademetric}}
\end{figure}

The resulting $\sqrt{|g|}$ for P$(2,2.5)$ and EP$(3,3.5,0)$ is shown
in Fig.~\ref{fig:pademetric}.  It is evident that most of the mismatch
volume is concentrated near the smallest $\beta$'s and $\eta$'s in
parameter space. This is encouraging, because it means that the
extension of the effective template family to high masses and high
$\eta$'s (necessary, as we have seen, to match several target models
with very high FF) will be relatively cheap with respect to the size
of the template bank (this picture, however, changes when we
introduce frequency-domain cuts for the EP models).  
With the $\sqrt{|g|}$'s we then computed the number of
P and EP templates necessary to cover the parameter ranges
$\beta:(4,24)$, $\eta:(0.15,1.00)$, and $\beta:(4,24)$, $\eta:(0.1,1.00)$
which span comfortably all the  
projected images of the target spaces onto the P and EP template
spaces, respectively. [Note the ranges 
include also BBHs where one of the BH has a mass less than 
$5 M_\odot$.] We obtained
\begin{equation}
\mathcal{N}_P \simeq 3260 \left( \frac{0.02}{1-\mathrm{MM}} \right), \quad \quad
\mathcal{N}_E \simeq 6700 \left( \frac{0.02}{1-\mathrm{MM}} \right),
\end{equation}
where MM is the required minimum match (analog to the parameter
$\mathrm{MM}_\psi$ of the preceding section).  By comparison, these numbers are
reduced to respectively $1230$ and $3415$ if we restrict $\eta$ to the
physical range.

The number $\mathcal{N}_E$ does not include the effect of multiple
ending frequencies (cuts). We estimate the number of distinct
$f_\mathrm{cut}$ needed for each $\beta$ by an argument similar to the
one used for the Fourier-domain effective templates (see Sec.\
\ref{sec6}); it turns out that more cuts are required for higher
masses. The resulting number of templates is $\mathcal{N}_{Ec} \simeq
51,000$ for $\mathrm{MM} = 0.98$, which is comparable to the result
for the effective Fourier-domain templates.

If we assume that the distance between the time-domain templates and
the target models is representative of the distance to the true
physical signal, we can guess that $\mathrm{FF} \gtrsim 0.95$ for P
and $\mathrm{FF} \gtrsim 0.97$ for EP with cuts. Under these
hypotheses, 6,500 P templates can buy us a (worst-case)
$\mathrm{MM}_\mathrm{T} \simeq 0.94$, corresponding to a loss in event
rate of $\sim 17\%$. For 51,000 EP templates, we get
$\mathrm{MM}_\mathrm{T} \simeq 0.95$, corresponding to a loss in event
rate of $\sim 14\%$.

Before ending this section we would like to point out another 
time-domain detection-template family which can be consider 
kindred of the Fourier-domain detection-template family 
introduced in Sec.~\ref{sec6}, see Eq.~(\ref{fourier}). 
We can use, for example, the following expression suggested 
by PN calculations [see, e.g., Ref.~\cite{BIWW}]
\beq
\label{time}
h_{\rm eff}(t) = {\cal A}^{\rm T}_{\rm eff}(t)\,e^{i \psi^{\rm T}_{\rm
eff}(t)}\,, \eeq where \beq {\cal A}^{\rm T}_{\rm eff}(t) =
(t_c-t)^{7/16}\,\left [1 - \alpha^{\rm T}\,(t_c-t)^{-1/4}\right ]\,
\theta(t_{\rm cut}-t)\,, \eeq 
\bea {\psi}_{\rm eff}(t) &=& \phi_c +
(t_c-t)^{5/8}\,\left [ \psi^{\rm T}_0 + \psi^{\rm
T}_{1/2}\,(t_c-t)^{-1/8}+ \psi^{\rm T}_{1}\,(t_c-t)^{-1/4} \right
. \nonumber \\ && \left. + \psi^{\rm T}_{3/2}\,(t_c-t)^{-3/8} +
\psi^{\rm T}_{2}\,(t_c-t)^{-1/2} + \cdots \right ]\,,
\label{phasingT}
\eea 
where $ \phi_c, t_c, \alpha^{\rm T}, \psi^{\rm T}_0, \psi^{\rm
T}_1, \psi^{\rm T}_{3/2}$ and $\psi^{\rm T}_2 $ are arbitrary
parameters whose range of values are determined 
maximizing the matches with the target two-body models.

\section{Summary}
\label{sec8}

This paper deals with the problem of detecting GWs from the most
promising sources for ground-based GW interferometers: comparable-mass
BBHs with total mass $M=10\mbox{--}40 M_\odot$ moving on quasicircular
orbits. The detection of these sources poses a delicate problem, 
because their transition from the adiabatic phase to the plunge, 
at least in the nonspinning case, is expected to occur in the 
LIGO and VIRGO frequency bands. Of course, the true GW signals 
from these inspirals should be obtained from exact
solutions of the Einstein equations for two bodies of comparable
mass. However, the theoretical templates used to search for these
signals will be, at best, finite-order approximations to the exact
solutions, usually derived in the PN formalism. Because the
perturbative PN approach begins to fail during the final stages of the
inspiral, when strong curvature and nonlinear effects can no longer be
neglected, various PN resummation methods have been introduced~\cite{BD1,BD2,EOB3PN} 
to improve the convergence of the PN series.

In the first part of this paper [see Sec.~\ref{sec3}, ~\ref{sec4} and
\ref{sec5}] we studied and compared in detail all the PN models of the
relativistic two-body dynamics currently available, including PN
Taylor-expanded and resummed models both in the adiabatic
approximation and in the nonadiabatic case. We noticed the following
features [see Tables~\ref{VIa}, \ref{VIb}]. 
At least for PN orders $n \leq 2.5$, the \emph{target} models T, P, and E 
have low cross matches, if the 2.5PN Taylor flux is used. 
For example, for almost all the masses in our range, we found $\mathrm{maxmax} \, \mathrm{FF} \leq
0.9$; the matches were much better only for P against E (and
viceversa). However, if the 2PN Taylor flux is used the overlaps are rather high. 
At 3PN order we found much higher matches between
T, P, and E, and also with the nonadiabatic model H, at least for 
masses $M \leq 30M_\odot$, and restricting to $\tilde{z}_1=0=\tilde{z}_2$.
These results make sense because at 3PN order the various approximations to the binding energy
and the flux seem to be much closer to each other than at lower
orders. This ``closeness'' of the different analytical approaches,
which at 3PN order are also much closer to some examples of numerical
quasiequilibrium BBH models \cite{GGB}, was recently pointed out in
Refs.~\cite{LB,DGG}. On the other hand, the extraction of BBH
parameters from a true measured signal, if done using the 3PN models,
would still give a range of rather different estimates. However, we want to point 
out that for quite high masses, e.g., $M =40 M_\odot$, the 3PN models 
can have again lower overlaps, also from the point of view of detection.

In addition, by studying the frequency-domain amplitude of the GW
signals that end inside the LIGO frequency band [see
Figs.~\ref{AmplT}, \ref{AmplP}, \ref{AmplE}, \ref{AmplH}], we
understood that if high matches are required it is crucial to
reproduce their deviations from the Newtonian amplitude evolution,
$f^{-7/6}$ (on the contrary, the Newtonian formula seems relatively
adequate to model the PN amplitude for GW frequencies below the
instantaneous GW frequency at the endpoint of orbital evolution).

Finally, the introduction of the HT, HP and L models in Secs.~\ref{subsec4.1}, 
\ref{subsec4.2} provided another
example of two-body nonadiabatic dynamics, quite different from the E
models. In the H models, the conservative dynamics does not have an
ISCO [see the discussion below Eq.\ \eqref{rel}] at 2PN and 3PN orders. As a consequence, 
the transition to the plunge is due to secular radiation-reaction 
effects, and it is pushed to much higher frequencies. This means that, for the H models,
the GW signals for BBHs of total mass $M=10\mbox{--}40 M_\odot$ end
outside the LIGO frequency band, and the frequency-domain amplitude
does not deviate much from the Newtonian result, at least until very
high frequencies [see Fig.~\ref{AmplH}]. The L models do not provide 
the waveforms during the late inspiral and plunge. 
This is due to the fact that because of the appearance of unphysical effects, e.g., 
the binding energy starts to increase with time instead of continuing 
decreasing, we are obliged to stop the evolution before the two BHs 
enter the last stages of inspiral.
It is important to point out that differently from the nonadiabatic E models, the 
nonadiabatic H and L models give rather different predictions when 
used at various PN orders. So, from these point of view they are less 
reliable and robust than the E models. 

In the second part of this paper [Secs.~\ref{sec6}, ~\ref{sec7}] we
pursued the following strategy. We assumed that the target models
spanned a region in signal space that (almost) included the true GW
signal. We were then able to provide a few detection template families
(either chosen among the time-domain target models, or built directly
from polynomial amplitude and phasings in the frequency domain) that
approximate quite well all the targets [$\mathrm{FF} \geq 0.95$ for
almost all the masses in our range, with much better $\mathrm{FFs}$
for most masses]. We speculate that the effectualness of the detection
model in approximating the targets is indicative of its effectualness
in approximating the true signals.

The Fourier-domain detection template family, discussed in
Sec.~\ref{sec6}, is simple and versatile. It uses a PN polynomial
structure for the frequency-domain amplitude and phasing, but it does
not constrain the coefficients to the PN functional dependencies on
the physical parameters. In this sense this bank follows the basic
idea that underlies the Fast Chirp Transform~\cite{JP}.  However,
because for the masses that we consider the GW signal can end within
the LIGO frequency band, we were forced to modify the Newtonian-order
formula for the amplitude, introducing a cutoff frequency and a
parameter to modify the shape of the amplitude curve (the parameter
$\alpha$). As discussed at the end of  Sec.~\ref{subsec6.7} 
the good performance of the two and three-dimensional families confirms 
also results obtained in Refs.~\cite{DIS3}, \cite{CA} and \cite{EPT}. 

We showed that our Fourier-domain detection template space has a FF
higher than $0.97$ for the T, P and E models, and $\gtrsim 0.96$ for
most of the 3PN HT and HP models; we then speculate that it will match true
BBH waveforms with FF $\sim 0.96$. We have computed the number of
templates required to give MM $\simeq 0.96$ (about $10^4$). The total
$\mathrm{MM}_\mathrm{T}$ should be larger than ${\rm FF} \cdot {\rm
MM} \sim 0.92$, which corresponds to a loss of event rate of $1-{\rm
MM}_\mathrm{T}^3 \approx 22\%$. This performance could be improved at
the price of introducing a larger number of templates, with the rough
scaling law of ${\cal N}=10^4 [0.04 /(0.96-{\rm MM})]^2$.

In Sec.~\ref{subsec6.6} we investigated where the less reliable 
2PN H and L models, and the E models at 3PN order further 
expanded considering $\tilde{z}_1 \neq 0$, lie in the detection
template space. 
The Fourier-domain template family has FF in the range [0.85,0.95] 
with the 2PN H models, and FF mostly higher than 0.95, but with several 
exceptions which can be as low as 0.93 with the L models. 
The E models with $\tilde{z}_1 \neq 0$ are matched by the detection template family with FF almost 
always higher than 0.95. 
The E models with $\tilde{z}_1 \neq 0$ and the L models are (almost) covered by the 
region delimiting the adiabatic models and the E models with $\tilde{z}_1 = 0$. 
However, these models require lower cutoff frequencies, which
will increase the number of templates up to a factor of $3$.
The 2PN H models sit outside this region and if we want to include 
them the number of templates should be doubled. 

The time-domain detection template families, discussed in
Sec.~\ref{sec7}, followed a slightly different philosophy. The idea in
this case was to provide a template bank that, for some choices of the
parameters, could coincide with one of the approximate two-body
models. Quite interestingly, this can be achieved by relaxing the
physical hypothesis that $0 \leq \eta \leq 0.25$. However, the good
performances of these banks are less systematic, and harder to
generalize than the performance of the Fourier-domain effective bank. 
As suggested at the end of Sec.~\ref{sec7} [see Eq.~(\ref{time})], the time-domain
bank could be improved by using a parametrization of the time-domain
amplitude and phase similar to the one used for the Fourier-domain templates. 
The detection template families based on the extension of the P$(2,2.5)$ and
EP$(3,3.5)$ to nonphysical values of $\eta$ were shown to have FF
respectively $\gtrsim 0.95$ and $\gtrsim 0.97$ for all the PN target
models, and considerably higher for most models and masses. We have
computed the number of P templates needed to obtain a $\mathrm{MM} =
0.99$ (about 6,500) and of EP templates to obtain a $\mathrm{MM}
= 0.98$ (about 51,000).  The expected total $\mathrm{MM}_\mathrm{T}$ is
then respectively $\gtrsim 0.94$ and $\gtrsim 0.95$, corresponding to losses
in event rate of $\lesssim 17$\% and $\lesssim 14$\%. The MMs scale roughly as
$[0.01/(1-\mathrm{MM})]$ for P and $[0.02/(1-\mathrm{MM})]^2$ for EP
(because of the additional frequency-cut parameter).

We notice that the number of templates that we estimate for the
Fourier- and time-domain detection template families is higher than
the number of templates we would obtain using only one PN model.
However, the number of \emph{independent shapes} that enters the
expression for the $\rho_*$ threshold [see Eq.~\ref{eq:falses}] does
not coincide with the number of templates that are laid down within a
discrete template bank to achieve a given MM; indeed, if MM is close
to one, these are almost guaranteed to be to yield S/N statistics that
are strongly correlated.  A rough estimate of the number of
independent shapes can be obtained taking a coarse-grained grid in
template space. For example by setting MM=0 in Eq.~(\ref{eq:metric}),
the number of independent shapes would be given roughly by the volume
of the template space.  As explained at the end of
Sec.~\ref{subsec2.2}, if we wish to keep the same false-alarm
probability, we have to increase the threshold by $\sim 3\%$ if we
increase the number of independent shapes by one order of
magnitude. This effect will cause a further loss in event rates
\cite{BS}.

Finally, in Sec.~\ref{subsec6.7} we extended the detection template family 
in the Fourier domain by requiring that it embeds the targets 
in a signal space of higher dimension (with more parameters). 
We investigated the three dimensional case and we found, as expected, 
the maxmax matches increase. In particular, the match  of the phasings 
are nearly perfect: always higher than $0.994$ for the two-body 
models which are farthest apart in the detection template space. 
Moreover, by projecting the points in the three-dimensional space back 
to the two-dimensional space, we can get nearly the same projections 
we would have got from matching directly the PN waveforms with 
the two-parameter--phasing model.
The analysis done in Sec.~\ref{subsec6.7} could suggest ways of systematically
expand the Fourier-domain templates.
Trying to guess the functional directions in which the true signals might 
lie with respect to the targets was the most delicate challenge of 
our investigation. However, our suggestions are
not guaranteed to produce templates that will capture the true signal,
and they should be considered as indications. 
When numerical relativity provides the first good examples of waveforms emitted in
the last stages of the binary inspiral and plunge, it will be very interesting 
to investigate whether the matches with our detection template families 
are high and in which region of the detection template space do they sit.

\acknowledgments We wish to thank Kashif Alvi, Luc Blanchet, David Chernoff, Teviet Creighton, Thibault Damour, Scott Hughes, Albert Lazzarini, Bangalore Sathyaprakash, Kip Thorne, Massimo Tinto and Andrea Vicer\`e for very useful discussions and interactions. We also thank Thibault Damour, Bangalore Sathyaprakash, and especially Kip Thorne for a very careful reading of this manuscript and for stimulating comments; we thank Luc Blanchet for useful discussions on the relation between MECO and ISCO analyzed in Sec.~\ref{subsec4.1} and David Chernoff for having shared with us its code for the computation of the L model. We acknowledge support from NSF grant PHY-0099568. For A.\ B., this research was also supported by Caltech's Richard Chase Tolman fund.

\newcolumntype{d}{D{.}{.}{1.3}}
\newcolumntype{e}{D{.}{.}{2.2}}
\newcolumntype{f}{D{.}{.}{1.2}}

\begingroup
\squeezetable
\begin{table*}
\begin{tabular}{lr|def|def|def|def|def}
\input tableVIa.tex
\end{tabular}
\caption[]{(Continued into Table \ref{VIb}.) Fitting factors between
several PN models, at 2PN and 3PN orders. For three choices of BBH
masses, this table shows the maxmax matches [see Eq.\
\eqref{eq:maxmaxd}] between the \emph{search} models at the top of the
columns and the \emph{target} models at the left end of the rows,
\emph{maximized over the intrinsic parameters of the search models in
the columns}. For each intersection, the three numbers mm, $M = m_1 +
m_2$ and $\eta = m_1 m_2 / M^2$ denote the maximized match and the
search-model mass parameters at which the maximum is attained. In
computing these matches, the parameter $\eta$ of the search models was
restricted to its physical range $0 < \eta \leq 1/4$.  The arbitrary
flux parameter $\hat{\theta}$ was always set equal to zero.

These matches represent the fitting factors [see Eq.\ \eqref{eq:ffd}]
for the projection of the target models onto the search models. The
reader will notice that the values shown are not symmetric across the
diagonal: for instance, the match for the search model T$(2,2.5)$
against the target model P$(2,2.5)$ is higher than the converse. This
is because the matches represent the inner product
\eqref{eq:innerproduct} between two different pairs of model
parameters: in the first case, the target parameters $(m_1 =
15M_\odot, m_2 = 15M_\odot)_P \equiv (M = 30M_\odot, \eta = 0.25)_P$
are mapped to the maximum-match search parameters $(M = 39.7 M_\odot,
\eta=0.24)_T$; in the second case, the target parameters $(m_1 =
15M_\odot, m_2 = 15M_\odot)_T \equiv (M = 30M_\odot, \eta = 0.25)_T$
are mapped to the maximum-match parameters $(M = 25.37M_\odot, \eta =
0.24)_P$ [so the symmetry of the inner product
\eqref{eq:innerproduct} is reflected by the fact that the search
parameters $(M = 25.3M_\odot, \eta = 0.24)_P$ are mapped into the
target parameters $(M = 30M_\odot, \eta = 0.25)_T$]. \label{VIa}}
\end{table*}
\endgroup

\begingroup
\squeezetable
\begin{table*}
\begin{tabular}{lr|def|def|def|def|def|def}
\input tableVIb.tex
\end{tabular}
\caption[]{(Continued from Table \ref{VIa}.)  Fitting factors between
several PN models, at 2PN and 3PN orders. Please see the caption to
Table\ \ref{VIa}.\label{VIb}}
\end{table*}
\endgroup

\begingroup
\squeezetable
\begin {table*} 
\begin {tabular}{lr|rrr|rrr|rrr|rrr|rrr|rrr}
\input tableVIII.tex
\end{tabular}
\caption{Fitting factors between T and ET models, at 2PN and 3PN
orders, and for different choices of the arbitrary flux parameter
$\hat{\theta}$. For three choices of BBH masses, this table shows the
maxmax matches [see Eq.\ \eqref{eq:maxmaxd}] between the \emph{search}
models at the top of the columns and the \emph{target} models at the
left end of the rows, \emph{maximized over the mass parameters of the
models in the columns}. For each intersection, the three numbers mm,
$M$ and $\eta$ denote the maximized match and the search-model mass
parameters at which the maximum is attained. The matches can be
interpreted as the fitting factors for the projection of the target
models onto the search models. See the caption to Table 
\protect\ref{VIb} for further details.
\label{VIII}} 
\end{table*}
\endgroup

\clearpage

\begingroup
\squeezetable
\begin {table*} 
\begin {tabular}{lr|ccc|ccc|ccc|ccc|ccc}
\input tableIX.tex
\end{tabular}
\caption{Fitting factors for the projection of EP$(3,3.5,0)$ templates
onto themselves, for various choices of the parameters ${z}_1$ and
${z}_2$. The values quoted are obtained by maximizing the maxmax (mm)
match over the mass parameters of the (search) models in the columns,
while keeping the mass parameters of the (target) models in the rows
fixed to their quoted values, $(15+15)M_\odot$, $(15+5)M_\odot$
$(5+5)M_\odot$.  The three numbers shown at each intersection are the
maximized match and the search parameters at which the maximum was
attained.  In labeling rows and columns we use the notation
EP$(3, 3.5, \hat{\theta}, {z}_1, {z}_2)$. 
See the caption to Table \protect\ref{VIb} for further details.\label{IX}}
\end{table*}
\endgroup

\setlength{\LTcapwidth}{\textwidth}

\begingroup
\squeezetable 
\begin{longtable*}{l|r|rrrrrr||r|rrrrrr}
& \multicolumn{14}{c}{FF for projection onto the Fourier-domain detection template families}
\\
PN model 
&  & 
\multicolumn{1}{c}{$f_{\rm end}$} & \multicolumn{1}{c}{mn} & 
\multicolumn{1}{c}{$\psi_0$} & \multicolumn{1}{c}{$\psi_{3/2}$} &
\multicolumn{1}{c}{$\alpha f_{\rm cut}^{2/3}$} &
\multicolumn{1}{c||}{$f_{\rm cut}$}   
& &
\multicolumn{1}{c}{$f_{\rm end}$} & 
\multicolumn{1}{c}{mn} & \multicolumn{1}{c}{$\psi_0$} & \multicolumn{1}{c}{$\psi_{3/2}$} &
\multicolumn{1}{c}{$\alpha f_{\rm cut}^{2/3}$} & \multicolumn{1}{c}{$f_{\rm cut}$} \\
\hline \hline
\endhead
\input tableETB1.tex
\end{longtable*}
\endgroup


\setlength{\LTcapwidth}{\textwidth}

\newcolumntype{d}{D{.}{.}{1.3}}
\newcolumntype{e}{D{.}{.}{2.2}}
\newcolumntype{f}{D{.}{.}{1.2}}

\begingroup
\squeezetable 
\begin{longtable*}{l|r|defdef||r|defdef}
& \multicolumn{14}{c}{FF for projection onto P$(2, 2.5)$, for $0 < \eta < 1$} \\
& &
\multicolumn{1}{c}{mm} & \multicolumn{1}{c}{$M$} & \multicolumn{1}{c}{$\eta$} &
\multicolumn{1}{c}{mn} & \multicolumn{1}{c}{$M$} & \multicolumn{1}{c||}{$\eta$} &
&
\multicolumn{1}{c}{mm} & \multicolumn{1}{c}{$M$} & \multicolumn{1}{c}{$\eta$}
& \multicolumn{1}{c}{mn} & \multicolumn{1}{c}{$M$} & \multicolumn{1}{c}{$\eta$}
\\
\hline \hline
\endhead
\input tableXV.tex
\end{longtable*}
\endgroup

\newcolumntype{d}{D{.}{.}{1.3}}
\newcolumntype{e}{D{.}{.}{2.2}}
\newcolumntype{f}{D{.}{.}{1.2}}
\newcolumntype{g}{D{.}{.}{3.1}}

\begingroup
\squeezetable
\begin{longtable*}{l|r|defdefg||r|defdefg}
& \multicolumn{16}{c}{FF for projection onto EP$(3, 3.5, 0)$, for $0 < \eta < 1$} \\
& &
\multicolumn{1}{c}{mm} & \multicolumn{1}{c}{$M$} & \multicolumn{1}{c}{$\eta$} &
\multicolumn{1}{c}{mmc} & \multicolumn{1}{c}{$M$} & \multicolumn{1}{c}{$\eta$} &
\multicolumn{1}{c||}{$f_{\rm cut}$}
& &
\multicolumn{1}{c}{mm} & \multicolumn{1}{c}{$M$} & \multicolumn{1}{c}{$\eta$} &
\multicolumn{1}{c}{mmc} & \multicolumn{1}{c}{$M$} & \multicolumn{1}{c}{$\eta$} &
\multicolumn{1}{c}{$f_{\rm cut}$} \\
\hline \hline
\endhead
\input tableXVI.tex
\end{longtable*}
\endgroup

\renewcommand{\prd}{\emph{Phys. Rev. D}}

\end{document}

%% file: tableL.tex
 & & \multicolumn{3}{c}{T(2, 2)}& \multicolumn{3}{c}{T(3, 3.5, 0)}&
 \multicolumn{3}{c}{P(2, 2.5)}& \multicolumn{3}{c}{P(3, 3.5, 0)}&
 \multicolumn{3}{c}{EP(2, 2.5)}& \multicolumn{3}{c}{EP(3, 3.5, 0)}&
 \multicolumn{3}{c}{HT(3, 3.5, 0)}\\ & & \multicolumn{1}{c}{mm} &
 \multicolumn{1}{c}{$M$} & \multicolumn{1}{c}{$\eta$}&
 \multicolumn{1}{c}{mm} & \multicolumn{1}{c}{$M$} &
 \multicolumn{1}{c}{$\eta$}& \multicolumn{1}{c}{mm} &
 \multicolumn{1}{c}{$M$} & \multicolumn{1}{c}{$\eta$}&
 \multicolumn{1}{c}{mm} & \multicolumn{1}{c}{$M$} &
 \multicolumn{1}{c}{$\eta$}& \multicolumn{1}{c}{mm} &
 \multicolumn{1}{c}{$M$} & \multicolumn{1}{c}{$\eta$}&
 \multicolumn{1}{c}{mm} & \multicolumn{1}{c}{$M$} &
 \multicolumn{1}{c}{$\eta$}& \multicolumn{1}{c}{mm} &
 \multicolumn{1}{c}{$M$} & \multicolumn{1}{c}{$\eta$}\\
\hline \hline 
L(2, 1) & (20+20)$M_\odot$ & 0.994 & 78.83 & 0.05 & 0.998 & 61.24 &
 0.09 & 0.999 & 52.76 & 0.13 & 0.998 & 57.96 & 0.11 & 0.935 & 70.76 &
 0.05 & 0.9446 & 72.04 & 0.06 & 0.994 & 49.53 & 0.14 \\ &
 (15+15)$M_\odot$ & 0.991 & 55.16 & 0.06 & 0.995 & 44.50 & 0.10 &
 0.999 & 39.96 & 0.13 & 0.998 & 43.57 & 0.11 & 0.912 & 46.67 & 0.09 &
 0.916 & 50.90 & 0.07 & 0.994 & 37.08 & 0.15 \\ & (15+5)$M_\odot$ &
 0.981 & 35.51 & 0.05 & 0.991 & 29.03 & 0.08 & 0.995 & 26.02 & 0.10 &
 0.994 & 27.99 & 0.09 & 0.942 & 27.46 & 0.09 & 0.941 & 28.85 & 0.08 &
 0.994 & 22.89 & 0.13 \\ & (5+5)$M_\odot$ & 0.956 & 10.68 & 0.20 &
 0.965 & 11.49 & 0.18 & 0.971 & 11.33 & 0.19 & 0.964 & 11.89 & 0.17 &
 0.964 & 11.03 & 0.19 & 0.960 & 11.69 & 0.17 & 0.966 & 11.32 & 0.18 \\
\hline 

%% file: tableTL.tex
 & & \multicolumn{3}{c}{L$(2,0)$} & \multicolumn{3}{c}{T$(2,0)$} &
 \multicolumn{3}{c}{L$(2,1)$} & \multicolumn{3}{c}{T$(2,1)$} \\ & &
 \multicolumn{1}{c}{mm} & \multicolumn{1}{c}{$M$} &
 \multicolumn{1}{c}{$\eta$} & \multicolumn{1}{c}{mm} &
 \multicolumn{1}{c}{$M$} & \multicolumn{1}{c}{$\eta$} &
 \multicolumn{1}{c}{mm} & \multicolumn{1}{c}{$M$} &
 \multicolumn{1}{c}{$\eta$} & \multicolumn{1}{c}{mm} &
 \multicolumn{1}{c}{$M$} & \multicolumn{1}{c}{$\eta$} \\
\hline \hline 
& (15+15)$M_{\odot}$ & & & & 0.884 & 42.02 & 0.237 & & & & & & \\
L$(2,0)$& (15+5)$M_{\odot}$ & & & & 0.769 & 24.71 & 0.201 & & & & & &
\\ & (5+5)$M_{\odot}$ & & & & 0.996 & 21.70 & 0.068 & & & & & & \\
\hline & (15+15)$M_{\odot}$ & 0.834 & 23.44 & 0.247 & & & & & & & & &
\\ T$(2,0)$& (15+5)$M_{\odot}$ & 0.823 & 14.90 & 0.250 & & & & & & & &
& \\ & (5+5)$M_{\odot}$ & 0.745 & 9.11 & 0.250 & & & & & & & & & \\
\hline & (15+15)$M_{\odot}$ & & & & & & & & & & 0.837 & 60.52 &
0.236\\ L$(2,1)$& (15+5)$M_{\odot}$ & & & & & & & & & & 0.844 & 55.70
& 0.052\\ & (5+5)$M_{\odot}$ & & & & & & & & & & 0.626 & 11.47 & 0.238
\\ \hline & (15+15)$M_{\odot}$ & & & & & & & 0.663 & 19.38 & 0.250 & &
& \\ T$(2,1)$& (15+5)$M_{\odot}$ & & & & & & & 0.672 & 13.56 & 0.250 &
& & \\ & (5+5)$M_{\odot}$ & & & & & & & 0.631 & 9.22 & 0.243 & & & \\
\hline

%% file: tableVIa.tex
 & & \multicolumn{3}{c}{T(2, 2)}& \multicolumn{3}{c}{T(2, 2.5)}& \multicolumn{3}{c}{T(3, 3.5, 0)}& \multicolumn{3}{c}{P(2, 2.5)}& \multicolumn{3}{c}{P(3, 3.5, 0)}\\ 
 & & \multicolumn{1}{c}{mm} & \multicolumn{1}{c}{$M$} & \multicolumn{1}{c}{$\eta$}& \multicolumn{1}{c}{mm} & \multicolumn{1}{c}{$M$} & \multicolumn{1}{c}{$\eta$}& \multicolumn{1}{c}{mm} & \multicolumn{1}{c}{$M$} & \multicolumn{1}{c}{$\eta$}& \multicolumn{1}{c}{mm} & \multicolumn{1}{c}{$M$} & \multicolumn{1}{c}{$\eta$}& \multicolumn{1}{c}{mm} & \multicolumn{1}{c}{$M$} & \multicolumn{1}{c}{$\eta$}\\ 
\hline \hline 
T(2, 2) & (20+20)$M_\odot$ &  & &  &   0.924 &  54.47 &   0.23  &   0.999 &  40.47 &   0.24  &   0.977 &  39.13 &   0.25  &   0.999 &  41.93 &   0.24  \\ 
 & (15+15)$M_\odot$ &  & &  &   0.873 &  39.46 &   0.24  &   0.999 &  30.35 &   0.24  &   0.980 &  29.69 &   0.25  &   0.998 &  31.54 &   0.23  \\ 
 & (15+5)$M_\odot$ &  & &  &   0.885 &  29.45 &   0.10  &   0.998 &  19.64 &   0.19  &   0.992 &  18.07 &   0.22  &   0.998 &  20.23 &   0.18  \\ 
 & (5+5)$M_\odot$ &  & &  &   0.988 &  21.28 &   0.06  &   0.998 &  10.61 &   0.22  &   0.994 &  10.54 &   0.22  &   0.999 &  11.16 &   0.20  \\ 
\hline 
T(2, 2.5) & (20+20)$M_\odot$ &   0.882 &  31.44 &   0.25  &  & &  &   0.870 &  31.54 &   0.25  &   0.824 &  30.25 &   0.25  &   0.893 &  33.09 &   0.25  \\ 
 & (15+15)$M_\odot$ &   0.845 &  24.85 &   0.25  &  & &  &   0.835 &  25.21 &   0.25  &   0.796 &  25.35 &   0.25  &   0.863 &  26.20 &   0.25  \\ 
 & (15+5)$M_\odot$ &   0.848 &  15.34 &   0.25  &  & &  &   0.865 &  15.74 &   0.25  &   0.870 &  15.85 &   0.25  &   0.894 &  15.90 &   0.25  \\ 
 & (5+5)$M_\odot$ &   0.801 &   9.41 &   0.25  &  & &  &   0.823 &   9.51 &   0.25  &   0.826 &   9.51 &   0.25  &   0.849 &   9.61 &   0.25  \\ 
\hline 
T(3, 3.5, 0) & (20+20)$M_\odot$ &   0.999 &  39.57 &   0.24  &   0.916 &  54.63 &   0.23  &  & &  &   0.989 &  39.03 &   0.24  &   0.997 &  41.56 &   0.23  \\ 
 & (15+15)$M_\odot$ &   0.999 &  29.71 &   0.24  &   0.855 &  39.46 &   0.24  &  & &  &   0.992 &  29.25 &   0.25  &   1.000 &  31.97 &   0.21  \\ 
 & (15+5)$M_\odot$ &   0.999 &  20.98 &   0.16  &   0.877 &  29.20 &   0.10  &  & &  &   0.997 &  18.82 &   0.20  &   1.000 &  20.81 &   0.17  \\ 
 & (5+5)$M_\odot$ &   0.991 &   9.67 &   0.25  &   0.986 &  19.49 &   0.07  &  & &  &   0.998 &   9.90 &   0.24  &   1.000 &  10.57 &   0.22  \\ 
\hline 
P(2, 2.5) & (20+20)$M_\odot$ &   0.970 &  40.47 &   0.24  &   0.879 &  56.77 &   0.23  &   0.991 &  41.80 &   0.22  &  & &  &   0.999 &  46.01 &   0.18  \\ 
 & (15+15)$M_\odot$ &   0.967 &  30.15 &   0.24  &   0.816 &  39.66 &   0.24  &   0.998 &  32.66 &   0.20  &  & &  &   0.999 &  34.02 &   0.19  \\ 
 & (15+5)$M_\odot$ &   0.989 &  23.77 &   0.12  &   0.792 &  20.56 &   0.20  &   0.996 &  21.55 &   0.15  &  & &  &   0.998 &  21.83 &   0.15  \\ 
 & (5+5)$M_\odot$ &   0.989 &   9.67 &   0.25  &   0.882 &  13.04 &   0.15  &   0.998 &  10.08 &   0.24  &  & &  &   0.997 &  10.75 &   0.21  \\ 
\hline 
P(3, 3.5, 0) & (20+20)$M_\odot$ &   0.999 &  38.33 &   0.24  &   0.923 &  51.51 &   0.24  &   0.997 &  38.97 &   0.24  &   0.971 &  37.70 &   0.25  &  & &  \\ 
 & (15+15)$M_\odot$ &   0.997 &  28.47 &   0.25  &   0.979 &  51.01 &   0.10  &   0.997 &  28.96 &   0.25  &   0.961 &  28.88 &   0.25  &  & &  \\ 
 & (15+5)$M_\odot$ &   0.997 &  19.53 &   0.18  &   0.825 &  20.89 &   0.19  &   1.000 &  19.12 &   0.19  &   0.998 &  18.32 &   0.21  &  & &  \\ 
 & (5+5)$M_\odot$ &   0.949 &   9.80 &   0.24  &   0.988 &  17.70 &   0.09  &   0.993 &   9.75 &   0.25  &   0.991 &   9.75 &   0.25  &  & &  \\ 
\hline 
EP(2, 2.5) & (20+20)$M_\odot$ &   0.954 &  38.10 &   0.25  &   0.936 &  51.14 &   0.24  &   0.933 &  39.10 &   0.25  &   0.878 &  38.22 &   0.25  &   0.962 &  39.94 &   0.25  \\ 
 & (15+15)$M_\odot$ &   0.965 &  29.34 &   0.25  &   0.895 &  37.45 &   0.25  &   0.960 &  29.60 &   0.25  &   0.903 &  29.56 &   0.25  &   0.975 &  30.15 &   0.25  \\ 
 & (15+5)$M_\odot$ &   0.988 &  20.79 &   0.16  &   0.769 &  21.97 &   0.19  &   0.983 &  20.22 &   0.18  &   0.969 &  19.54 &   0.19  &   0.980 &  20.85 &   0.17  \\ 
 & (5+5)$M_\odot$ &   0.996 &   9.70 &   0.25  &   0.980 &  20.46 &   0.07  &   0.997 &  10.29 &   0.23  &   0.995 &  10.22 &   0.23  &   0.997 &  10.83 &   0.21  \\ 
\hline 
EP(3, 3.5, 0) & (20+20)$M_\odot$ &   0.946 &  37.11 &   0.25  &   0.949 &  48.90 &   0.24  &   0.930 &  37.84 &   0.25  &   0.867 &  36.72 &   0.25  &   0.954 &  38.80 &   0.24  \\ 
 & (15+15)$M_\odot$ &   0.955 &  28.78 &   0.24  &   0.913 &  35.38 &   0.24  &   0.948 &  28.89 &   0.25  &   0.893 &  28.82 &   0.25  &   0.968 &  29.50 &   0.25  \\ 
 & (15+5)$M_\odot$ &   0.992 &  18.51 &   0.20  &   0.808 &  22.15 &   0.18  &   0.985 &  18.92 &   0.20  &   0.970 &  18.34 &   0.21  &   0.983 &  19.63 &   0.19  \\ 
 & (5+5)$M_\odot$ &   0.968 &   9.65 &   0.25  &   0.985 &  18.41 &   0.08  &   0.994 &   9.76 &   0.25  &   0.992 &   9.77 &   0.25  &   0.998 &  10.16 &   0.23  \\ 
\hline 
HT(2, 2) & (20+20)$M_\odot$ &   0.777 &  21.39 &   0.25  &   0.890 &  27.58 &   0.25  &   0.768 &  21.61 &   0.25  &   0.732 &  21.63 &   0.25  &   0.789 &  22.57 &   0.25  \\ 
 & (15+15)$M_\odot$ &   0.674 &  20.20 &   0.24  &   0.780 &  21.83 &   0.25  &   0.673 &  21.02 &   0.25  &   0.657 &  21.03 &   0.25  &   0.687 &  21.07 &   0.25  \\ 
 & (15+5)$M_\odot$ &   0.616 &  15.88 &   0.20  &   0.666 &  18.84 &   0.18  &   0.625 &  17.37 &   0.18  &   0.645 &  16.10 &   0.22  &   0.631 &  17.14 &   0.18  \\ 
 & (5+5)$M_\odot$ &   0.796 &   9.62 &   0.25  &   0.935 &  10.00 &   0.25  &   0.833 &   9.73 &   0.25  &   0.834 &   9.74 &   0.25  &   0.856 &   9.75 &   0.25  \\ 
\hline 
HT(3, 3.5, 0) & (20+20)$M_\odot$ &   0.812 &  32.35 &   0.25  &   0.925 &  44.91 &   0.24  &   0.795 &  34.76 &   0.25  &   0.737 &  32.98 &   0.25  &   0.812 &  37.10 &   0.24  \\ 
 & (15+15)$M_\odot$ &   0.848 &  27.97 &   0.25  &   0.919 &  33.30 &   0.25  &   0.835 &  28.70 &   0.25  &   0.788 &  28.78 &   0.25  &   0.875 &  29.07 &   0.25  \\ 
 & (15+5)$M_\odot$ &   0.998 &  23.08 &   0.13  &   0.788 &  21.15 &   0.20  &   0.999 &  21.25 &   0.16  &   0.994 &  19.77 &   0.18  &   0.999 &  21.81 &   0.15  \\ 
 & (5+5)$M_\odot$ &   0.952 &   9.65 &   0.25  &   0.828 &  10.36 &   0.24  &   0.984 &   9.76 &   0.25  &   0.984 &   9.77 &   0.25  &   0.992 &   9.99 &   0.24  \\ 
\hline 
HP(2, 2.5) & (20+20)$M_\odot$ &   0.756 &  18.71 &   0.25  &   0.853 &  23.74 &   0.24  &   0.752 &  18.96 &   0.25  &   0.725 &  19.09 &   0.25  &   0.769 &  19.70 &   0.25  \\ 
 & (15+15)$M_\odot$ &   0.631 &  17.87 &   0.24  &   0.714 &  18.06 &   0.25  &   0.634 &  17.86 &   0.25  &   0.630 &  18.46 &   0.25  &   0.642 &  18.53 &   0.25  \\ 
 & (15+5)$M_\odot$ &   0.582 &  14.33 &   0.25  &   0.631 &  16.88 &   0.20  &   0.587 &  14.54 &   0.25  &   0.600 &  16.40 &   0.18  &   0.589 &  17.88 &   0.15  \\ 
 & (5+5)$M_\odot$ &   0.731 &   9.41 &   0.25  &   0.869 &   9.75 &   0.25  &   0.755 &   9.51 &   0.25  &   0.755 &   9.54 &   0.25  &   0.765 &   9.54 &   0.25  \\ 
\hline 
HP(3, 3.5, 0) & (20+20)$M_\odot$ &   0.748 &  32.36 &   0.25  &   0.879 &  42.53 &   0.25  &   0.733 &  32.51 &   0.25  &   0.679 &  30.72 &   0.25  &   0.756 &  34.48 &   0.25  \\ 
 & (15+15)$M_\odot$ &   0.789 &  27.41 &   0.24  &   0.915 &  31.80 &   0.25  &   0.782 &  27.43 &   0.25  &   0.741 &  27.43 &   0.25  &   0.817 &  28.60 &   0.25  \\ 
 & (15+5)$M_\odot$ &   0.998 &  21.75 &   0.15  &   0.792 &  20.41 &   0.21  &   1.000 &  20.57 &   0.17  &   0.995 &  19.29 &   0.19  &   0.999 &  21.17 &   0.16  \\ 
 & (5+5)$M_\odot$ &   0.912 &   9.62 &   0.25  &   0.990 &  16.20 &   0.10  &   0.959 &   9.73 &   0.25  &   0.961 &   9.76 &   0.25  &   0.982 &   9.76 &   0.25  \\ 
\hline 

%% file: tableVIb.tex
 & & \multicolumn{3}{c}{EP(2, 2.5)}& \multicolumn{3}{c}{EP(3, 3.5, 0)}& \multicolumn{3}{c}{HT(2, 2)}& \multicolumn{3}{c}{HT(3, 3.5, 0)}& \multicolumn{3}{c}{HP(2, 2.5)}& \multicolumn{3}{c}{HP(3, 3.5, 0)}\\ 
 & & \multicolumn{1}{c}{mm} & \multicolumn{1}{c}{$M$} & \multicolumn{1}{c}{$\eta$}& \multicolumn{1}{c}{mm} & \multicolumn{1}{c}{$M$} & \multicolumn{1}{c}{$\eta$}& \multicolumn{1}{c}{mm} & \multicolumn{1}{c}{$M$} & \multicolumn{1}{c}{$\eta$}& \multicolumn{1}{c}{mm} & \multicolumn{1}{c}{$M$} & \multicolumn{1}{c}{$\eta$}& \multicolumn{1}{c}{mm} & \multicolumn{1}{c}{$M$} & \multicolumn{1}{c}{$\eta$}& \multicolumn{1}{c}{mm} & \multicolumn{1}{c}{$M$} & \multicolumn{1}{c}{$\eta$}\\ 
\hline \hline 
T(2, 2) & (20+20)$M_\odot$ &   0.953 &  41.67 &   0.24  &   0.952 &  43.00 &   0.24  &   0.951 &  80.34 &   0.24  &   0.855 &  56.69 &   0.24  &   0.965 &  90.12 &   0.24  &   0.859 &  74.80 &   0.25  \\ 
 & (15+15)$M_\odot$ &   0.962 &  30.41 &   0.24  &   0.991 &  35.32 &   0.17  &   0.899 &  58.93 &   0.24  &   0.997 &  33.03 &   0.20  &   0.922 &  67.38 &   0.24  &   0.998 &  33.67 &   0.20  \\ 
 & (15+5)$M_\odot$ &   0.988 &  19.11 &   0.20  &   0.992 &  20.93 &   0.17  &   0.924 &  69.96 &   0.05  &   0.998 &  19.38 &   0.19  &   0.876 &  57.94 &   0.07  &   0.999 &  19.81 &   0.18  \\ 
 & (5+5)$M_\odot$ &   0.997 &  10.33 &   0.23  &   0.998 &  11.09 &   0.20  &   0.788 &   9.93 &   0.25  &   0.998 &  10.92 &   0.21  &   0.727 &  10.19 &   0.25  &   0.999 &  11.19 &   0.20  \\ 
\hline 
T(2, 2.5) & (20+20)$M_\odot$ &   0.908 &  31.37 &   0.25  &   0.929 &  32.98 &   0.25  &   0.959 &  58.39 &   0.24  &   0.928 &  35.74 &   0.24  &   0.955 &  67.85 &   0.24  &   0.892 &  36.87 &   0.23  \\ 
 & (15+15)$M_\odot$ &   0.861 &  24.52 &   0.25  &   0.893 &  25.58 &   0.25  &   0.932 &  53.46 &   0.17  &   0.926 &  26.82 &   0.25  &   0.920 &  51.38 &   0.24  &   0.921 &  27.99 &   0.24  \\ 
 & (15+5)$M_\odot$ &   0.822 &  15.40 &   0.25  &   0.867 &  15.81 &   0.25  &   0.790 &  16.59 &   0.25  &   0.903 &  15.81 &   0.25  &   0.839 &  51.91 &   0.07  &   0.955 &  16.03 &   0.25  \\ 
 & (5+5)$M_\odot$ &   0.814 &   9.52 &   0.25  &   0.839 &   9.59 &   0.25  &   0.941 &   9.63 &   0.25  &   0.838 &   9.52 &   0.25  &   0.872 &   9.80 &   0.25  &   0.866 &   9.61 &   0.25  \\ 
\hline 
T(3, 3.5, 0) & (20+20)$M_\odot$ &   0.925 &  40.09 &   0.24  &   0.918 &  42.90 &   0.24  &   0.940 &  80.76 &   0.24  &   0.833 &  57.71 &   0.24  &   0.958 &  89.85 &   0.24  &   0.840 &  73.84 &   0.25  \\ 
 & (15+15)$M_\odot$ &   0.955 &  29.98 &   0.24  &   0.937 &  30.78 &   0.24  &   0.887 &  58.83 &   0.24  &   0.996 &  32.67 &   0.20  &   0.914 &  66.56 &   0.24  &   0.758 &  31.32 &   0.24  \\ 
 & (15+5)$M_\odot$ &   0.983 &  19.68 &   0.18  &   0.985 &  20.97 &   0.16  &   0.926 &  69.81 &   0.05  &   0.999 &  19.47 &   0.19  &   0.887 &  60.02 &   0.07  &   1.000 &  19.79 &   0.18  \\ 
 & (5+5)$M_\odot$ &   0.992 &   9.99 &   0.24  &   0.997 &  10.40 &   0.22  &   0.826 &   9.83 &   0.25  &   0.993 &  10.48 &   0.22  &   0.749 &  10.07 &   0.25  &   0.995 &  10.81 &   0.21  \\ 
\hline 
P(2, 2.5) & (20+20)$M_\odot$ &   0.866 &  41.72 &   0.24  &   0.859 &  43.14 &   0.24  &   0.912 &  83.09 &   0.24  &   0.795 &  65.45 &   0.24  &   0.934 &  92.91 &   0.24  &   0.805 &  82.71 &   0.25  \\ 
 & (15+15)$M_\odot$ &   0.898 &  30.06 &   0.24  &   0.963 &  38.21 &   0.14  &   0.857 &  62.07 &   0.24  &   0.992 &  33.28 &   0.19  &   0.890 &  69.31 &   0.24  &   0.709 &  59.88 &   0.25  \\ 
 & (15+5)$M_\odot$ &   0.966 &  20.48 &   0.17  &   0.966 &  21.86 &   0.15  &   0.907 &  70.42 &   0.05  &   0.993 &  20.08 &   0.17  &   0.904 &  64.71 &   0.06  &   0.997 &  20.29 &   0.17  \\ 
 & (5+5)$M_\odot$ &   0.995 &   9.79 &   0.25  &   0.994 &  10.43 &   0.22  &   0.825 &   9.81 &   0.25  &   0.990 &  10.51 &   0.22  &   0.748 &  10.05 &   0.25  &   0.992 &  10.83 &   0.21  \\ 
\hline 
P(3, 3.5, 0) & (20+20)$M_\odot$ &   0.960 &  40.10 &   0.23  &   0.953 &  41.06 &   0.24  &   0.943 &  76.61 &   0.24  &   0.835 &  53.85 &   0.24  &   0.961 &  86.56 &   0.24  &   0.842 &  70.76 &   0.25  \\ 
 & (15+15)$M_\odot$ &   0.965 &  29.33 &   0.24  &   0.966 &  30.14 &   0.24  &   0.893 &  56.29 &   0.24  &   0.993 &  31.83 &   0.20  &   0.920 &  63.91 &   0.24  &   0.996 &  32.41 &   0.20  \\ 
 & (15+5)$M_\odot$ &   0.982 &  18.87 &   0.20  &   0.983 &  20.29 &   0.17  &   0.926 &  68.98 &   0.05  &   0.996 &  19.15 &   0.19  &   0.886 &  58.97 &   0.07  &   0.999 &  19.45 &   0.19  \\ 
 & (5+5)$M_\odot$ &   0.973 &   9.74 &   0.25  &   0.998 &   9.85 &   0.25  &   0.849 &   9.81 &   0.25  &   0.992 &  10.02 &   0.24  &   0.761 &  10.04 &   0.25  &   0.993 &  10.46 &   0.22  \\ 
\hline 
EP(2, 2.5) & (20+20)$M_\odot$ &  & &  &   0.996 &  41.72 &   0.24  &   0.953 &  75.09 &   0.24  &   0.929 &  47.51 &   0.24  &   0.948 &  84.61 &   0.24  &   0.907 &  59.72 &   0.24  \\ 
 & (15+15)$M_\odot$ &  & &  &   0.999 &  32.66 &   0.21  &   0.908 &  56.68 &   0.24  &   0.889 &  32.89 &   0.24  &   0.915 &  64.87 &   0.24  &   0.997 &  33.00 &   0.20  \\ 
 & (15+5)$M_\odot$ &  & &  &   0.999 &  21.35 &   0.16  &   0.909 &  70.41 &   0.05  &   0.992 &  19.52 &   0.19  &   0.858 &  64.23 &   0.06  &   0.986 &  20.00 &   0.18  \\ 
 & (5+5)$M_\odot$ &  & &  &   0.999 &  10.75 &   0.21  &   0.807 &   9.84 &   0.25  &   0.997 &  10.69 &   0.21  &   0.733 &  10.08 &   0.25  &   0.998 &  10.99 &   0.20  \\ 
\hline 
EP(3, 3.5, 0) & (20+20)$M_\odot$ &   0.995 &  38.25 &   0.25  &  & &  &   0.958 &  72.99 &   0.24  &   0.918 &  45.74 &   0.24  &   0.956 &  81.66 &   0.24  &   0.896 &  59.30 &   0.25  \\ 
 & (15+15)$M_\odot$ &   0.992 &  28.77 &   0.25  &  & &  &   0.938 &  70.37 &   0.14  &   0.999 &  31.41 &   0.21  &   0.922 &  61.77 &   0.24  &   1.000 &  32.11 &   0.21  \\ 
 & (15+5)$M_\odot$ &   0.999 &  18.53 &   0.20  &  & &  &   0.905 &  69.04 &   0.05  &   0.998 &  18.97 &   0.20  &   0.858 &  61.43 &   0.06  &   0.994 &  19.26 &   0.19  \\ 
 & (5+5)$M_\odot$ &   0.982 &   9.74 &   0.25  &  & &  &   0.832 &  10.00 &   0.24  &   0.996 &  10.24 &   0.23  &   0.748 &  10.06 &   0.25  &   0.997 &  10.61 &   0.22  \\ 
\hline 
HT(2, 2) & (20+20)$M_\odot$ &   0.794 &  21.34 &   0.25  &   0.815 &  22.35 &   0.25  &  & &  &   0.840 &  24.31 &   0.25  &   0.968 &  46.75 &   0.25  &   0.835 &  25.77 &   0.25  \\ 
 & (15+15)$M_\odot$ &   0.651 &  18.40 &   0.24  &   0.674 &  19.03 &   0.24  &  & &  &   0.377 &  37.58 &   0.25  &   0.936 &  36.99 &   0.24  &   0.392 &  47.22 &   0.25  \\ 
 & (15+5)$M_\odot$ &   0.624 &  14.96 &   0.25  &   0.632 &  15.15 &   0.25  &  & &  &   0.608 &  17.70 &   0.17  &   0.965 &  17.85 &   0.22  &   0.612 &  17.35 &   0.18  \\ 
 & (5+5)$M_\odot$ &   0.817 &   9.72 &   0.25  &   0.845 &   9.74 &   0.25  &  & &  &   0.845 &   9.74 &   0.25  &   0.841 &   9.97 &   0.25  &   0.865 &   9.76 &   0.25  \\ 
\hline 
HT(3, 3.5, 0) & (20+20)$M_\odot$ &   0.904 &  34.61 &   0.24  &   0.920 &  37.64 &   0.24  &   0.903 &  65.68 &   0.24  &  & &  &   0.873 &  74.44 &   0.25  &   0.999 &  41.41 &   0.23  \\ 
 & (15+15)$M_\odot$ &   0.891 &  27.49 &   0.25  &   0.926 &  28.59 &   0.25  &   0.883 &  49.56 &   0.24  &  & &  &   0.867 &  59.23 &   0.24  &   1.000 &  31.02 &   0.23  \\ 
 & (15+5)$M_\odot$ &   0.986 &  20.73 &   0.16  &   0.986 &  21.99 &   0.15  &   0.919 &  71.02 &   0.05  &  & &  &   0.886 &  61.90 &   0.07  &   1.000 &  20.34 &   0.17  \\ 
 & (5+5)$M_\odot$ &   0.964 &   9.75 &   0.25  &   0.993 &   9.79 &   0.25  &   0.834 &   9.83 &   0.25  &  & &  &   0.749 &  10.07 &   0.25  &   1.000 &  10.35 &   0.23  \\ 
\hline 
HP(2, 2.5) & (20+20)$M_\odot$ &   0.762 &  18.74 &   0.25  &   0.784 &  19.44 &   0.25  &   0.973 &  36.64 &   0.21  &   0.794 &  20.75 &   0.24  &  & &  &   0.801 &  21.53 &   0.25  \\ 
 & (15+15)$M_\odot$ &   0.595 &  16.37 &   0.24  &   0.617 &  16.40 &   0.24  &   0.931 &  27.84 &   0.21  &   0.329 &  40.09 &   0.25  &  & &  &   0.343 &  48.60 &   0.25  \\ 
 & (15+5)$M_\odot$ &   0.577 &  16.04 &   0.20  &   0.599 &  14.32 &   0.25  &   0.957 &  22.10 &   0.14  &   0.589 &  15.53 &   0.21  &  & &  &   0.593 &  15.59 &   0.21  \\ 
 & (5+5)$M_\odot$ &   0.741 &   9.50 &   0.25  &   0.754 &   9.53 &   0.25  &   0.975 &  11.46 &   0.18  &   0.755 &   9.52 &   0.25  &  & &  &   0.770 &   9.61 &   0.25  \\ 
\hline 
HP(3, 3.5, 0) & (20+20)$M_\odot$ &   0.832 &  31.43 &   0.25  &   0.840 &  35.15 &   0.25  &   0.850 &  60.63 &   0.25  &   0.974 &  37.71 &   0.25  &   0.806 &  72.61 &   0.25  &  & &  \\ 
 & (15+15)$M_\odot$ &   0.831 &  26.96 &   0.25  &   0.860 &  28.03 &   0.25  &   0.852 &  46.65 &   0.24  &   0.975 &  28.95 &   0.25  &   0.842 &  55.71 &   0.24  &  & &  \\ 
 & (15+5)$M_\odot$ &   0.986 &  20.13 &   0.17  &   0.986 &  21.50 &   0.15  &   0.922 &  70.24 &   0.05  &   1.000 &  19.64 &   0.18  &   0.884 &  60.67 &   0.07  &  & &  \\ 
 & (5+5)$M_\odot$ &   0.933 &   9.72 &   0.25  &   0.971 &   9.75 &   0.25  &   0.857 &   9.80 &   0.25  &   0.991 &   9.75 &   0.25  &   0.758 &  10.03 &   0.25  &  & &  \\ 
\hline 

%% file: tableVIII.tex
 & & \multicolumn{3}{c}{T$(2,2.5)$} & \multicolumn{3}{c}{ET$(2,2.5)$} & \multicolumn{3}{c}{T$(3,3.5,+2)$} & \multicolumn{3}{c}{T$(3,3.5,-2)$} & \multicolumn{3}{c}{ET$(3,3.5,+2)$} & \multicolumn{3}{c}{ET$(3,3.5,-2)$} \\ 
 & & \multicolumn{1}{c}{mm} & \multicolumn{1}{c}{$M$} & \multicolumn{1}{c}{$\eta$} & \multicolumn{1}{c}{mm} & \multicolumn{1}{c}{$M$} & \multicolumn{1}{c}{$\eta$} & \multicolumn{1}{c}{mm} & \multicolumn{1}{c}{$M$} & \multicolumn{1}{c}{$\eta$} & \multicolumn{1}{c}{mm} & \multicolumn{1}{c}{$M$} & \multicolumn{1}{c}{$\eta$} & \multicolumn{1}{c}{mm} & \multicolumn{1}{c}{$M$} & \multicolumn{1}{c}{$\eta$} & \multicolumn{1}{c}{mm} & \multicolumn{1}{c}{$M$} & \multicolumn{1}{c}{$\eta$} \\ 
\hline \hline 
& (15+15)$M_{\odot}$  &   &   &    &  0.914 &  27.58 &  0.248  &   &   &    &   &   &    &   &   &    &   &   &  \\ 
T$(2,2.5)$& (15+5)$M_{\odot}$  &   &   &    &  0.916 &  16.81 &  0.249  &   &   &    &   &   &    &   &   &    &   &   &  \\ 
& (5+5)$M_{\odot}$  &   &   &    &  0.900 &  10.13 &  0.241  &   &   &    &   &   &    &   &   &    &   &   &   \\  
 \hline 
& (15+15)$M_{\odot}$  &  0.922 &  33.93 &  0.241  &   &   &    &   &   &    &   &   &    &   &   &    &   &   &  \\ 
ET$(2,2.5)$& (15+5)$M_{\odot}$  &  0.971 &  33.17 &  0.076  &   &   &    &   &   &    &   &   &    &   &   &    &   &   &  \\ 
& (5+5)$M_{\odot}$  &  0.984 &  13.57 &  0.147  &   &   &    &   &   &    &   &   &    &   &   &    &   &   &   \\  
 \hline 
& (15+15)$M_{\odot}$  &   &   &    &   &   &    &   &   &    &  0.995 &  29.83 &  0.243  &  0.963 &  30.52 &  0.240  &  0.974 &  30.32 &  0.240\\ 
T$(3,3.5,+2)$& (15+5)$M_{\odot}$  &   &   &    &   &   &    &   &   &    &  1.000 &  19.06 &  0.204  &  0.984 &  20.03 &  0.186  &  0.974 &  20.09 &  0.182\\ 
& (5+5)$M_{\odot}$  &   &   &    &   &   &    &   &   &    &  0.981 &   9.96 &  0.250  &  0.991 &  10.16 &  0.242  &  0.972 &   9.94 &  0.250 \\  
 \hline 
& (15+15)$M_{\odot}$  &   &   &    &   &   &    &  0.998 &  30.94 &  0.242  &   &   &    &  0.951 &  31.27 &  0.239  &  0.960 &  30.59 &  0.241\\ 
T$(3,3.5,-2)$& (15+5)$M_{\odot}$  &   &   &    &   &   &    &  1.000 &  20.93 &  0.173  &   &   &    &  0.985 &  20.89 &  0.173  &  0.983 &  20.27 &  0.181\\ 
& (5+5)$M_{\odot}$  &   &   &    &   &   &    &  0.999 &  10.61 &  0.226  &   &   &    &  0.994 &  10.26 &  0.240  &  0.993 &  10.19 &  0.241 \\  
 \hline 
& (15+15)$M_{\odot}$  &   &   &    &   &   &    &  0.951 &  30.39 &  0.240  &  0.931 &  29.76 &  0.241  &   &   &    &  0.994 &  30.06 &  0.241\\ 
ET$(3,3.5,+2)$& (15+5)$M_{\odot}$  &   &   &    &   &   &    &  0.981 &  20.16 &  0.186  &  0.985 &  18.97 &  0.207  &   &   &    &  1.000 &  19.23 &  0.201\\ 
& (5+5)$M_{\odot}$  &   &   &    &   &   &    &  0.996 &  10.22 &  0.240  &  0.985 &   9.96 &  0.250  &   &   &    &  0.979 &   9.95 &  0.250 \\  
 \hline 
& (15+15)$M_{\odot}$  &   &   &    &   &   &    &  0.963 &  30.94 &  0.240  &  0.953 &  30.30 &  0.241  &  0.999 &  31.07 &  0.238  &   &   &  \\ 
ET$(3,3.5,-2)$& (15+5)$M_{\odot}$  &   &   &    &   &   &    &  0.983 &  20.65 &  0.179  &  0.980 &  20.32 &  0.182  &  1.000 &  20.83 &  0.175  &   &   &  \\ 
& (5+5)$M_{\odot}$  &   &   &    &   &   &    &  0.987 &  10.27 &  0.240  &  0.996 &  10.21 &  0.241  &  1.000 &  10.51 &  0.230  &   &   &  \\ 
 \hline 

%% file: tableIX.tex
 & & \multicolumn{3}{c}{EP$(3,3.5,2,-4,0)$} & \multicolumn{3}{c}{EP$(3,3.5,2,0,-4)$} & \multicolumn{3}{c}{EP$(3,3.5,2,0,0)$} & \multicolumn{3}{c}{EP$(3,3.5,2,0,4)$} & \multicolumn{3}{c}{EP$(3,3.5,2,4,0)$} \\ 
 & & \multicolumn{1}{c}{mm} & \multicolumn{1}{c}{$M$} & \multicolumn{1}{c}{$\eta$} & \multicolumn{1}{c}{mm} & \multicolumn{1}{c}{$M$} & \multicolumn{1}{c}{$\eta$} & \multicolumn{1}{c}{mm} & \multicolumn{1}{c}{$M$} & \multicolumn{1}{c}{$\eta$} & \multicolumn{1}{c}{mm} & \multicolumn{1}{c}{$M$} & \multicolumn{1}{c}{$\eta$} & \multicolumn{1}{c}{mm} & \multicolumn{1}{c}{$M$} & \multicolumn{1}{c}{$\eta$} \\ 
\hline \hline 
& (15+15)$M_{\odot}$  &   &   &    &  0.995 &  30.93 &  0.238  &  0.994 &  30.85 &  0.240  &  0.995 &  30.87 &  0.239  &  0.952 &  31.17 &  0.242\\ 
EP$(3,3.5,2,-4,0)$& (15+5)$M_{\odot}$  &   &   &    &  0.998 &  20.61 &  0.177  &  0.999 &  20.71 &  0.176  &  0.999 &  20.60 &  0.177  &  0.993 &  21.59 &  0.162\\ 
& (5+5)$M_{\odot}$  &   &   &    &  0.999 &  10.22 &  0.240  &  0.999 &  10.22 &  0.240  &  0.999 &  10.22 &  0.240  &  0.996 &  10.46 &  0.231 \\  
 \hline 
& (15+15)$M_{\odot}$  &  0.983 &  30.12 &  0.241  &   &   &    &  0.999 &  30.47 &  0.240  &  0.999 &  30.43 &  0.241  &  0.987 &  30.88 &  0.240\\ 
EP$(3,3.5,2,0,-4)$& (15+5)$M_{\odot}$  &  0.999 &  19.28 &  0.201  &   &   &    &  1.000 &  20.06 &  0.186  &  1.000 &  20.03 &  0.187  &  0.999 &  20.70 &  0.175\\ 
& (5+5)$M_{\odot}$  &  0.993 &  10.01 &  0.249  &   &   &    &  0.996 &  10.19 &  0.241  &  0.996 &  10.19 &  0.241  &  0.998 &  10.22 &  0.240 \\  
 \hline 
& (15+15)$M_{\odot}$  &  0.983 &  30.12 &  0.241  &  0.999 &  30.47 &  0.241  &   &   &    &  0.999 &  30.42 &  0.241  &  0.987 &  30.88 &  0.240\\ 
EP$(3,3.5,2,0,0)$& (15+5)$M_{\odot}$  &  0.999 &  19.26 &  0.202  &  1.000 &  20.06 &  0.186  &   &   &    &  1.000 &  20.03 &  0.187  &  0.999 &  20.70 &  0.175\\ 
& (5+5)$M_{\odot}$  &  0.993 &   9.99 &  0.250  &  1.000 &  10.00 &  0.250  &   &   &    &  0.996 &  10.19 &  0.241  &  0.998 &  10.22 &  0.240 \\  
 \hline 
& (15+15)$M_{\odot}$  &  0.982 &  30.12 &  0.241  &  0.999 &  30.54 &  0.240  &  0.999 &  30.54 &  0.240  &   &   &    &  0.987 &  30.88 &  0.240\\ 
EP$(3,3.5,2,0,4)$& (15+5)$M_{\odot}$  &  0.999 &  19.35 &  0.200  &  1.000 &  20.05 &  0.187  &  1.000 &  19.98 &  0.188  &   &   &    &  0.998 &  20.73 &  0.175\\ 
& (5+5)$M_{\odot}$  &  0.993 &  10.01 &  0.249  &  1.000 &  10.00 &  0.250  &  0.996 &  10.19 &  0.241  &   &   &    &  0.998 &  10.22 &  0.240 \\  
 \hline 
& (15+15)$M_{\odot}$  &  0.929 &  29.60 &  0.240  &  0.968 &  30.11 &  0.242  &  0.968 &  30.16 &  0.240  &  0.967 &  30.15 &  0.240  &   &   &  \\ 
EP$(3,3.5,2,4,0)$& (15+5)$M_{\odot}$  &  0.992 &  18.42 &  0.219  &  0.998 &  19.29 &  0.201  &  0.998 &  19.36 &  0.199  &  0.998 &  19.29 &  0.201  &   &   &  \\ 
& (5+5)$M_{\odot}$  &  0.970 &  10.17 &  0.241  &  0.993 &   9.99 &  0.250  &  0.993 &   9.99 &  0.250  &  0.993 &   9.99 &  0.250  &   &   &   \\  
 \hline 

%% file: tableETB1.tex
                     &   (20+20)$M_{\odot}$ &   221.4 &  0.983 &  23891. & $-554.63$   &   $ 0.949$   &   240.7 & (20+5)$M_{\odot}$ &   341.2 &  0.992 &  77508. &   $-1041.30$   &   $ 0.897$   &   347.0 \\ 
                     &   (20+15)$M_{\odot}$ &   252.4 &  0.987 &  30200. & $-606.41$   &   $ 0.975$   &   272.5 & (10+10)$M_{\odot}$ &   442.8 &  0.992 &  72639. &   $-768.78$   &   $ 0.632$   &   331.4 \\ 
T$(2, 2)$&(15+15)$M_{\odot}$ &   295.2 &  0.989 &  38126. & $-653.61$   &   $ 0.968$   &   313.5 & (15+5)$M_{\odot}$ &   431.3 &  0.993 &  96191. &   $-1030.20$   &   $ 0.831$   &   440.8 \\ 
                     &   (20+10)$M_{\odot}$ &   291.7 &  0.989 &  41735. & $-677.51$   &   $ 1.002$   &   314.2 & (10+5)$M_{\odot}$ &   583.4 &  0.993 &  130600. &   $-1019.10$   &   $ 1.001$   &   805.3 \\ 
                     &   (15+10)$M_{\odot}$ &   352.7 &  0.991 &  52565. & $-713.54$   &   $ 0.968$   &   387.1 & (5+5)$M_{\odot}$ &   885.6 &  0.989 &  225060. &   $-1056.80$   &   $ 0.531$   &   894.4 \\ 

 \hline 
                     &   (20+20)$M_{\odot}$ &   161.2 &  0.970 &  19807. & $  62.32$   &   $ 0.691$   &   224.4 & (20+5)$M_{\odot}$ &   281.6 &  0.987 &  71552. &   $-188.92$   &   $ 0.227$   &   312.7 \\ 
                     &   (20+15)$M_{\odot}$ &   185.9 &  0.975 &  25398. & $  57.59$   &   $ 0.347$   &   220.3 & (10+10)$M_{\odot}$ &   322.4 &  0.983 &  66783. &   $ -37.92$   &   $ 0.490$   &   630.9 \\ 
T$(2, 2.5)$&(15+15)$M_{\odot}$ &   214.9 &  0.979 &  32787. & $  40.11$   &   $ 0.210$   &   245.0 & (15+5)$M_{\odot}$ &   345.6 &  0.988 &  89296. &   $-166.70$   &   $ 0.107$   &   373.8 \\

%% file: tableXV.tex
 & (20+20)$M_\odot$ &   0.997 &  35.53 &   0.35 &   0.994 &  35.55 &   0.35 & (20+5)$M_\odot$ &   0.988 &  22.04 &   0.21 &   0.979 &  22.51 &   0.20 \\ 
 & (20+15)$M_\odot$ &   0.997 &  32.43 &   0.31 &   0.994 &  31.69 &   0.32 & (10+10)$M_\odot$ &   0.996 &  19.29 &   0.28 &   0.990 &  18.74 &   0.30 \\ 
T(2, 2) & (15+15)$M_\odot$ &   0.997 &  28.45 &   0.29 &   0.993 &  27.54 &   0.32 & (15+5)$M_\odot$ &   0.993 &  18.31 &   0.23 &   0.985 &  17.94 &   0.24 \\ 
 & (20+10)$M_\odot$ &   0.996 &  27.72 &   0.27 &   0.992 &  26.83 &   0.30 & (10+5)$M_\odot$ &   0.992 &  14.55 &   0.24 &   0.989 &  14.86 &   0.23 \\ 
 & (15+10)$M_\odot$ &   0.995 &  23.68 &   0.28 &   0.988 &  22.95 &   0.30 & (5+5)$M_\odot$ &   0.994 &  10.60 &   0.23 &   0.992 &  10.73 &   0.22 \\ 
\hline 
 & (20+20)$M_\odot$ &   0.821 &  18.77 &   0.94 &   0.962 &  22.59 &   0.65 & (20+5)$M_\odot$ &   0.958 &  11.66 &   0.63 &   0.987 &  12.81 &   0.53 \\ 
 & (20+15)$M_\odot$ &   0.862 &  16.60 &   0.94 &   0.966 &  19.40 &   0.68 & (10+10)$M_\odot$ &   0.948 &   9.96 &   0.89 &   0.984 &  10.71 &   0.77 \\ 
T(2, 2.5) & (15+15)$M_\odot$ &   0.891 &  14.26 &   0.97 &   0.969 &  16.61 &   0.71 & (15+5)$M_\odot$ &   0.965 &   9.72 &   0.69 &   0.989 &  10.26 &   0.62 \\

%% file: tableXVI.tex
 & (20+20)$M_\odot$ &   0.984 &  51.05 &   0.14 &   0.984 &  51.38 &   0.14 &   171.7 & (20+5)$M_\odot$ &   0.981 &  25.34 &   0.16 &   0.981 &  25.32 &   0.16 &   347.1 \\* 
 & (20+15)$M_\odot$ &   0.981 &  44.12 &   0.14 &   0.981 &  44.11 &   0.15 &   199.7 & (10+10)$M_\odot$ &   0.984 &  22.16 &   0.21 &   0.985 &  22.15 &   0.21 &   395.5 \\* 
T(2, 2) & (15+15)$M_\odot$ &   0.977 &  37.15 &   0.16 &   0.978 &  37.16 &   0.16 &   236.7 & (15+5)$M_\odot$ &   0.986 &  20.90 &   0.18 &   0.987 &  20.89 &   0.18 &   424.3 \\* 
 & (20+10)$M_\odot$ &   0.974 &  35.85 &   0.15 &   0.974 &  35.62 &   0.16 &   246.7 & (10+5)$M_\odot$ &   0.992 &  16.17 &   0.20 &   0.999 &  16.20 &   0.20 &   368.4 \\* 
 & (15+10)$M_\odot$ &   0.976 &  27.95 &   0.20 &   0.976 &  27.82 &   0.20 &   316.6 & (5+5)$M_\odot$ &   0.996 &  11.05 &   0.21 &   0.999 &  11.12 &   0.21 &   553.1 \\ 
\hline 
 & (20+20)$M_\odot$ &   0.948 &  24.94 &   0.57 &   0.985 &  24.09 &   0.60 &   202.5 & (20+5)$M_\odot$ &   0.975 &  14.60 &   0.41 &   0.975 &  14.52 &   0.42 &   567.2 \\* 
 & (20+15)$M_\odot$ &   0.956 &  21.50 &   0.58 &   0.990 &  20.77 &   0.62 &   241.9 & (10+10)$M_\odot$ &   0.983 &  11.72 &   0.65 &   0.995 &  11.48 &   0.69 &   415.0 \\* 
T(2, 2.5) & (15+15)$M_\odot$ &   0.965 &  18.08 &   0.62 &   0.986 &  18.76 &   0.57 &   279.0 & (15+5)$M_\odot$ &   0.983 &  11.38 &   0.51 &   0.994 &  11.50 &   0.50 &   448.1 \\